\documentclass[a4paper,11pt]{article}

\usepackage{jcappub} 
\usepackage{hyperref, cleveref}
\usepackage{amsfonts, amsmath, amssymb, latexsym}
\usepackage{graphicx}

\usepackage[T1]{fontenc} 
\usepackage{graphicx}
\usepackage{dcolumn}
\usepackage{bm}

\usepackage{xcolor}
\usepackage{footnote}
\usepackage{comment}

\usepackage{siunitx}
\DeclareSIUnit \parsec {pc}

\newcommand{\ar}[1]{\textcolor{red}{#1}}
\newcommand{\gp}[1]{\textcolor{orange}{#1}}

\title{Fully non-Gaussian Scalar-Induced Gravitational Waves}

\author[a,b]{Gabriele Perna,}
\author[a]{Chiara Testini,}
\author[c,d,a]{Angelo Ricciardone,}
\author[a,b,e,f]{Sabino Matarrese}

\affiliation[a]{Dipartimento di Fisica e Astronomia ``Galileo Galilei'', Universit\`a degli Studi di Padova,\\ Via Marzolo 8, I-35131, Padova, Italy}
\affiliation[b]{INFN, Sezione di Padova,\\ Via Marzolo 8, I-35131, Padova, Italy}
\affiliation[c]{Dipartimento di Fisica ``Enrico Fermi'', Universit\`a di Pisa, \\ Largo Bruno Pontecorvo 3, Pisa I-56127, Italy}
\affiliation[d]{INFN, Sezione di Pisa,\\ Largo Bruno Pontecorvo 3, Pisa I-56127, Italy}
\affiliation[e]{INAF - Osservatorio Astronomico di Padova, \\ Vicolo dell'Osservatorio 5, I-35122 Padova, Italy}
\affiliation[f]{Gran Sasso Science Institute, \\Viale F. Crispi 7, I-67100 L'Aquila, Italy}

\emailAdd{gabriele.perna@phd.unipd.it}
\emailAdd{chiara.testini@studenti.unipd.it}
\emailAdd{angelo.ricciardone@unipi.it}
\emailAdd{sabino.matarrese@pd.infn.it}

\abstract{Scalar-induced Gravitational Waves (SIGWs) represent a particular class of primordial signals which are sourced at second-order in perturbation theory whenever a scalar fluctuation of the metric is present. They form a guaranteed Stochastic Gravitational Wave Background (SGWB) that, depending on the amplification of primordial scalar fluctuations, can be detected by GW detectors. The amplitude and the frequency shape of the scalar-induced SGWB can be influenced by the statistical properties of the scalar density perturbations. In this work we study the intuitive physics behind SIGWs and we analyze the imprints of local non-Gaussianity of the primordial curvature perturbation on the GW spectrum. We consider all the relevant non-Gaussian contributions up to fifth-order in the scalar seeds without any hierarchy, and we derive the related GW energy density $\Omega_{\rm GW}(f)$. We perform a Fisher matrix analysis to understand to which accuracy non-Gaussianity can be constrained with the LISA detector, which will be sensitive in the milli-Hertz frequency band. We find that LISA, neglecting the impact of astrophysical foregrounds, will be able to measure the amplitude, the width and the peak of the spectrum with an accuracy up to $\mathcal{O}(10^{-4})$, while non-Gaussianity can be measured up to $\mathcal{O}(10^{-3})$. Finally, we discuss the implications of our non-Gaussianity expansion on the fraction of Primordial Black Holes.}

\begin{document}
\maketitle
\flushbottom

\section{Introduction}

Primordial Gravitational Waves (GWs) have the potential to probe the entire cosmological history, as a consequence of the weakness of the gravitational interaction: as soon as they are emitted, they decouple from the thermal bath and evolve almost independently on any intervening matter after the emission \cite{Contaldi:2016koz, Bartolo:2019oiq, Bartolo:2019yeu, Ricciardone:2021kel, Schulze:2023ich, LISACosmologyWorkingGroup:2022kbp}. Ongoing and planned observations of GWs span over 21 orders of magnitudes in frequency, starting from the Cosmic Microwave Background (CMB) at the lowest frequencies, up to interferometric detectors passing through Pulsar Timing Array (PTA) experiments \cite{2021MNRAS.505.4396K}. \par
Several mechanisms can generate GWs in the early universe \cite{Maggiore:2007ulw, Guzzetti:2016mkm, Cai:2017cbj, Caprini:2018mtu}. While all these mechanisms reside on some source modelling assumptions, there is a primordial guaranteed source of GWs: those generated by second-order scalar fluctuations, called Scalar-Induced GWs (SIGWs). 
SIGWs were firstly introduced in \cite{1967PThPh..37..831T, 1993PhRvD..47.1311M, 1994PhRvL..72..320M, 1998PhRvD..58d3504M, 2005PhRvD..71d3508C, 2007PhRvD..75l3518A}. When a second-order perturbation of the metric and of the stress-energy tensor is taken into account, the equation of motion, that at first order describes the dynamic of tensor modes as free propagating GW, gets a source term \cite{2016arXiv160501615C}. This is due to the fact that, going beyond linear order, scalar, vector and tensor perturbations are not independent among each other and thus, combination of first-order scalar perturbations may act as sources of GW (and viceversa \cite{Bari:2021xvf, Bari:2022grh}). Recently, also SIGWs sourced by scalar-tensor perturbations have been analyzed \cite{Bari:2023rcw, Picard:2023sbz}. \par

As shown in the past, such a SIGWB signal is sensitive to the underlying statistics of the sourcing curvature fluctuations, and in particular, to primordial non-Gaussianity (nG) \cite{Gangui:1993tt, Matarrese:2000iz, Bartolo:2001cw, Maldacena:2002vr, Bartolo:2003jx, Bartolo:2004if, Celoria:2018euj}. In all the works, mainly contributions up to $f_{\rm{NL}}$ on the scalar-induced GW spectrum have been explored \cite{Nakama:2016gzw, Garcia-Bellido:2017aan, Unal:2018yaa, Cai:2018dig, Cai:2019amo, Ragavendra:2020sop, Yuan:2020iwf, 2021JCAP...10..080A, Abe:2022xur} (see e.g., \cite{Yuan:2023ofl, Li:2023xtl} for recent papers which include contributions up to $g_{\rm{NL}}$ and \cite{Li:2024zwx} for a study on the bispectrum and trispectrum of induced GWs). However, contributions of the same order are expected when higher-order local nG terms (i.e., $h_{\rm{NL}},i_{\rm{NL}}$) are taken into account. So, we compute the GW spectral energy density considering all the relevant non-Gaussian contributions up to fifth-order in the scalar seeds, without assuming any hierarchy among the non-Gaussian parameters.  We derive analytically the various contributions starting from the 6- and 8-point correlation functions exploiting the Wick's theorem, and then we perform numerical high dimensional integrations via Monte Carlo methods \cite{2021JCoPh.43910386L}.

Primordial fluctuations, both tensor and density ones, are generated and stretched on super-horizon scales during inflation. The time of emission for the SIGWB is assumed to be when they re-enter the horizon in radiation domination. We assume a log-normal enhancement of the scalar power spectrum at the scales corresponding to these re-entering modes, which are much smaller than the one probed by CMB. We consider power spectrum amplitudes of $\mathcal{O}(10^{-2}, 10^{-3})$. Such an enhancement of the amplitude, far from CMB scales, is responsible also for the formation of Primordial Black Holes (PBHs) \cite{Carr:1974nx, Green:2014faa, Sato-Polito:2019hws, Bartolo:2018evs, Escriva:2020tak} and it could be generated by several physical mechanisms, such as features in the gravitational potential \cite{Yokoyama:1998pt, Cheng:2016qzb, Garcia-Bellido:2017mdw, Cheng:2018yyr, Dalianis:2018frf, Gao:2018pvq, Byrnes:2018txb, Tada:2019amh, Xu:2019bdp, Atal:2019erb, Mishra:2019pzq, Liu:2020oqe, Kefala:2020xsx}, modified gravity \cite{Kannike:2017bxn, Pi:2017gih, Fu:2019ttf, Dalianis:2019vit, Cheong:2019vzl, Fu:2019vqc, Lin:2020goi, Cheong:2020rao}, multi-field inflation \cite{Garcia-Bellido:1996mdl, Kawasaki:1997ju, Frampton:2010sw, Clesse:2015wea, Inomata:2017okj, Espinosa:2017sgp, Inomata:2018cht, Kawasaki:2019hvt, Palma:2020ejf, Fumagalli:2020adf, Braglia:2020eai}, curvaton scenarios \cite{Kawasaki:2012wr, Kohri:2012yw, Ando:2017veq, Ando:2018nge}, models with parametric resonance \cite{Cai:2018tuh, Cai:2019bmk, Cai:2019jah, Chen:2019zza, Chen:2020uhe}, etc (see \cite{2023arXiv231019857B} for a review). In these cases, PBHs can be generated in the very early stages of the universe, before any astrophysical object exists, if the presence of large initial fluctuations gives rise to regions where the gravitational potential exceeds (the kinetic energy of) the expansion of the universe. 
Since such PBHs form only in rare, large fluctuations, the number of PBHs is also very sensitive to the change in the shape of the tail of the fluctuation distribution,  which is regulated by the amount of non-Gaussianity. Similarly to SIGWs, also in the calculation of the abundance of PBH, truncating the expansion at order $f_{\rm{NL}}$, does not give accurate results \cite{Young:2013oia}. Further modifications can also derive from quantum diffusion which generates non-Gaussian exponential tails \cite{Ezquiaga:2019ftu,Ezquiaga:2022qpw}. 
Since SIGWs are generated approximately at the same epoch, one observable can be exploited to constrain the parameter space of the other \cite{Saito:2008jc, ZhengRuiFeng:2021zoz, Zhang:2021vak, Yuan:2021qgz, Yi:2022ymw, 2023PhRvD.107d3515Z}.  See also \cite{papanikolaou2024new} for a recent paper discussing a complementary way to constrain nG using PBH clustering properties.

SIGWs have been proposed as a possible explanation of the recent Pulsar Timing Array (PTA) collaboration detected signal \cite{2023ApJ...951L...8A, Figueroa:2023zhu, Franciolini:2023pbf, Balaji:2023ehk} and are among possible targets of the future space-based interferometer LISA \cite{LISACosmologyWorkingGroup:2022jok, 2023arXiv231019857B, Colpi:2024xhw}. For this reason, in the second part of the paper, we focus on understanding if the signal coming from SIGWs can be used to estimate the accuracy with which non-Gaussianity can be probed with LISA. In particular, accounting for the latest detector specifications, we perform a Fisher matrix analysis on the signal parameters \cite{Caprini:2019pxz, Flauger:2020qyi, Hartwig:2023pft}. Finally, we discuss the interplay between our new results and the PBH abundance, when all the nG contributions are taken into account, following \cite{Young:2022phe, Ferrante:2022mui}.

\vspace{0.1cm}

The paper is composed as follows. In Section \ref{inducedGWchapter} the intuitive physics behind scalar-induced GW and the construction of the power spectra are analyzed, starting from second-order perturbations of the metric. Studying the solution of the equation of motion of tensor perturbation, it is computed how the power spectrum of scalar-induced GW, which is proportional to the 4-point correlation function of primordial curvature perturbation. Then we analyze the imprints of local type non-Gaussianity of the primordial fluctuations on the GW spectrum, considering all the relevant non-Gaussian contributions up to fifth-order in the scalar seeds. Section \ref{resultsection} collects all the analytical and numerical results obtained, reporting the details for performing the integrations considering a lognormal parametrization for the primordial curvature power spectrum. The resulting spectra of GWs are then compared with the sensitivity of the LISA detector. In Section \ref{Fishersection} we address the detectability of the signal, trying to infer how well LISA will constrain signal and non-Gaussian parameters. In Section \ref{pbh section} we discuss PBH implications within our choice for the parameters of the model. Finally, in Section \ref{conclusionsection} we discuss the main results  and we summarize the achievements. Some Appendices are added to describe in details all the steps that has been done for the computation of the power spectra and the Feynman diagrams associated to each contribution in the spectrum.

\section{Second-order scalar-induced gravitational waves}
\label{inducedGWchapter}
The detectability of the SIGWB depends on the amplitude of the primordial density fluctuations since it comes from second-order term in cosmological perturbation theory \cite{2020JCAP...08..017D}. The GW spectral energy density today $\Omega_{\rm GW,0}$ induced by primordial scalar perturbations represents the observable quantity. In order to evaluate this, it is necessary to compute the Power Spectrum (PS) $\mathcal{P}_h$ associated to the induced GW starting from the general solution of the Equation of Motion (EoM) of second-order tensor perturbations derived from Einstein's equation. For the general formalism, the main references are \cite{1998PhRvD..58d3504M, 2007PhRvD..76h4019B, 1993PhRvD..47.1311M, 2005PhRvD..71d3508C, 2020JCAP...08..017D, 2021Univ....7..398D, 2007PhRvD..75l3518A, 2021JCAP...10..080A, 1997CQGra..14.2585B, Acquaviva:2002ud}. \par

\subsection{Second-order equation of motion}

To compute the induced tensor modes we follow the approach of \cite{2007PhRvD..76h4019B}, where the perturbed metric reads
\begin{equation}
    \begin{aligned}
    ds^2 = \hspace{0.2cm} a^2(\eta)\Bigg\{&-\left(1+2\Phi^{(1)}+\Phi^{(2)}\right)d\eta^2\\
    &+ 2\omega_i^{(2)}d\eta dx^i + \Big[(1-2\Psi^{(1)}-\Psi^{(2)})\delta_{ij}
    + \frac{1}{2}h_{ij}\Big]dx^idx^j\Bigg\}\,,
    \end{aligned}
\end{equation}
where first-order vector and tensor perturbations are ignored and $h_{ij} \equiv h_{ij}^{(2)}$.
The next step is to apply the projector tensor $\hat{\mathcal{T}}_{ij}^{lm}$ to the second-order Einstein equations
\begin{equation}
    \hat{\mathcal{T}}_{ij}^{lm}G_{l m}^{(2)} = 8\pi G \hat{\mathcal{T}}_{ij}^{lm}T^{(2)}_{l m}\,.
\end{equation}
The projector tensor extracts the transverse, traceless part of any tensor and eliminates the terms involving $\Phi^{(2)}$, $\Psi^{(2)}$, $\omega^{(2)}_i$ and the scalar and vector parts of the anisotropic stress $\Pi^{(2)i}\hspace{0.1cm}_j$ in the second-order Einstein equations.
In the absence of anisotropic stress $\Psi = \Phi$, the evolution equation for tensor modes results
\begin{equation}
    h_{ij}'' + 2\mathcal{H}h_{ij}' - \nabla^2h_{ij} = -4\hat{\mathcal{T}}_{ij}^{lm}\mathcal{S}_{lm}\,,
\end{equation}
with the source term $\mathcal{S}_{ij}(\textbf{x},\eta)$ given by \cite{2021JCAP...10..080A}
\begin{equation}
    \begin{aligned}
    \mathcal{S}_{ij}(\textbf{x},\eta) \equiv&\hspace{0.1cm} 4\Phi\partial_i\partial_j\Phi + \frac{2(1+3w)}{3(1+w)}\partial_i\Phi\partial_j\Phi - \frac{4}{3(1+w)\mathcal{H}^2}\left[\partial_i\Phi'\partial_j\Phi' + \mathcal{H}\partial_i\Phi\partial_j\Phi' + \mathcal{H}\partial_i\Phi'\partial_j\Phi\right]\,,
    \end{aligned}
    \label{source1}
\end{equation}
where $\Phi \equiv \Phi^{(1)} \equiv \Phi(\textbf{x},\eta)$ and $w$ is the equation of state parameter. Going to Fourier space we can write the tensor metric perturbation as
\begin{equation}
    h_{ij}(\textbf{x}, \eta) = \sum_{\lambda = +,\times}\int \frac{d^3\textbf{k}}{(2\pi)^{3/2}}e^{i\textbf{k}\cdot\textbf{x}}h_{\textbf{k}, \lambda}(\eta)\varepsilon_{ij}^{\lambda}(\hat{\textbf{k}})\,,
    \label{ft}
\end{equation}
where $\lambda = +,\times$ are the two GW polarizations and $\varepsilon_{ij}^{\lambda}(\hat{\textbf{k}})$ the polarization tensors. 
The scalar perturbation $\Phi(\textbf{k}, \eta)$ can be split into a transfer function, $\phi(k\eta)$ and a primordial curvature fluctuation $\mathcal{R}(\textbf{k})$  \cite{2007PhRvD..76h4019B, 2021Univ....7..398D, 2021JCAP...10..080A}. The curvature fluctuation is conserved on super-horizon scales and hence it provides a well-defined initial conditions in order to describe primordial perturbations \cite{2005JCAP...05..004L}. In the Newtonian gauge, the split of $\Phi(\textbf{k}, \eta)$, representing the Newtonian gravitational potential, can be written as
\begin{align}
    \Phi(\textbf{k},\eta) = \frac{3+3w}{5+3w}\phi(k\eta)\mathcal{R}(\textbf{k}) = \frac{b+2}{2b+3}\phi(k\eta)\mathcal{R}(\textbf{k})\,.
    \label{primordial}
\end{align}
In the last equation $b = (1-3w)/(1+3w)$ is a parameter, often used for convenience in the literature, which tells how much the equation of state of the universe differs from the one associated to radiation $w=1/3 \rightarrow b=0$: $b<0$ and $b>0$ correspond to a stiffer and a softer fluid respectively. The transfer function $\phi(k\eta)$ encodes the linear evolution of perturbations after horizon re-entry (see \cite{Domenech:2021ztg} for a recent review). \par
The primordial curvature fluctuation $\mathcal{R}(\textbf{k})$ is characterized by the power spectrum
\begin{equation}
    \langle \mathcal{R}(\textbf{k}) \mathcal{R}(\textbf{q}) \rangle = \delta^{(3)}(\textbf{k}+\textbf{q})P_{\mathcal{R}}(k)\,,
    \label{spettro}
\end{equation}
with $P_{\mathcal{R}}(k)$ parametrized by $P_{\mathcal{R}}(k) = \Delta_{\mathcal{R}}(k_0)\left(\frac{k}{k_0}\right)^{n_s-1}$, with $k_0$ the pivot scale and $n_s-1$ the spectral index. The latest constraints from Planck 2018 gave $\Delta^2_{\mathcal{R}}(k_0) \sim 2.2 \times 10^{-9}$  and  $n_s = 0.9649 \pm 0.0042$ \cite{2020A&A...641A..10P}.\par
In Fourier space, the EoM for the GW amplitude $h$ for each polarization state $\lambda$ becomes
\begin{equation}
    h''_{\lambda}(\textbf{k}, \eta) + 2\mathcal{H}h'_{\lambda}(\textbf{k}, \eta) + k^2h_{\lambda}(\textbf{k}, \eta) = 4\mathcal{S}_{\lambda}(\textbf{k}, \eta)\,,
    \label{eom1}
\end{equation}
where $\mathcal{S}_{\lambda}(\textbf{k}, \eta)$ encloses the FT of the source (\ref{source1}) given by
\begin{equation}
    \begin{aligned}
    \mathcal{S}_{\lambda}(\textbf{k}, \eta) \equiv& -\hspace{0.1cm}\varepsilon^{\lambda}_{lm}(\hat{\textbf{k}})S_{lm}(\textbf{k},\eta)\\  
    =& \int\frac{d^3\textbf{q}}{(2\pi)^{3/2}}Q_{\lambda}(\textbf{k},\textbf{q})f(|\textbf{k}-\textbf{q}|,q,\eta)\mathcal{R}(\textbf{q})\mathcal{R}(\textbf{k}-\textbf{q})\,.
    \end{aligned}
    \label{ftsource}
\end{equation}
In the last relation the projection factor
\begin{equation}
    Q_{\lambda}(\textbf{k},\textbf{q}) \equiv \varepsilon_{\lambda}^{ij}(\hat{\textbf{k}})q_iq_j
    \label{projection}
\end{equation}
encloses the angle between the two vectors $\textbf{k}$ and $\textbf{q}$. Taking $\hat{\textbf{k}}$ as a unit vector along the $\hat{z}$ direction and considering $\textbf{q} = q(\sin\theta \cos\varphi, \sin\theta \sin\varphi, \cos\theta)$ a generic vector with modulus $q$, equation (\ref{projection}) becomes
\begin{equation}
    Q_{\lambda}(\textbf{k},\textbf{q}) = \frac{q^2}{\sqrt{2}}\sin^2(\theta)\times
    \begin{cases}
    \cos(2\varphi), \hspace{0.2cm} \lambda = + \\
    \sin(2\varphi), \hspace{0.2cm}\lambda = \times
    \end{cases}\,.
    \label{proj}
\end{equation}
Recalling that $\mathcal{H} = 2/(\eta(1+3w))$ and defining $\textbf{p} = \textbf{k}-\textbf{q}$ to simplify the notation, the growing mode $f(p, q, \eta)$ in \eqref{ftsource} can be written as
\begin{equation}
\begin{aligned}
    f(p,q,\eta) =& \, \frac{3(1+w)}{(5+3w)^2}\left[2(5+3w)\phi(p\eta)\phi(q\eta)
    + \frac{4}{\mathcal{H}^2}\phi'(p\eta)\phi'(q\eta) \right.\nonumber\\
    &\phantom{\frac{3(1+w)}{(5+3w)^2}\,} \left.+ \frac{4}{\mathcal{H}}\left(\phi(p\eta)\phi'(q\eta)+\phi'(p\eta)\phi(q\eta)\right)\right] \\
    =& \,\frac{3(1+w)}{(5+3w)^2}\left[2(5+3w)\phi(p\eta)\phi(q\eta) + \eta^2(1+3w)^2\phi'(p\eta)\phi'(q\eta) \nonumber\right.\\
    &\phantom{\,\frac{3(1+w)}{(5+3w)^2}}\left.+ 2\eta(1+3w)\left(\phi(p\eta)\phi'(q\eta)+\phi'(p\eta)\phi(q\eta)\right)\right]\,.
    \end{aligned}
    \label{grow}
\end{equation}
This quantity encloses the transfer function $\phi(k\eta)$ and the dependence on the equation of state of the cosmic fluid $w$, arising when the splitting of the field \eqref{primordial} is included in the expression of the source term $S_{ij}(\textbf{k},\eta)$.

\subsection{Power spectrum}
\label{Sub::1.2}
Having a complete expression for the equation of motion, the next step consists in finding a solution for $h_\lambda$, in order to build the corresponding power spectrum, as anticipated before. Using the Green's method, in fact, we can build the particular solution of the evolution equation of induced tensor modes. After obtaining the solutions of the homogeneous equation $h_{\textbf{k},\lambda}'' + 2\mathcal{H}h_{\textbf{k},\lambda}' + k^2h_{\textbf{k},\lambda} = 0$, it reads (here we follow the notation of \cite{2021JCAP...10..080A})
\begin{equation}
    h_{\lambda}(\textbf{k}, \eta) = \frac{4}{a(\eta)} \int_{\eta_i}^{\eta}d\Bar{\eta}\, G_{\textbf{k}}(\eta,\Bar{\eta})a(\Bar{\eta})\mathcal{S}_{\lambda}(\textbf{k},\Bar{\eta})\,,
    \label{particular}
\end{equation}
where $G_{\textbf{k}}(\eta,\Bar{\eta})$ is the Green's function of the homogeneous solution. As reported in \cite{2007PhRvD..75l3518A}, equation \eqref{particular} is obtained assuming $h(\textbf{k}, \eta_i)= 0 $ and $h'(\textbf{k}, \eta_i)= 0 $ as initial conditions. This choice is consistent with the assumption that, neglecting possible non trivial cross-correlated spectrum by tensor-scalar-scalar couplings during inflation, primordial first-order tensor modes are uncorrelated with the first-order scalar modes and hence uncorrelated with the induced second-order tensor modes. This justifies to consider that there are no GWs before horizon re-entry \cite{2021Univ....7..398D}. Any
GW with primordial origin may be simply added to the solution \eqref{particular}.

Using now (\ref{ftsource}), the solution (\ref{particular}) can be rewritten as
\begin{equation}
    \begin{aligned}
    h_{\lambda}(\textbf{k}, \eta) = \frac{4}{a(\eta)} \int_{\eta_i}^{\eta}d\Bar{\eta} \, G_{\textbf{k}}(\eta,\Bar{\eta})a(\Bar{\eta})\int\frac{d^3q}{(2\pi)^{3/2}}Q_{\lambda}(\textbf{k},\textbf{q})f(|\textbf{k}-\textbf{q}|,q,\Bar{\eta})\mathcal{R}(\textbf{q})\mathcal{R}(\textbf{k}-\textbf{q})\,.
    \end{aligned}
\end{equation}
Furthermore, it is possible to define
\begin{equation}
    I(|\textbf{k}-\textbf{q}|,q,\eta) = \int_{\eta_i}^{\eta}d\Bar{\eta}\,G_{\textbf{k}}(\eta,\Bar{\eta})\frac{a(\Bar{\eta})}{a(\eta)}f(|\textbf{k}-\textbf{q}|,q,\Bar{\eta})\,,
    \label{kernel}
\end{equation}
which is called \textit{kernel} or \textit{transfer function}. The particular solution of the evolution equation of the scalar-induced tensor modes finally becomes
\begin{equation}
    h_{\lambda}(\textbf{k}, \eta) = 4\int\frac{d^3q}{(2\pi)^{3/2}}Q_{\lambda}(\textbf{k},\textbf{q})I(|\textbf{k}-\textbf{q}|,q,\eta)\mathcal{R}(\textbf{q})\mathcal{R}(\textbf{k}-\textbf{q}) \,.
\end{equation}
Hence the 2-point correlation function of the induced GW yields
\begin{equation}
    \begin{aligned}
    \langle h_{\lambda_{1}}(\textbf{k}_1,\eta_1)h_{\lambda_{2}}(\textbf{k}_2,\eta_2)\rangle \hspace{0.1cm} =& \hspace{0.1cm} 16\int\frac{d^3q_1}{(2\pi)^{3/2}}\int\frac{d^3q_2}{(2\pi)^{3/2}}Q_{\lambda_{1}}(\textbf{k}_1,\textbf{q}_1)I(|\textbf{k}_1-\textbf{q}_1|,q_1,\eta_1)\nonumber\\
    &\hspace{1.3cm}\times Q_{\lambda_{2}}(\textbf{k}_2,\textbf{q}_2)I(|\textbf{k}_2-\textbf{q}_2|,q_2,\eta_2)\nonumber\\
    &\hspace{1.3cm}\times \langle\mathcal{R}(\textbf{q}_1)\mathcal{R}(\textbf{k}_1-\textbf{q}_1)\mathcal{R}(\textbf{q}_2)\mathcal{R}(\textbf{k}_2-\textbf{q}_2)\rangle
    \end{aligned}
    \label{source}
\end{equation}
where $\langle\dots\rangle$ is the ensemble average that can be computed exploiting Wick's theorem \cite{lim}. As shown in the last equation, the 2-point correlation function shows that the induced GW spectrum depends on the 4-point function of the primordial curvature perturbation $\mathcal{R}$.\par
Assuming statistical homogeneity and isotropy on the curvature perturbation and $\langle \mathcal{R} \rangle = 0$, the 4-point correlation function can be decomposed into a \textit{disconnected} contribution and a \textit{connected} one as
\begin{equation}
\label{Eq::Connected_and_Disconnected}
\langle\mathcal{R}(\textbf{k}_1)\mathcal{R}(\textbf{k}_2)\mathcal{R}(\textbf{k}_3)\mathcal{R}(\textbf{k}_4)\rangle = \langle\mathcal{R}(\textbf{k}_1)\mathcal{R}(\textbf{k}_2)\mathcal{R}(\textbf{k}_3)\mathcal{R}(\textbf{k}_4)\rangle_d +\langle\mathcal{R}(\textbf{k}_1)\mathcal{R}(\textbf{k}_2)\mathcal{R}(\textbf{k}_3)\mathcal{R}(\textbf{k}_4)\rangle_c\,.
\end{equation}
The \textit{connected} component is defined in terms of the primordial connected trispectrum $T_{\mathcal{R}}$, defined as
\begin{equation}
    \langle\mathcal{R}(\textbf{k}_1)\mathcal{R}(\textbf{k}_2)\mathcal{R}(\textbf{k}_3)\mathcal{R}(\textbf{k}_4)\rangle_c =\hspace{0.1cm} \delta^{(3)}(\textbf{k}_1+\textbf{k}_2+\textbf{k}_3+\textbf{k}_4) T_{\mathcal{R}}(\textbf{k}_1,\textbf{k}_2,\textbf{k}_3,\textbf{k}_4)\,.
\end{equation}
Thus the corresponding connected 2-point function of the induced GW becomes
\begin{equation}
    \begin{aligned}
    \langle h_{\lambda_{1}}(\textbf{k}_1,\eta_1)h_{\lambda_{2}}(\textbf{k}_2,\eta_2)\rangle_{c} \hspace{0.1cm} =& \hspace{0.2cm}\delta^{(3)}(\textbf{k}_1 + \textbf{k}_2)\delta^{\lambda_{1}\lambda_{2}}\delta^{\eta_{1}\eta_{2}}P_{h,\lambda}(k,\eta)\big|_{c} \\
    =& \hspace{0.1cm}16\int\frac{d^3q_1}{(2\pi)^{3/2}}\int\frac{d^3q_2}{(2\pi)^{3/2}}Q_{\lambda_{1}}(\textbf{k}_1,\textbf{q}_1)I(|\textbf{k}_1-\textbf{q}_1|,q_1,\eta_1)Q_{\lambda_{2}}(\textbf{k}_2,\textbf{q}_2)\\ 
    &\hspace{0.7cm}\times I(|\textbf{k}_2-\textbf{q}_2|,q_2,\eta_2)\delta^{(3)}(\textbf{k}_1+\textbf{k}_2)T_\mathcal{R}(\textbf{q}_1,\textbf{k}_1-\textbf{q}_1,\textbf{q}_2,\textbf{k}_2-\textbf{q}_2)\,.
    \end{aligned}
\end{equation}
Applying now the effects of the delta and imposing $\textbf{k}_1 = -\textbf{k}_2 = \textbf{k}$, $\lambda_1 = \lambda_2 = \lambda$ and $\eta_1 = \eta_2 = \eta$, the connected PS reads
\begin{equation}
    \begin{aligned}
    P_{h,\lambda}(k,\eta)\big|_{c} =& \hspace{0.1cm} 16\int\frac{d^3q_1}{(2\pi)^{3/2}}\int\frac{d^3q_2}{(2\pi)^{3/2}}Q_{\lambda}(\textbf{k},\textbf{q}_1)I(|\textbf{k}-\textbf{q}_1|,q_1,\eta)Q_{\lambda}(\textbf{k},\textbf{q}_2)\\
    &\hspace{1.1cm}\times I(|\textbf{k}-\textbf{q}_2|,q_2,\eta)T_\mathcal{R}(\textbf{q}_1,\textbf{k}-\textbf{q}_1,-\textbf{q}_2,\textbf{q}_2-\textbf{k})\,.
    \end{aligned}
\end{equation}
When $\mathcal{R}$ is a Gaussian field this contribution vanishes, since the only connected n-point function of a Gaussian field is the 2-point one (in the next section, when local nG is considered, the contributions to the connected trispectrum will be computed, instead).

The \textit{disconnected} part, as already mentioned, can be directly decomposed using the Wick's theorem. Considering all the possible combinations it reads
\begin{equation}
    \begin{aligned}
    \langle\mathcal{R}(\textbf{q}_1)\mathcal{R}(\textbf{k}_1-\textbf{q}_1)\mathcal{R}(\textbf{q}_2)\mathcal{R}(\textbf{k}_2-\textbf{q}_2)\rangle_{d} =\hspace{0.1cm} &\langle\mathcal{R}(\textbf{q}_1)\mathcal{R}(\textbf{k}_1-\textbf{q}_1)\rangle\langle\mathcal{R}(\textbf{q}_2)\mathcal{R}(\textbf{k}_2-\textbf{q}_2)\rangle \\
    &+\langle\mathcal{R}(\textbf{q}_1)\mathcal{R}(\textbf{q}_2)\rangle\langle\mathcal{R}(\textbf{k}_1-\textbf{q}_1)\mathcal{R}(\textbf{k}_2-\textbf{q}_2)\rangle\\
    &+\langle\mathcal{R}(\textbf{q}_1)\mathcal{R}(\textbf{k}_2-\textbf{q}_2)\rangle\langle\mathcal{R}(\textbf{k}_1-\textbf{q}_1)\mathcal{R}(\textbf{q}_2)\rangle\,.
    \end{aligned}
\end{equation}
Exploiting the properties of the deltas and the projection factors $Q_{\lambda}(\textbf{k},\textbf{q})$, we obtain the disconnected 2-point function of the induced GW
\begin{equation}
    \begin{aligned}
    \langle h_{\lambda_{1}}(\textbf{k}_1,\eta_1)&h_{\lambda_{2}}(\textbf{k}_2,\eta_2)\rangle_{d} \hspace{0.1cm} = \hspace{0.1cm} \delta^{(3)}(\textbf{k}_1 + \textbf{k}_2)\delta^{\lambda_{1}\lambda_{2}}\delta^{\eta_{1}\eta_{2}}P_{h,\lambda}(k,\eta)\big|_{d} \\
    =& \hspace{0.1cm} 16\int\frac{d^3q_2}{(2\pi)^{3}}Q_{\lambda_{1}}(\textbf{k}_1,\textbf{q}_2)I(q_2,|\textbf{k}_1+\textbf{q}_2|,\eta_1)Q_{\lambda_{2}}(\textbf{k}_2,\textbf{q}_2)I(|\textbf{k}_2-\textbf{q}_2|,q_2,\eta_2)\\
    & \hspace{1.1cm}\times \delta^{(3)}(\textbf{k}_1+\textbf{k}_2)P_{\mathcal{R}}(q_2)P_{\mathcal{R}}(|\textbf{k}_1+\textbf{q}_2|)\\
    &+ 16\int\frac{d^3q_2}{(2\pi)^{3}}Q_{\lambda_{1}}(\textbf{k}_1,\textbf{q}_2)I(|\textbf{k}_1+\textbf{q}_2|,q_2,\eta_1)Q_{\lambda_{2}}(\textbf{k}_2,\textbf{q}_2)I(|\textbf{k}_2-\textbf{q}_2|,q_2,\eta_2)\\
    &\hspace{1.1cm}\times \delta^{(3)}(\textbf{k}_1+\textbf{k}_2)P_{\mathcal{R}}(q_2)P_{\mathcal{R}}(|\textbf{k}_1+\textbf{q}_2|)\,.
    \end{aligned}
\end{equation}
Rewriting the last equation and calling $\textbf{q}_2 \rightarrow \textbf{q}$, the disconnected PS finally reads
\begin{equation}
\label{Eq::Ph_Gaussian}
    \begin{aligned}
    P_{h,\lambda}(k,\eta)\big|_{d} = 32\int\frac{d^3q}{(2\pi)^{3}}&Q_{\lambda}^2(\textbf{k},\textbf{q})I^2(q,|\textbf{k}-\textbf{q}|,\eta)P_{\mathcal{R}}(q)P_{\mathcal{R}}(|\textbf{k}-\textbf{q}|) \,.
    \end{aligned}
\end{equation}
This contribution is the only one which survives when the simplest case of a Gaussian primordial scalar field $\mathcal{R}$ is considered. We further specify, as underlined in \cite{PhysRevD.99.041301}, that the leading order GW power spectrum arises from a 1-loop Feynman diagram, in turn originating from the tree-level contributions to $P_{\mathcal{R}}$, as we further discuss in Appendix \ref{Sec::Feynman_Diagrams}. 

\par

In full generality it is possible to define the dimensionless power spectrum $\Delta^2(k)$ associated to $h_\lambda$ as
\begin{equation}
    \Delta^2(k) = \frac{k^3}{2\pi^2}P(k)\,,
\end{equation} 
i.e. the logarithmic interval contribution to the variance $\sigma^2$ of the stochastic field \cite{Maggiore:2018sht} 
\begin{equation}
    \sigma^2 = \langle\delta^2(\textbf{x})\rangle = \int \frac{d^3k}{(2\pi)^3} P(k) = \int \frac{dk}{k} \Delta^2(k)\,.
\end{equation}
The 2-point correlation function of scalar-induced GW becomes
\begin{equation}
    \begin{aligned}
    \langle h_{\lambda_{1}}(\textbf{k}_1,\eta)h_{\lambda_{2}}(\textbf{k}_2,\eta)\rangle =& \hspace{0.1cm}\delta^{(3)}(\textbf{k}_1 +\textbf{k}_2)\delta^{\lambda_{1}\lambda_{2}}P_{h,\lambda}(k,\eta)\\
    =&\hspace{0.1cm}\delta^{(3)}(\textbf{k}_1 +\textbf{k}_2)\delta^{\lambda_{1}\lambda_{2}}\frac{2\pi^2}{k^3}\Delta^2_{h,\lambda}(k,\eta)\,.
    \end{aligned}
\end{equation}
As a last step, we obtain $\Omega_{\rm GW}(k,\eta)$, which constitutes the observationally relevant quantity for the SGWB \cite{2021JCAP...10..080A,2018PhRvD..97l3532K}. The spatially averaged energy density of GW on sub-horizon scales results \cite{2018JCAP...09..012E}
\begin{equation}
    \rho_{\rm GW}(\eta) = \int d \ln k \rho_{\rm GW}(k,\eta) =  \frac{M_{Pl}^2}{16a^2(\eta)}\langle\overline{\left(\nabla h_{ij}\right)^2}\rangle\,,
\end{equation}
where $M_{Pl} = (8\pi G)^{-1/2}$ is the reduced Planck mass, $\rho_{\rm GW}(\eta)$ is the total energy density of GW filling the universe (or GW power) and the over-bar denotes a time average. Hence \cite{2021JCAP...10..080A}
\begin{equation}
    \Omega_{\rm GW}(k,\eta) \equiv \frac{\rho_{\rm GW}(k,\eta)}{\rho_{tot}(\eta)} = \frac{1}{48} \left(\frac{k}{a(\eta)H(\eta)}\right)^2\sum_{\lambda=+,\times}\overline{\Delta_{h,\lambda}^2(k,\eta)}\,,
    \label{omega}
\end{equation}
with $\Omega_{\rm GW}(k,\eta)$ the fractional energy density in GW per logarithmic wavenumber.
The oscillation average of the dimensionless power spectrum $\overline{\Delta_{h,\lambda}^2(k,\eta)}$ is reflected in the oscillation average of the kernel (\ref{kernel}) $\overline{I(|\textbf{k}-\textbf{q}|,q,\eta)}$, since it encloses all the time dependencies of the induced GW. The observable spectrum today $\Omega_{\rm GW,0}(k)$, assuming that the emission of GW occurs after the reheating phase, is obtained considering that GW redshift as radiation with the expansion of the universe and hence \cite{Caprini:2018mtu}
\begin{equation}
    \Omega_{\rm GW,0}(k)h^2 = \Omega_{rad,0}h^2\left(\frac{g_{*,0}}{g_{*,e}}\right)^{1/3}\Omega_{\rm GW,e}(k)\,,
    \label{altempoattuale}
\end{equation}
observed at the present-day frequency
\begin{equation}
    f = \frac{k/2\pi a_e}{\sqrt{H_e M_{Pl}}}\left(\Omega_{rad,0}H_0^2M_{Pl}^2\right)^{1/4}\left(\frac{g_{*,0}}{g_{*,e}}\right)^{1/12}
\end{equation}
for a given comoving momentum $k$. In this expression $h$ is the reduced Hubble constant $h = H_0/100\hspace{0.1cm} \si{\km \,\s^{-1}/\mega\parsec}$, $g_{*}$ is the number of relativistic degrees of freedom (at present day and at the time of emission) in the energy density and $\Omega_{rad,0}$ represents the present-day amount of radiation, which from current observations is $\Omega_{rad,0}h^2 \sim 4.2 \times 10^{-5}$ \cite{2020A&A...641A...6P}.

\subsection{Local model of non-Gaussianity}
\label{SubSec::Local_nG}
The imprint of primordial nG on the spectrum of SIGW has been already studied in previous works (see e.g. \cite{2018PhRvD..97l3532K, Cai:2018dig, 2019JCAP...10..059C, Hajkarim:2019nbx, Atal:2021jyo, Yuan:2020iwf, Domenech:2021and, 2023JCAP...03..057G, Liu:2023ymk,  Yuan:2023ofl, Li:2023xtl}). In this work we consider the effect of local primordial nG mainly following \cite{Unal:2018yaa, 2021JCAP...10..080A}. One of the main assumptions often made when considering local nG is to assume a hierarchical scaling for the nG parameters, with $f_{\rm NL} \ll 1$, $g_{\rm NL} \propto O(f_{\rm NL}^2)$ and so on. As a consequence, some contributions to the final SIGW spectrum are neglected. In order to be self-consistent up to the 8-point correlation function\footnote{This sentence will be clear in the following, after we perform the full calculation of the tensor power spectrum contributions.} we relax the hierarchical scaling of the nG parameters and we consider an expansion of $\mathcal{R}$ up to the fifth order
\begin{equation}
\label{Eq::Phi_Exp}
\begin{aligned}
    \mathcal{R}(\textbf{x}) =& \hspace{0.1cm}\mathcal{R}_g(\textbf{x}) + \frac{3}{5}f_{\rm NL}(\mathcal{R}_g^2(\textbf{x})-\langle\mathcal{R}_g^2\rangle)+ \frac{9}{25}g_{\rm NL}\mathcal{R}_g^3(\textbf{x})\\
    &+\frac{27}{125}h_{\rm NL}(\mathcal{R}_g^4(\textbf{x})-3\langle\mathcal{R}_g^2\rangle^2)+\frac{81}{625}i_{\rm NL}\mathcal{R}_g^5(\textbf{x})\,, 
\end{aligned}
\end{equation}
where $f_{\rm NL}$, $g_{\rm NL}$, $h_{\rm NL}$ and $i_{\rm NL}$ are free parameters of the expansion and $\langle\mathcal{R}_g^2\rangle$ represents the variance of the Gaussian field $\mathcal{R}_g$. The terms $-3f_{\rm NL}\langle\mathcal{R}_g^2\rangle/5$ and $-81h_{\rm NL}\langle\mathcal{R}^2_g\rangle^2/125$ are added to maintain the mean value of $\mathcal{R}(\textbf{x})$ unchanged, meaning that $\langle\mathcal{R}\rangle = \langle\mathcal{R}_g\rangle = 0$. As it will be clear in the following, the expansion is truncated at fifth order in order to account for all the possible contributions to the 8-point correlation function, arising when considering next-to-next-to leading order corrections to the spectrum. 
The factors $\left(3/5\right)^\alpha$ are related to the local nG expansion of the Bardeen potential $\Phi$. After expanding $\Phi$ in terms of the nG parameters one can match with the expansion of $\mathcal{R}$ in matter domination, where $\Phi = \frac{3}{5} \mathcal{R}$ at first order \cite{Bartolo:2004if}. It is then always possible to redefine the nG parameters in the $\mathcal{R}$ expansion reabsorbing the $\frac{3}{5}$ factors, thus obtaining\footnote{In the following we will keep calling the nG parameters as $f_{\rm NL}$, $g_{\rm NL}$, \dots. Actually to make order we could call the ones in \eqref{Eq::Phi_Exp} as $f^\Phi_{\rm NL}$, $g^\Phi_{\rm NL}$, while the ones in \eqref{Eq::R_Exp} as $f^{\mathcal{R}}_{\rm NL}$, $g^{\mathcal{R}}_{\rm NL}$, \dots. As stated in the main text the two are related by
\begin{equation}
    f^{\mathcal{R}}_{\rm NL} = \frac{3}{5} f^\Phi_{\rm NL}\,,\quad g^{\mathcal{R}}_{\rm NL} = \frac{9}{25} g^\Phi_{\rm NL}\,,\quad\dots
\end{equation}
In the following we will actually consider the coefficients associated to ``$\mathcal{R}$'', dropping the label just to ease the notation. } 
\begin{equation}
\label{Eq::R_Exp}
    \mathcal{R}(\textbf{x}) = \hspace{0.1cm}\mathcal{R}_g(\textbf{x}) + f_{\rm NL}(\mathcal{R}_g^2(\textbf{x})-\langle\mathcal{R}_g^2\rangle) + g_{\rm NL}\mathcal{R}_g^3(\textbf{x})+ h_{\rm NL}(\mathcal{R}_g^4(\textbf{x})-3\langle\mathcal{R}_g^2\rangle^2)+i_{\rm NL}\mathcal{R}_g^5(\textbf{x})\,.
\end{equation}

Substituting this latter equation in \eqref{Eq::Connected_and_Disconnected}, and repeating the same calculation reported in Subsection \ref{Sub::1.2} one can obtain the nG corrections to \eqref{Eq::Ph_Gaussian}. We will address in the following the nG corrections, starting with those at order $\mathcal{A}_{\mathcal{R}}^3$ and subsequently examining the ones at order $\mathcal{A}_{\mathcal{R}}^4$. Hence, considering again that $\langle h_{\lambda_{1}}(\textbf{k}_1,\eta_1)h_{\lambda_{2}}(\textbf{k}_2,\eta_2)\rangle = \delta^{(3)}(\textbf{k}_1 + \textbf{k}_2)\delta^{\lambda_{1}\lambda_{2}}\delta^{\eta_{1}\eta_{2}}P_{h,\lambda}(k,\eta)$, and imposing $\textbf{k}_1 = -\textbf{k}_2 = \textbf{k}$, $\eta_1 = \eta_2 = \eta$ and $\lambda_1 = \lambda_2 = \lambda$, we obtain (the Gaussian contribution is written again for completeness) 
\begin{equation}
    \begin{aligned}
    P_{h,\lambda}(k,\eta)\big|_{\rm Gaussian} = \hspace{0.1cm} 2^5 \int \frac{d^3 q}{(2\pi)^3}Q^2_{\lambda}(\textbf{k},\textbf{q})I^2(|\textbf{k}-\textbf{q}|,q,\eta) P_{\mathcal{R}_g}(q)P_{\mathcal{R}_g}(|\textbf{k}-\textbf{q}|)\,;
    \end{aligned}
    \label{g}
\end{equation}

\begin{equation}
    \begin{aligned}
    P_{h,\lambda}(k,\eta)\big|_{\rm t} =& \hspace{0.1cm} 2^8 f_{\rm NL}^2 \int \frac{d^3 q_1}{(2\pi)^3} \int \frac{d^3 q_2}{(2\pi)^3}Q_{\lambda}(\textbf{k},\textbf{q}_1)Q_{\lambda}(\textbf{k},\textbf{q}_2)I(|\textbf{k}-\textbf{q}_1|,q_1,\eta)I(|\textbf{k}-\textbf{q}_2|,q_2,\eta)\\
    &\hspace{3.5cm}\times  P_{\mathcal{R}_g}(q_2)P_{\mathcal{R}_g}(|\textbf{k}-\textbf{q}_2|)P_{\mathcal{R}_g}(|\textbf{q}_1-\textbf{q}_2|)\,;
    \end{aligned}
    \label{t}
\end{equation}

\begin{equation}
    \begin{aligned}
    P_{h,\lambda}(k,\eta)\big|_{\rm u} =& \hspace{0.1cm} 2^8 f_{\rm NL}^2 \int \frac{d^3 q_1}{(2\pi)^3} \int \frac{d^3 q_2}{(2\pi)^3}Q_{\lambda}(\textbf{k},\textbf{q}_1)Q_{\lambda}(\textbf{k},\textbf{q}_2) I(|\textbf{k}-\textbf{q}_1|,q_1,\eta)I(|\textbf{k}-\textbf{q}_2|,q_2,\eta)\\
    &\hspace{3.5cm}\times P_{\mathcal{R}_g}(q_1)P_{\mathcal{R}_g}(q_2)P_{\mathcal{R}_g}(|\textbf{k}-(\textbf{q}_1+\textbf{q}_2)|)\,;
    \end{aligned}
    \label{u}
\end{equation}
\begin{equation}
    \begin{aligned}
    P_{h,\lambda}(k,\eta)\big|_{\rm s} =& \hspace{0.1cm} 2^8 f_{\rm NL}^2 \int \frac{d^3 q_1}{(2\pi)^3} \int \frac{d^3 q_2}{(2\pi)^3}Q_{\lambda}(\textbf{k},\textbf{q}_1)Q_{\lambda}(\textbf{k},\textbf{q}_2)I(|\textbf{k}-\textbf{q}_1|,q_1,\eta)I(|\textbf{k}-\textbf{q}_2|,q_2,\eta) \\
    &\hspace{3.5cm}\times P_{\mathcal{R}_g}(q_1)P_{\mathcal{R}_g}(q_2)P_{\mathcal{R}_g}(k)\,;
    \end{aligned}
    \label{s}
\end{equation}
\begin{equation}
    \begin{aligned}
    P_{h,\lambda}(k,\eta)\big|_{\rm g_{\rm NL}} =& \hspace{0.1cm} 3 \cdot 2^7 g_{\rm NL} \int \frac{d^3 q_1}{(2\pi)^3} \int \frac{d^3 q_2}{(2\pi)^3}Q_{\lambda}(\textbf{k},\textbf{q}_1)
    Q_{\lambda}(\textbf{k},\textbf{q}_2) I(|\textbf{k}-\textbf{q}_1|,q_1,\eta)\\
    &\hspace{2.5cm} \times I(|\textbf{k}-\textbf{q}_2|,q_2,\eta) P_{\mathcal{R}_g}(q_1)P_{\mathcal{R}_g}(q_2)P_{\mathcal{R}_g}(|\textbf{k}-\textbf{q}_1|)\,;
    \end{aligned}
    \label{gnl}
\end{equation}
\begin{equation}
    \begin{aligned}
    P_{h,\lambda}(k,\eta)\big|_{\rm hybrid} = \hspace{0.1cm} 2^7 f_{\rm NL}^2 \int& \frac{d^3 q_1}{(2\pi)^3}\int \frac{d^3 q_2}{(2\pi)^3}Q^2_{\lambda}(\textbf{k},\textbf{q}_1)I^2(|\textbf{k}-\textbf{q}_1|,q_1,\eta)\\
    & \times P_{\mathcal{R}_g}(|\textbf{k}-\textbf{q}_1|)P_{\mathcal{R}_g}(q_2)P_{\mathcal{R}_g}(|\textbf{q}_1-\textbf{q}_2|)\,;
    \end{aligned}
    \label{ibrid}
\end{equation}
\begin{equation}
    \begin{aligned}
    P_{h,\lambda}(k,\eta)\big|_{\rm new} =& \hspace{0.1cm} 3\cdot 2^7 g_{\rm NL} \int \frac{d^3 q_1}{(2\pi)^3}\int \frac{d^3 q_2}{(2\pi)^3}Q^2_{\lambda}(\textbf{k},\textbf{q}_1) I^2(|\textbf{k}-\textbf{q}_1|,q_1,\eta)\\
    &\hspace{3cm}\times P_{\mathcal{R}_g}(q_1)P_{\mathcal{R}_g}(q_2)P_{\mathcal{R}_g}(|\textbf{k}-\textbf{q}_1|)\\
    =& \hspace{0.1cm} 3\cdot 2^7 g_{\rm NL} \langle\mathcal{R}_g^2\rangle \int \frac{d^3 q_1}{(2\pi)^3}Q^2_{\lambda}(\textbf{k},\textbf{q}_1) I^2(|\textbf{k}-\textbf{q}_1|,q_1,\eta) P_{\mathcal{R}_g}(q_1)P_{\mathcal{R}_g}(|\textbf{k}-\textbf{q}_1|)\\
    =& \hspace{0.1cm}3\cdot2^2\hspace{0.1cm} g_{\rm NL}\langle\mathcal{R}_g^2\rangle P_{h,\lambda}(k,\eta)\big|_{\rm Gaussian}\,.
    \end{aligned}
    \label{new}
\end{equation}

\vspace{0.3cm}

The six nG contributions at order $\mathcal{A}_{\mathcal{R}}^3$, actually come from the 2-loop correction to the leading order tensor power spectrum. We obtain three additional terms with respect to \cite{2021JCAP...10..080A}, that we label as $g_{\rm NL}$, ``s'' and ``new'' (note that we follow the notation of \cite{2023JCAP...03..057G}). The former two, as we show in the following, just vanish \cite{Unal:2018yaa, 2023JCAP...03..057G}; the ``new'' term, on the other hand, still contributes to the final spectrum. It is the only contribution sensitive to the sign of the nG parameter, since it is linearly dependent on it and thus in principle could compensate the $f_{\rm NL}^2$ ones. Furthermore, as it can be appreciated from Figures \ref{Fig::fNL2} and \ref{Fig::gNL}, the $g_{\rm NL}$ contribution is almost one order of magnitude greater than the $f_{\rm NL}^2$ one, at the peak. Hence, imposing that $\Omega_{\rm GW}^{f_{\rm NL}^2}/\Omega_{\rm GW}^{g_{\rm NL}}\sim 1$, roughly we can write $f_{\rm NL}\sim  \sqrt{10\, g_{\rm NL}}$, so that the $f_{\rm NL}^2$ contribution is comparable with the $g_{\rm NL}$ one. Assuming for example $f_{\rm NL} \sim 10^{-1}$, with $g_{\rm NL}\sim 10^{-3}$, (and so in the case of a hierarchy between the nG parameters) the ``new'' term could still provide an important contribution at this order. We conclude noting that this latter contribution can be rewritten in terms of the Gaussian one, exploiting the definition of variance $\langle\mathcal{R}_g^2\rangle = \int d^3k P_{\mathcal{R}_g}(k)/(2\pi)^3$. This suggests the presence of a possible degeneracy between $g_{\rm NL}$ and $\mathcal{A}_{\mathcal
R}$ at this order, which can have a relevant impact on the Fisher matrix.  
At order $\mathcal{A}_{\mathcal{R}}^4$, we get the 3-loop corrections to the leading order tensor power spectrum. We label them in terms of the dependence on the nG parameters as follows

\begin{align}
    P_{h,\lambda}(k,\eta)\big|_{\rm f^2_{\rm NL}g_{\rm NL}} =& \hspace{0.1cm} 3 \cdot 2^{9} f_{\rm NL}^2g_{\rm NL} \int \frac{d^3 q_1}{(2\pi)^3}\int \frac{d^3 q_2}{(2\pi)^3}\int\frac{d^3 q_3}{(2\pi)^3}Q_{\lambda}(\textbf{k},\textbf{q}_1)Q_{\lambda}(\textbf{k},\textbf{q}_2)\nonumber\\
    &\hspace{2cm}\times I(|\textbf{k}-\textbf{q}_1|,q_1,\eta)I(|\textbf{k}-\textbf{q}_2|,q_2,\eta)\nonumber\\
    &\hspace{2cm}\times\Big[\hspace{0.1cm}P_{\mathcal{R}_g}(k)P_{\mathcal{R}_g}(q_1)P_{\mathcal{R}_g}(q_2)P_{\mathcal{R}_g}(q_3)\nonumber\\
    &\hspace{2.3cm}+ P_{\mathcal{R}_g}(k)P_{\mathcal{R}_g}(q_1)P_{\mathcal{R}_g}(q_3)P_{\mathcal{R}_g}(|\textbf{q}_2-\textbf{q}_3|)\hspace{0.1cm}\nonumber\\
    &\hspace{2.3cm}+ P_{\mathcal{R}_g}(q_1)P_{\mathcal{R}_g}(|\textbf{q}_1-\textbf{q}_2|)P_{\mathcal{R}_g}(q_3)P_{\mathcal{R}_g}(|\textbf{k}-\textbf{q}_1-\textbf{q}_3|)\hspace{0.1cm}\nonumber\\
    &\hspace{2.3cm}+ P_{\mathcal{R}_g}(q_1)P_{\mathcal{R}_g}(|\textbf{k}-\textbf{q}_2|)P_{\mathcal{R}_g}(|\textbf{q}_1-\textbf{q}_2|)P_{\mathcal{R}_g}(q_3)\hspace{0.1cm}\nonumber\\
    &\hspace{2.3cm}+ P_{\mathcal{R}_g}(q_1)P_{\mathcal{R}_g}(q_3)P_{\mathcal{R}_g}(|\textbf{k}-\textbf{q}_1-\textbf{q}_3|)P_{\mathcal{R}_g}(|\textbf{q}_2-\textbf{q}_3|)\hspace{0.1cm}\nonumber\\
    &\hspace{2.3cm}+ P_{\mathcal{R}_g}(q_1)P_{\mathcal{R}_g}(q_3)P_{\mathcal{R}_g}(|\textbf{k}-\textbf{q}_1-\textbf{q}_2+\textbf{q}_3|)P_{\mathcal{R}_g}(|\textbf{q}_2-\textbf{q}_3|)\hspace{0.1cm}\nonumber\\
    &\hspace{2.3cm}+ P_{\mathcal{R}_g}(q_1)P_{\mathcal{R}_g}(|\textbf{k}-\textbf{q}_1|)P_{\mathcal{R}_g}(|\textbf{q}_1-\textbf{q}_2|)P_{\mathcal{R}_g}(q_3)\hspace{0.1cm}\nonumber\\
    &\hspace{2.3cm}+ P_{\mathcal{R}_g}(q_1)P_{\mathcal{R}_g}(|\textbf{q}_1-\textbf{q}_2|)P_{\mathcal{R}_g}(q_3)P_{\mathcal{R}_g}(|\textbf{k}-\textbf{q}_2-\textbf{q}_3|)\hspace{0.1cm}\nonumber\\
    &\hspace{2.3cm}+ P_{\mathcal{R}_g}(q_1)P_{\mathcal{R}_g}(q_3)P_{\mathcal{R}_g}(|\textbf{k}-\textbf{q}_3|)P_{\mathcal{R}_g}(|\textbf{q}_2-\textbf{q}_3|)\hspace{0.1cm}\Big]\nonumber\allowdisplaybreaks\\\allowdisplaybreaks
    =& \hspace{0.1cm} 3 \cdot 2^{9} f_{\rm NL}^2g_{\rm NL} \int \frac{d^3 q_1}{(2\pi)^3}\int \frac{d^3 q_2}{(2\pi)^3}\int\frac{d^3 q_3}{(2\pi)^3}Q_{\lambda}(\textbf{k},\textbf{q}_1)Q_{\lambda}(\textbf{k},\textbf{q}_2)\nonumber\\
    &\hspace{2cm}\times I(|\textbf{k}-\textbf{q}_1|,q_1,\eta)I(|\textbf{k}-\textbf{q}_2|,q_2,\eta)\nonumber\\
    &\hspace{2cm}\times\Big[\hspace{0.1cm}P_{\mathcal{R}_g}(k)P_{\mathcal{R}_g}(q_1)P_{\mathcal{R}_g}(q_2)P_{\mathcal{R}_g}(q_3)\nonumber\\
    &\hspace{2.3cm}+ P_{\mathcal{R}_g}(k)P_{\mathcal{R}_g}(q_1)P_{\mathcal{R}_g}(q_3)P_{\mathcal{R}_g}(|\textbf{q}_2-\textbf{q}_3|)\hspace{0.1cm}\nonumber\\
    &\hspace{2.3cm}+ P_{\mathcal{R}_g}(q_1)P_{\mathcal{R}_g}(|\textbf{q}_1-\textbf{q}_2|)P_{\mathcal{R}_g}(q_3)P_{\mathcal{R}_g}(|\textbf{k}-\textbf{q}_1-\textbf{q}_3|)\hspace{0.1cm}\nonumber\\
    &\hspace{2.3cm}+ 2 P_{\mathcal{R}_g}(q_1)P_{\mathcal{R}_g}(q_3)P_{\mathcal{R}_g}(|\textbf{k}-\textbf{q}_1-\textbf{q}_3|)P_{\mathcal{R}_g}(|\textbf{q}_2-\textbf{q}_3|)\hspace{0.1cm}\nonumber\\
    &\hspace{2.3cm}+ P_{\mathcal{R}_g}(q_1)P_{\mathcal{R}_g}(|\textbf{q}_1-\textbf{q}_2|)P_{\mathcal{R}_g}(q_3)P_{\mathcal{R}_g}(|\textbf{k}-\textbf{q}_2-\textbf{q}_3|)\hspace{0.1cm}\nonumber\\
    &\hspace{2.3cm}+ P_{\mathcal{R}_g}(q_1)P_{\mathcal{R}_g}(q_3)P_{\mathcal{R}_g}(|\textbf{k}-\textbf{q}_3|)P_{\mathcal{R}_g}(|\textbf{q}_2-\textbf{q}_3|)\hspace{0.1cm}\Big]\nonumber\\
    &+ 3\cdot 2 g_{\rm NL} \langle\mathcal{R}_g^2\rangle P_{h,\lambda}(k,\eta)\big|_{\rm u}\nonumber\\
    &+ 3\cdot 2 g_{\rm NL} \langle\mathcal{R}_g^2\rangle P_{h,\lambda}(k,\eta)\big|_{\rm t}\,;
    \label{fnl2-gnl}
\end{align}

\begin{equation}
    \begin{aligned}
    P_{h,\lambda}(k,\eta)\big|_{\rm f_{\rm NL}^2g_{\rm NL},d} =& \hspace{0.1cm} 3 \cdot 2^{8} f_{\rm NL}^2g_{\rm NL} \int \frac{d^3 q_1}{(2\pi)^3}\int \frac{d^3 q_2}{(2\pi)^3}\int\frac{d^3 q_3}{(2\pi)^3}Q^2_{\lambda}(\textbf{k},\textbf{q}_1)I^2(|\textbf{k}-\textbf{q}_1|,q_1,\eta)\\
    &\hspace{1.5cm}\times P_{\mathcal{R}_g}(q_1)P_{\mathcal{R}_g}(q_2)P_{\mathcal{R}_g}(q_3)P_{\mathcal{R}_g}(|\textbf{k}-\textbf{q}_1-\textbf{q}_2|)\allowdisplaybreaks\\
    =& \hspace{0.1cm} 3 \cdot 2^{8}f_{\rm NL}^2g_{\rm NL} \langle\mathcal{R}_g^2\rangle \int \frac{d^3 q_1}{(2\pi)^3}\int \frac{d^3 q_2}{(2\pi)^3}Q^2_{\lambda}(\textbf{k},\textbf{q}_1)I^2(|\textbf{k}-\textbf{q}_1|,q_1,\eta)\\
    &\hspace{1.5cm}\times P_{\mathcal{R}_g}(q_1)P_{\mathcal{R}_g}(q_2)P_{\mathcal{R}_g}(|\textbf{k}-\textbf{q}_1-\textbf{q}_2|)\\
    =& \hspace{0.1cm} 3 \cdot 2^{8}f_{\rm NL}^2g_{\rm NL} \langle\mathcal{R}_g^2\rangle \int \frac{d^3 q_1}{(2\pi)^3}\int \frac{d^3 q_2}{(2\pi)^3}Q^2_{\lambda}(\textbf{k},\textbf{q}_1)I^2(|\textbf{k}-\textbf{q}_1|,q_1,\eta)\\
    &\hspace{1.5cm}\times P_{\mathcal{R}_g}(|\textbf{k}-\textbf{q}_1|)P_{\mathcal{R}_g}(q_2)P_{\mathcal{R}_g}(|\textbf{q}_1-\textbf{q}_2|)\\
    =& \hspace{0.1cm} 3\cdot2 \hspace{0.1cm} g_{\rm NL} \langle\mathcal{R}_g^2\rangle P_{h,\lambda}(k,\eta)\big|_{\rm hybrid}\,;
    \end{aligned}
    \label{fnl2-gnl-disc}
\end{equation}

\begin{equation}
    \begin{aligned}
    P_{h,\lambda}(k,\eta)\big|_{\rm g_{\rm NL}^2} =& \hspace{0.1cm} 3^2 \cdot 2^{7} g_{\rm NL}^2 \int \frac{d^3 q_1}{(2\pi)^3}\int \frac{d^3 q_2}{(2\pi)^3}\int\frac{d^3 q_3}{(2\pi)^3}Q_{\lambda}(\textbf{k},\textbf{q}_1)Q_{\lambda}(\textbf{k},\textbf{q}_2)\\
    &\hspace{2cm} \times I(|\textbf{k}-\textbf{q}_1|,q_1,\eta)I(|\textbf{k}-\textbf{q}_2|,q_2,\eta)\\
    &\hspace{2cm}\times\Big[\hspace{0.1cm}P_{\mathcal{R}_g}(q_1)P_{\mathcal{R}_g}(|\textbf{k}-\textbf{q}_1|)P_{\mathcal{R}_g}(q_3)P_{\mathcal{R}_g}(|\textbf{q}_1-\textbf{q}_2-\textbf{q}_3|)\\
    &\hspace{2.3cm}+ P_{\mathcal{R}_g}(q_1)P_{\mathcal{R}_g}(q_2)P_{\mathcal{R}_g}(q_3)P_{\mathcal{R}_g}(|\textbf{k}-\textbf{q}_3|)\hspace{0.1cm}\\
    &\hspace{2.3cm}+ P_{\mathcal{R}_g}(q_1)P_{\mathcal{R}_g}(q_2)P_{\mathcal{R}_g}(q_3)P_{\mathcal{R}_g}(|\textbf{k}-\textbf{q}_1-\textbf{q}_2+\textbf{q}_3|)\hspace{0.1cm}\\
    &\hspace{2.3cm}+3 P_{\mathcal{R}_g}(q_1)P_{\mathcal{R}_g}(|\textbf{k}-\textbf{q}_1|)P_{\mathcal{R}_g}(q_2)P_{\mathcal{R}_g}(q_3)\Big]\,;
    \end{aligned}
    \label{gnl2}
\end{equation}

\begin{equation}
    \begin{aligned}
    P_{h,\lambda}(k,\eta)\big|_{\rm g_{\rm NL}^2,d} =& \hspace{0.1cm} 3 \cdot 2^{6} g_{\rm NL}^2 \int \frac{d^3 q_1}{(2\pi)^3}\int \frac{d^3 q_2}{(2\pi)^3}\int\frac{d^3 q_3}{(2\pi)^3}Q^2_{\lambda}(\textbf{k},\textbf{q}_1)I^2(|\textbf{k}-\textbf{q}_1|,q_1,\eta)\\
    &\hspace{0.3cm}\times \Big[\hspace{0.1cm}3^2 P_{\mathcal{R}_g}(q_1)P_{\mathcal{R}_g}(|\textbf{k}-\textbf{q}_1|)P_{\mathcal{R}_g}(q_2)P_{\mathcal{R}_g}(q_3)\\
    &\hspace{0.6cm}+ 2P_{\mathcal{R}_g}(q_1)P_{\mathcal{R}_g}(q_2)P_{\mathcal{R}_g}(q_3)P_{\mathcal{R}_g}(|\textbf{k}-\textbf{q}_1-\textbf{q}_2+\textbf{q}_3|)\hspace{0.1cm}\Big]\\
    =& \hspace{0.1cm} 3^3 \cdot 2^{6} g_{\rm NL}^2 \langle\mathcal{R}_g^2\rangle^2\int \frac{d^3 q_1}{(2\pi)^3} Q^2_{\lambda}(\textbf{k},\textbf{q}_1)I^2(|\textbf{k}-\textbf{q}_1|,q_1,\eta)P_{\mathcal{R}_g}(q_1)P_{\mathcal{R}_g}(|\textbf{k}-\textbf{q}_1|)\\
    &+ 3 \cdot 2^{7} g_{\rm NL}^2 \int \frac{d^3 q_1}{(2\pi)^3}\int \frac{d^3 q_2}{(2\pi)^3}\int\frac{d^3 q_3}{(2\pi)^3}Q^2_{\lambda}(\textbf{k},\textbf{q}_1)I^2(|\textbf{k}-\textbf{q}_1|,q_1,\eta)\\
    &\hspace{1.7cm}\times P_{\mathcal{R}_g}(q_1)P_{\mathcal{R}_g}(q_2)P_{\mathcal{R}_g}(q_3)P_{\mathcal{R}_g}(|\textbf{k}-\textbf{q}_1-\textbf{q}_2+\textbf{q}_3|)\\
    =& \hspace{0.1cm} 3^3 \cdot 2^{6} g_{\rm NL}^2 \langle\mathcal{R}_g^2\rangle^2 \hspace{0.1cm} 2^{-5} P_{h,\lambda}(k,\eta)\big|_{\rm Gaussian}\\
    &+ 3 \cdot 2^{7} g_{\rm NL}^2 \int \frac{d^3 q_1}{(2\pi)^3}\int \frac{d^3 q_2}{(2\pi)^3}\int\frac{d^3 q_3}{(2\pi)^3}Q^2_{\lambda}(\textbf{k},\textbf{q}_1)I^2(|\textbf{k}-\textbf{q}_1|,q_1,\eta)\\
    &\hspace{1.7cm}\times P_{\mathcal{R}_g}(q_1)P_{\mathcal{R}_g}(q_2)P_{\mathcal{R}_g}(q_3)P_{\mathcal{R}_g}(|\textbf{k}-\textbf{q}_1-\textbf{q}_2+\textbf{q}_3|)\\
    =& \hspace{0.1cm} 3^3\cdot2\hspace{0.1cm} g_{\rm NL}^2 \langle\mathcal{R}_g^2\rangle^2 P_{h,\lambda}(k,\eta)\big|_{\rm Gaussian}\\
    &+ 3 \cdot 2^{7} g_{\rm NL}^2 \int \frac{d^3 q_1}{(2\pi)^3}\int \frac{d^3 q_2}{(2\pi)^3}\int\frac{d^3 q_3}{(2\pi)^3}Q^2_{\lambda}(\textbf{k},\textbf{q}_1)I^2(|\textbf{k}-\textbf{q}_1|,q_1,\eta)\\
    &\hspace{1.7cm}\times P_{\mathcal{R}_g}(q_1)P_{\mathcal{R}_g}(q_2)P_{\mathcal{R}_g}(q_3)P_{\mathcal{R}_g}(|\textbf{k}-\textbf{q}_1-\textbf{q}_2+\textbf{q}_3|)\,;
    \end{aligned}
    \label{gnl2-disc}
\end{equation}

\begin{equation}
    \begin{aligned}
    P_{h,\lambda}(k,\eta)\big|_{\rm f^4_{\rm NL}} =& \hspace{0.1cm} 2^{8} f_{\rm NL}^4 \int \frac{d^3 q_1}{(2\pi)^3}\int \frac{d^3 q_2}{(2\pi)^3}\int\frac{d^3 q_3}{(2\pi)^3}Q_{\lambda}(\textbf{k},\textbf{q}_1)Q_{\lambda}(\textbf{k},\textbf{q}_2) \\
    &\hspace{0.2cm} I(|\textbf{k}-\textbf{q}_1|,q_1,\eta)I(|\textbf{k}-\textbf{q}_2|,q_2,\eta)\\
    &\hspace{0.2cm}\times\Big[\hspace{0.1cm} 2P_{\mathcal{R}_g}(q_3)P_{\mathcal{R}_g}(|\textbf{q}_1-\textbf{q}_3|)P_{\mathcal{R}_g}(|\textbf{q}_2-\textbf{q}_3|)P_{\mathcal{R}_g}(|\textbf{k}-\textbf{q}_3|)\\
    &\hspace{0.2cm}+ P_{\mathcal{R}_g}(q_3)P_{\mathcal{R}_g}(|\textbf{q}_1-\textbf{q}_3|)P_{\mathcal{R}_g}(|\textbf{q}_2-\textbf{q}_3|)P_{\mathcal{R}_g}(|\textbf{k}-\textbf{q}_1-\textbf{q}_2+\textbf{q}_3|)\hspace{0.1cm}\Big]\,;
    \end{aligned}
    \label{fnl4}
\end{equation}

\begin{equation}
    \begin{aligned}
    P_{h,\lambda}(k,\eta)\big|_{\rm f_{\rm NL}^4,d} =& \hspace{0.1cm} 2^{7} f_{\rm NL}^4 \int \frac{d^3 q_1}{(2\pi)^3}\int \frac{d^3 q_2}{(2\pi)^3}\int\frac{d^3 q_3}{(2\pi)^3}Q^2_{\lambda}(\textbf{k},\textbf{q}_1)I^2(|\textbf{k}-\textbf{q}_1|,q_1,\eta)\\
    &\hspace{0.5cm}\times P_{\mathcal{R}_g}(q_2)P_{\mathcal{R}_g}(q_3)P_{\mathcal{R}_g}(|\textbf{q}_1-\textbf{q}_2|)P_{\mathcal{R}_g}(|\textbf{k}-\textbf{q}_1-\textbf{q}_3|)\,;
    \end{aligned}
    \label{fnl4-disc}
\end{equation}

\begin{equation}
    \begin{aligned}
    P_{h,\lambda}(k,\eta)\big|_{\rm f_{\rm NL}h_{\rm NL}} =& \hspace{0.1cm} 3 \cdot 2^{9} f_{\rm NL}\hspace{0.5mm}h_{\rm NL} \int \frac{d^3 q_1}{(2\pi)^3}\int \frac{d^3 q_2}{(2\pi)^3}\int\frac{d^3 q_3}{(2\pi)^3}Q_{\lambda}(\textbf{k},\textbf{q}_1)Q_{\lambda}(\textbf{k},\textbf{q}_2)\\
    &\hspace{2cm}\times I(|\textbf{k}-\textbf{q}_1|,q_1,\eta)I(|\textbf{k}-\textbf{q}_2|,q_2,\eta)\\
    &\hspace{1.6cm}\times\Big[\hspace{0.1cm}2\Big(P_{\mathcal{R}_g}(k)P_{\mathcal{R}_g}(q_1)P_{\mathcal{R}_g}(q_2)P_{\mathcal{R}_g}(q_3)\\
    &\hspace{2.3cm}+ P_{\mathcal{R}_g}(q_1)P_{\mathcal{R}_g}(|\textbf{k}-\textbf{q}_1|)P_{\mathcal{R}_g}(|\textbf{q}_1-\textbf{q}_2|)P_{\mathcal{R}_g}(q_3)\hspace{0.1cm}\\
    &\hspace{2.3cm}+ P_{\mathcal{R}_g}(q_1)P_{\mathcal{R}_g}(q_2)P_{\mathcal{R}_g}(q_3)P_{\mathcal{R}_g}(|\textbf{k}-\textbf{q}_1-\textbf{q}_3|)\hspace{0.1cm}\\
    &\hspace{2.3cm}+ P_{\mathcal{R}_g}(q_1)P_{\mathcal{R}_g}(q_2)P_{\mathcal{R}_g}(q_3)P_{\mathcal{R}_g}(|\textbf{k}-\textbf{q}_1-\textbf{q}_2|)\Big)\\
    &\hspace{2.3cm}+P_{\mathcal{R}_g}(q_1)P_{\mathcal{R}_g}(|\textbf{k}-\textbf{q}_1|)P_{\mathcal{R}_g}(q_3)P_{\mathcal{R}_g}(|\textbf{q}_2-\textbf{q}_3|)\hspace{0.1cm}\Big]\\
    =&\hspace{0.1cm} 3 \cdot 2^{9} f_{\rm NL}\hspace{0.5mm}h_{\rm NL} \int \frac{d^3 q_1}{(2\pi)^3}\int \frac{d^3 q_2}{(2\pi)^3}\int\frac{d^3 q_3}{(2\pi)^3}Q_{\lambda}(\textbf{k},\textbf{q}_1)Q_{\lambda}(\textbf{k},\textbf{q}_2)\\
    &\hspace{2cm}\times I(|\textbf{k}-\textbf{q}_1|,q_1,\eta)I(|\textbf{k}-\textbf{q}_2|,q_2,\eta)\\
    &\hspace{1.6cm}\times\Big[\hspace{0.1cm}2\Big(P_{\mathcal{R}_g}(k)P_{\mathcal{R}_g}(q_1)P_{\mathcal{R}_g}(q_2)P_{\mathcal{R}_g}(q_3)\\
    &\hspace{2.3cm}+ P_{\mathcal{R}_g}(q_1)P_{\mathcal{R}_g}(q_2)P_{\mathcal{R}_g}(q_3)P_{\mathcal{R}_g}(|\textbf{k}-\textbf{q}_1-\textbf{q}_3|)\Big)\hspace{0.1cm}\\
    &\hspace{2.3cm}+P_{\mathcal{R}_g}(q_1)P_{\mathcal{R}_g}(|\textbf{k}-\textbf{q}_1|)P_{\mathcal{R}_g}(q_3)P_{\mathcal{R}_g}(|\textbf{q}_2-\textbf{q}_3|)\hspace{0.1cm}\Big]\\
    &+3\cdot 2^2 f_{\rm NL}^{-1} h_{\rm NL}\langle\mathcal{R}_g^2\rangle P_{h,\lambda}(k,\eta)\big|_{\rm t}\\
    &+3\cdot 2^2 f_{\rm NL}^{-1} h_{\rm NL}\langle\mathcal{R}_g^2\rangle P_{h,\lambda}(k,\eta)\big|_{\rm u}\,;
    \end{aligned}
    \label{fnl-hnl}
\end{equation}

\begin{equation}
    \begin{aligned}
    P_{h,\lambda}(k,\eta)\big|_{\rm f_{\rm NL}h_{\rm NL},d} =& \hspace{0.1cm} 3 \cdot 2^{9} f_{\rm NL}\hspace{0.5mm}h_{\rm NL} \int \frac{d^3 q_1}{(2\pi)^3}\int \frac{d^3 q_2}{(2\pi)^3}\int\frac{d^3 q_3}{(2\pi)^3}Q^2_{\lambda}(\textbf{k},\textbf{q}_1)I^2(|\textbf{k}-\textbf{q}_1|,q_1,\eta)\\
    &\hspace{0.5cm}\times P_{\mathcal{R}_g}(q_1)P_{\mathcal{R}_g}(q_2)P_{\mathcal{R}_g}(q_3)P_{\mathcal{R}_g}(|\textbf{k}-\textbf{q}_1-\textbf{q}_2|)\\
    =& \hspace{0.1cm} 3 \cdot 2^{9} f_{\rm NL}\hspace{0.5mm}h_{\rm NL} \langle\mathcal{R}_g^2\rangle \int \frac{d^3 q_1}{(2\pi)^3}\int \frac{d^3 q_2}{(2\pi)^3}Q^2_{\lambda}(\textbf{k},\textbf{q}_1)I^2(|\textbf{k}-\textbf{q}_1|,q_1,\eta)\\
    &\hspace{0.5cm}\times P_{\mathcal{R}_g}(q_1)P_{\mathcal{R}_g}(q_2)P_{\mathcal{R}_g}(|\textbf{k}-\textbf{q}_1-\textbf{q}_2|)\\
    =& 3\cdot 2^2\hspace{0.1cm}f^{-1}_{\rm NL}h_{\rm NL}\langle\mathcal{R}_g^2\rangle P_{h,\lambda}(k,\eta)\big|_{\rm hybrid}\,;
    \end{aligned}
    \label{fnl-hnl-disc}
\end{equation}
\begin{equation}
    \begin{aligned}
    P_{h,\lambda}(k,\eta)\big|_{\rm i_{\rm NL}} =& \hspace{0.1cm} 5\cdot3\cdot2^{8} i_{\rm NL} \int \frac{d^3 q_1}{(2\pi)^3}\int \frac{d^3 q_2}{(2\pi)^3}\int\frac{d^3 q_3}{(2\pi)^3}Q_{\lambda}(\textbf{k},\textbf{q}_1)Q_{\lambda}(\textbf{k},\textbf{q}_2)\\
    &\hspace{3cm}I(|\textbf{k}-\textbf{q}_1|,q_1,\eta)I(|\textbf{k}-\textbf{q}_2|,q_2,\eta)\\
    &\hspace{3cm}\times P_{\mathcal{R}_g}(q_1)P_{\mathcal{R}_g}(|\textbf{k}-\textbf{q}_1|)P_{\mathcal{R}_g}(q_2)P_{\mathcal{R}_g}(q_3)\,;
    \end{aligned}
    \label{inl}
\end{equation}
\begin{equation}
    \begin{aligned}
    P_{h,\lambda}(k,\eta)\big|_{\rm i_{\rm NL},d} =& \hspace{0.1cm} 5\cdot3\cdot2^{7} i_{\rm NL} \int \frac{d^3 q_1}{(2\pi)^3}\int \frac{d^3 q_2}{(2\pi)^3}\int\frac{d^3 q_3}{(2\pi)^3}Q^2_{\lambda}(\textbf{k},\textbf{q}_1)I^2(|\textbf{k}-\textbf{q}_1|,q_1,\eta)\\
    &\hspace{0.5cm}\times P_{\mathcal{R}_g}(q_1)P_{\mathcal{R}_g}(|\textbf{k}-\textbf{q}_1|)P_{\mathcal{R}_g}(q_2)P_{\mathcal{R}_g}(q_3)\\
    =& \hspace{0.1cm} 5\cdot3\cdot2^{7}i_{\rm NL} \langle\mathcal{R}_g^2\rangle^2 \int \frac{d^3 q_1}{(2\pi)^3}Q^2_{\lambda}(\textbf{k},\textbf{q}_1)I^2(|\textbf{k}-\textbf{q}_1|,q_1,\eta) P_{\mathcal{R}_g}(q_1)P_{\mathcal{R}_g}(|\textbf{k}-\textbf{q}_1|)\\
    =& 5\cdot3\cdot2^2 \hspace{0.1cm} i_{\rm NL}\langle\mathcal{R}_g^2\rangle^2 P_{h,\lambda}(k,\eta)\big|_{\rm Gaussian}\,.
    \end{aligned}
    \label{inl-disc}
\end{equation}

\vspace{0.3cm}

In equations \eqref{fnl2-gnl-disc}, \eqref{gnl2-disc}, \eqref{fnl4-disc}, \eqref{fnl-hnl-disc} and \eqref{inl-disc} the letter ``d'' indicates that these terms are originated by the disconnected trispectrum of the primordial fluctuations. All the other terms correspond to connected contributions, instead. The additional internal momentum $\textbf{q}_3$ arises from the local model of non-Gaussianity. 



Again, also at this order some of the disconnected contribution can be rewritten in terms of the lower order power spectra $P_{h,\lambda}(k,\eta)\big|_{\rm Gaussian}$ or $P_{h,\lambda}(k,\eta)\big|_{\rm hybrid}$, with different numerical factors in front of them. It is also possible to rewrite some of the connected contributions in terms of the lower order ones. In the first step of equation \eqref{fnl-hnl}, for example, we can rewrite the second and the fourth line in the square brackets in terms of the ``u'' and ``t'' contributions.

Recalling now that the constants $f_{\rm NL}, g_{\rm NL}, h_{\rm NL}$ and $i_{\rm NL}$ are related to products of two, three, four and five Gaussian fields in the non-Gaussian expansion of $\mathcal{R}$ in Fourier space, respectively, it is straightforward to recover the initial terms in that expansion which, combined together, give rise to the expressions \eqref{g}-\eqref{inl-disc}. Since they all come from the 4-point correlator of the primordial curvature perturbation $\langle\mathcal{R}(\textbf{q}_1)\mathcal{R}(\textbf{k}_1-\textbf{q}_1)\mathcal{R}(\textbf{q}_2)\mathcal{R}(\textbf{k}_2-\textbf{q}_2)\rangle$, all terms need to have exactly 4 entries. Hence, at leading order, the unique combination contributing to the 4-point function comes from the product of four terms, each linearly dependent on the Gaussian field $\mathcal{R}_g$. The next-to-leading order corrections arise from all the possible combinations in the 4-point correlator of the nG scalar field giving rise to 6-point functions, i.e.
\begin{itemize}
    \item two terms containing the Gaussian field squared ($\propto f_{\rm NL}$) and two terms linearly dependent on it $ \rightarrow f_{\rm NL}^2$;
    \item one term containing the Gaussian field cubed ($\propto g_{\rm NL}$) and three terms linearly dependent on it $ \rightarrow g_{\rm NL}$. 
\end{itemize}
The next-to-next-to-leading order contributions, instead, arise from the 8-point correlation functions and originate from
\begin{itemize}
    \item two terms proportional to the Gaussian field squared ($\propto f_{\rm NL}$), one proportional to its cube ($\propto g_{\rm NL}$) and the last one linearly dependent on $\mathcal{R}_g$ $ \rightarrow f_{\rm NL}^2g_{\rm NL}$;
    \item two terms containing the cube of the Gaussian field ($\propto g_{\rm NL}$) and two terms linearly dependent on it $ \rightarrow g_{\rm NL}^2$;
    \item four terms proportional to the square of the Gaussian field ($\propto f_{\rm NL}$) $ \rightarrow f_{\rm NL}^4$;
    \item one term proportional to the fourth power of the Gaussian field ($\propto h_{\rm NL}$), one term containing the Gaussian field squared ($\propto f_{\rm NL}$) and two terms linearly dependent on it $ \rightarrow f_{\rm NL}\hspace{0.5mm}h_{\rm NL}$;
    \item one term proportional to the fifth power of the Gaussian field ($\propto i_{\rm NL}$) and three terms linearly dependent on it $ \rightarrow i_{\rm NL}$.
\end{itemize}
We remark that the terms just reported are all the possible ones contributing up to the 8-point function of the primordial scalar perturbations that arise when the local model of nG is considered and they do not require any further correction at the perturbative order considered for this work. An higher order term in the expression \eqref{Eq::R_Exp}, for example proportional to ${\rm A} \mathcal{R}_g^6(\textbf{x})$, would lead at least to a 10-point function of the primordial Gaussian fluctuations.\par

The goal of this work is  to provide a Fisher analysis for the LISA detectors, trying to infer the accuracy for the measurement of the non-Gaussian parameters. 
In the following the results of the different contributions to the spectral density are shown. Useful changes of variable to simplify the numerical evaluation are reported in Appendix \ref{manipulation}.

\section{Results}\label{resultsection}

Having the analytical expressions of all the components contributing to the spectrum of second-order scalar-induced GW \eqref{g}-\eqref{inl-disc}, we now study the relative importance of the different contributions to the spectrum. Similar arguments, treated with different notations, can be found in \cite{2021Univ....7..398D,2007PhRvD..75l3518A,2018PhRvD..97l3532K,2018JCAP...09..012E,2020JCAP...02..028B}. 
In this work we choose a primordial dimensionless power spectrum parametrized as a lognormal function \cite{Unal:2018yaa}
\begin{equation}
        \Delta^2_g(k) = \frac{\mathcal{A}_{\mathcal{R}}}{\sqrt{2\pi\sigma^2}} \exp\left(-\frac{\ln^2(k/k_{*})}{2\sigma^2} \right)\,.
        \label{lognormal}
\end{equation}
With this choice, the variance of the primordial Gaussian perturbation $\langle\mathcal{R}^2_g\rangle$ is normalized as $\int d \ln k\hspace{0.1cm} \Delta^2_g(k) = \mathcal{A}_{\mathcal{R}}$, i.e the amplitude of the spectrum. Here $k_*$ (or equivalently $f_*$) determines the position of the peak and $\sigma$ its width. We consider $\sigma = 1/10$ and $f_{*} \sim 0.005$ Hz, assuming that the spectrum is centered in the LISA band.
Here $\mathcal{A}_{\mathcal{R}}$ is taken to be of $\mathcal{O}(10^{-2}, 10^{-3})$, which is larger than the value coming from CMB experiments \cite{2020A&A...641A..10P}, but this choice is justified since at the scales of interest we have no tight constraints. We note that in the above parametrization the spectrum depends on the ratio $k/k_*$; hence the results are normalized with respect to the peak scale, simplifying the comparison between the analysis performed considering experiments working in different frequency ranges.

In the following we report the contributions of the scalar-induced GW spectra at all orders considered in this work. The integrals are computed numerically, first exploiting the suitable change of variables reported in Appendix \ref{manipulation} and then using the \texttt{Vegas} package \cite{2021JCoPh.43910386L}.
\subsection{``Gaussian'' component of the spectrum}

Starting from the ``Gaussian'' term of the spectrum, equation (\ref{g}), we obtain
\begin{equation}
\begin{aligned}
    \Omega_{\rm GW}(k,\eta)|_{\rm Gaussian} =& \hspace{0.1cm}\frac{1}{12}  \left(\frac{k}{a(\eta)H(\eta)}\right)^2 \int_{0}^{\infty} dt \int_{-1}^{1} ds \hspace{0.1cm}\frac{1}{u^2 v^2}\hspace{0.1cm}\overline{\Tilde{J}^2(u,v,x)}\hspace{0.1cm} \Delta^2_{g}(v k)\Delta^2_{g}(u k)\,.
    \label{eqgausscomp}
\end{aligned}
\end{equation}
In the last equation we introduced $\Tilde{J}(u,v,x) = v^2\sin^2\theta\Tilde{I}(u,v,x)$ and we performed the integration over the angle $\phi$, since nothing in the integrand depends on it (see Appendix \ref{manipulation}). 
\begin{figure}[t!]
    \centering
    \includegraphics[width=0.6\textwidth]{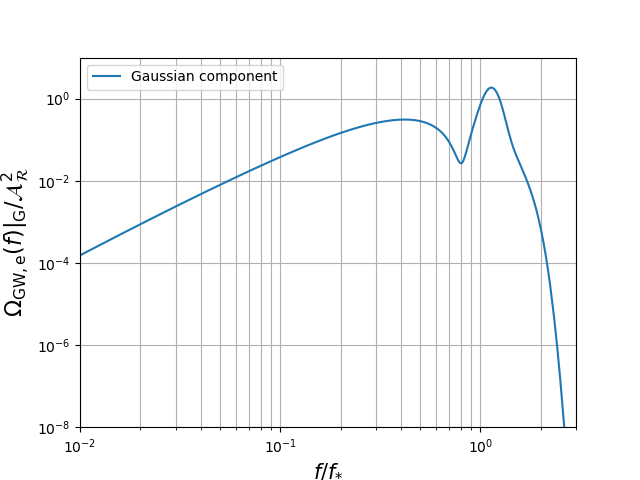}
    \caption{The plot shows the Gaussian component of the spectrum, obtained using a lognormal seed with $\sigma = 0.1$. The spectrum is normalized with respect to $\mathcal{A}_{\mathcal{R}}^2$.}
    \label{Fig::Gaussian}
\end{figure}
The resulting GW spectrum for this component, normalized to the amplitude of the primordial curvature perturbation $\mathcal{A}_{\mathcal{R}}^2$ is shown in Figure \ref{Fig::Gaussian}. The corresponding Feynman diagram, instead, is depicted in Figure \ref{Fig::Feynman_Gaussian}. Further details are reported in Appendix \ref{SubSec::Feynman_Gauss}.

\begin{figure}[t!]
    \centering
    \includegraphics[width=0.4\textwidth]{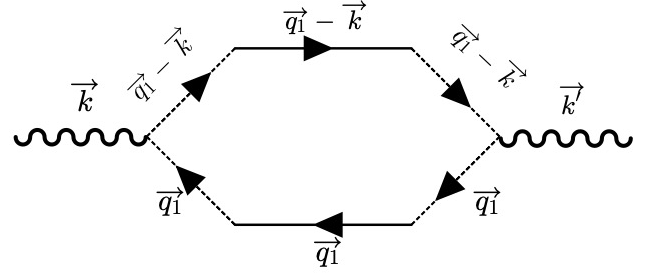}
    \caption{Feynman diagram corresponding to the Gaussian contribution.}
    \label{Fig::Feynman_Gaussian}
\end{figure}

\subsection{``t'' component of the spectrum}

The ``t'' component of the spectrum, coming from equation (\ref{t}), can only be computed numerically. However some useful changes of coordinates can be exploited to rewrite the integral in a more treatable form (see Appendix \ref{manipulation}). 
The spectral density for the ``t'' component is
\begin{equation}
\begin{aligned}
    \Omega_{\rm GW}(k,\eta)|_{\rm t} =& \hspace{0.1cm}\frac{1}{12\pi} \left(\frac{k}{a(\eta)H(\eta)}\right)^2 f_{\rm NL}^2 \int_{0}^{\infty} d t_1 \int_{-1}^{1} ds_1 \int_{0}^{\infty} d t_2 \int_{-1}^{1} ds_2\\
    &\hspace{2.5cm}\times\int_{0}^{2\pi}d\varphi_{12} \cos 2\varphi_{12}\frac{u_1v_1}{(u_2v_2)^2}\frac{1}{w_{a,12}^3}\overline{\Tilde{J}(u_1,v_1,x)\Tilde{J}(u_2,v_2,x)}\\
    &\hspace{2.5cm}\times\Delta^2_{g}(v_2k)\Delta_g^2(u_2k)\Delta^2_g(w_{a,12} k)\,,
    \label{eqccomp}
\end{aligned}
\end{equation}
where $w_{a,12} = |\textbf{q}_1-\textbf{q}_2|/k$ and $\varphi_{12} = \phi_1-\phi_2$. This new variable is introduced to ease the integration, since in this way the integrand depends only on the difference between the azimuthal angles.
In Figure \ref{Fig::fNL2}, we plot the GW spectrum for this component, normalized to the factor $f_{\rm NL}^2$ and to the amplitude $\mathcal{A}_{\mathcal{R}}^3$. The Feynman diagram for this contribution is reported on the left in the first row of Figure \ref{Fig::Feynman_fNL2}, while further details are reported in Appendix \ref{SubSec::Feynman_fNL2}.

\begin{figure}[t!]
    \centering
    \includegraphics[width=0.8\textwidth]{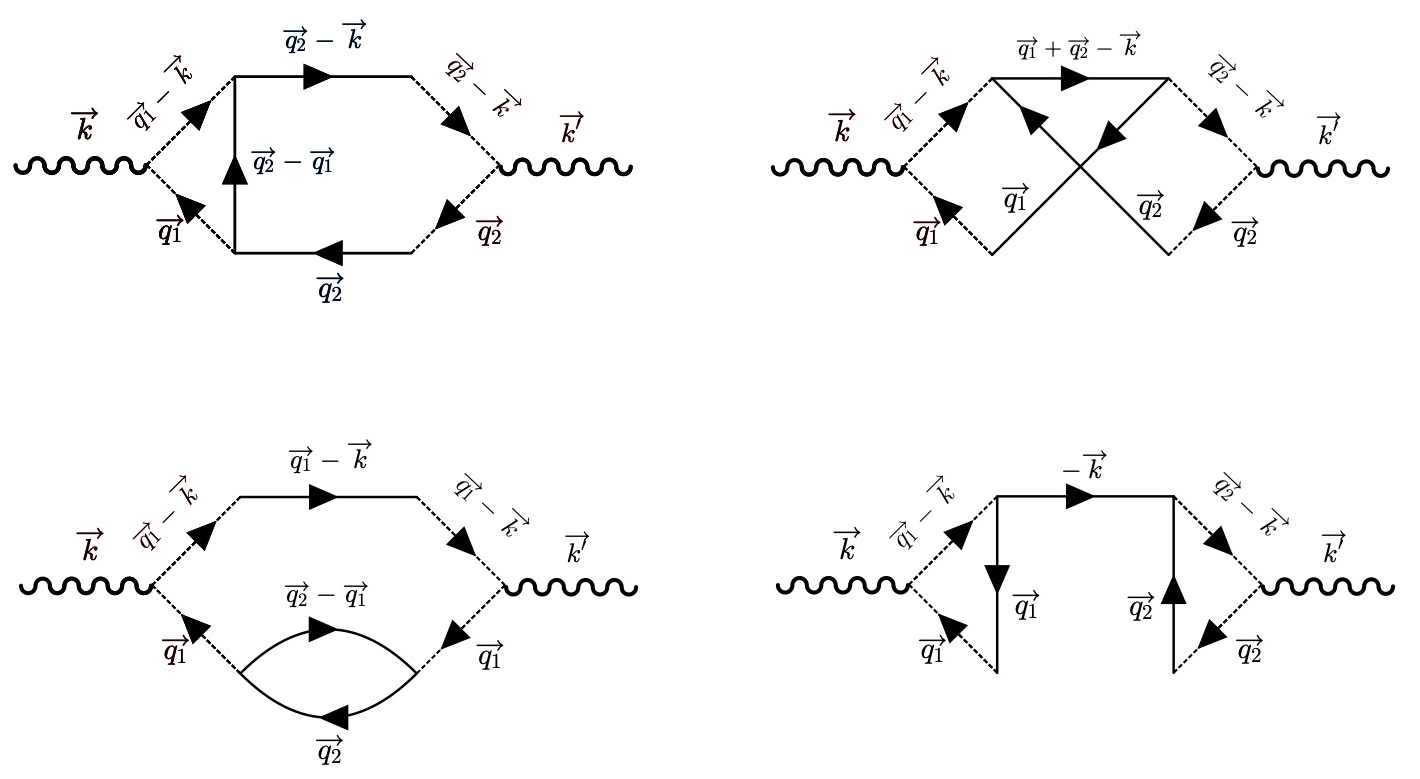}
    \caption{Feynman diagrams corresponding to the connected and disconnected $f_{\rm NL}^2$ contributions. On the top row, starting from the left, we have respectively the ``t'' and ``u'' components, while on the bottom we have the hybrid and the (vanishing) ``s''.}
    \label{Fig::Feynman_fNL2}
\end{figure}

\subsection{``u'' component of the spectrum}

Similarly, for the  ``u'' component of the spectrum we have
\begin{equation}
\begin{aligned}
    \Omega_{\rm GW}(k,\eta)|_{\rm u} =& \hspace{0.1cm}\frac{1}{12\pi} \left(\frac{k}{a(\eta)H(\eta)}\right)^2 f_{\rm NL}^2\int_{0}^{\infty} d t_1 \int_{-1}^{1} ds_1 \int_{0}^{\infty} d t_2 \int_{-1}^{1} ds_2\\
    &\hspace{2.5cm}\times\int_{0}^{2\pi}d\varphi_{12} \cos 2\varphi_{12} \frac{u_1u_2}{(v_1v_2)^2}\frac{1}{w_{b,12}^3}\overline{\Tilde{J}(u_1,v_1,x)\Tilde{J}(u_2,v_2,x)}\\
    &\hspace{2.5cm}\times\Delta^2_{g}(v_1k)\Delta_g^2(v_2k)\Delta^2_g(w_{b,12} k)\,,
    \label{eqzcomp}
\end{aligned}
\end{equation}
where $w_{b,12} = |\textbf{k}-(\textbf{q}_1+\textbf{q}_2)|/k$ and $\varphi_{12}= \phi_1-\phi_2$. In Figure \ref{Fig::fNL2} we plot the resulting GW spectrum for this component, normalized to $f_{\rm NL}^2$ and to the amplitude $\mathcal{A}_{\mathcal{R}}^3$. The Feynman diagram for this contribution is shown on the right in the first row of Figure \ref{Fig::Feynman_fNL2}. Further details are reported in Appendix \ref{SubSec::Feynman_fNL2}.

\subsection{``hybrid'' component of the spectrum}

For the ``hybrid'' component, coming from equation (\ref{ibrid}), a slightly different change of variables can be considered to further simplify the numerical evaluation of the integrals. 
Before going into the details of the calculation of this term, we point out that the momentum $\textbf{q}_2$ in this expression represents an \textit{internal momentum} in the trispectrum, which comes from the Fourier Transform of \eqref{Eq::R_Exp}. This additional integration of the two power spectra $P_{\mathcal{R}_g}(q_2)P_{\mathcal{R}_g}(|\textbf{q}_1-\textbf{q}_2|)$ suggests that this term originates from a 1-loop correction of the disconnected part of the trispectrum.
As discussed in Appendix C of \cite{2021JCAP...10..080A}, a suitable change of variables in this case is

\begin{equation}
    u_1 = \frac{|\textbf{k}-\textbf{q}_1|}{k}\,;\quad\quad  v_1 = \frac{q_1}{k}\,;\quad\quad  u_2 = \frac{|\textbf{q}_1-\textbf{q}_2|}{q_1}\,;\quad\quad  v_2 = \frac{q_2}{q_1}\,.
\end{equation}
Once the Jacobian of the transformation and the correct changes in the extremes of integration are performed, a suitable second change of variables can be considered\footnote{This change of variables is useful to exploit that the domain of the $t_i$ and $s_i$ variables is rectangular.}
\begin{equation}
    s_i =\hspace{0.1cm} u_i - v_i\quad\quad\text{and}\quad\quad   t_i = \hspace{0.1cm} u_i + v_i - 1\,.
\end{equation}
Hence the final spectrum yields 
\begin{equation}
\begin{aligned}
    \Omega_{\rm GW}(k,\eta)|_{\rm hybrid} =&\hspace{0.1cm} \frac{1}{12} \left(\frac{k}{a(\eta)H(\eta)}\right)^2 f_{\rm NL}^2
    \int_0^{\infty}dt_1\int_{-1}^1 ds_1\int_0^{\infty} dt_2\int_{-1}^1 ds_2\\
    &\times \frac{1}{(u_1 v_1 u_2 v_2)^2} \hspace{0.1cm} \overline{\Tilde{J}^2(u_1,v_1,x)}\hspace{0.1cm}\Delta_g^2(u_1 k)\Delta_g^2(v_2v_1 k)\Delta_g^2(u_2v_1 k)\,.
    \label{eqhybcomp}
\end{aligned}
\end{equation}
The Feynman diagram for this contribution is shown on the left in the second row of Figure \ref{Fig::Feynman_fNL2}. Further details are reported in Appendix \ref{SubSec::Feynman_fNL2}.

\begin{figure}
    \centering
    \includegraphics[width=0.6\textwidth]{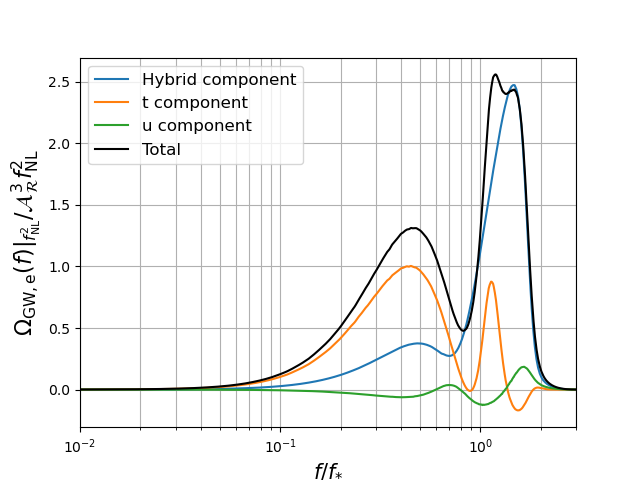}
    \caption{The plot shows the various terms at $\mathcal{O}(f_{\rm NL}^2)$. The t and u components correspond respectively to the C and Z components of \cite{2021JCAP...10..080A}. Their sum is reported in black.}
    \label{Fig::fNL2}
\end{figure}

The resulting GW spectra for the various components operating at $\mathcal{O}(f_{\rm NL}^2)$, normalized to the nG parameter $f_{\rm NL}^2$ and to the amplitude of the primordial curvature perturbation $\mathcal{A}_{\mathcal{R}}^3$ are shown in Figure \ref{Fig::fNL2}. We observe the presence of a particular feature, with the highest peak actually made of two contiguous peaks mainly deriving from the ``t'' component and the ``hybrid'' one. We label this behaviour as ``double peak'' feature in the following. 

\subsection{``s'' and ``\texorpdfstring{$g_{\rm NL}$}{g}'' components of the spectrum}

As described in \cite{2023JCAP...03..057G}, these two contributions are vanishing due to symmetry properties of the integrals. Thus, we report them just for completeness.
\begin{equation}
\begin{aligned}
    \Omega_{\rm GW}(k,\eta)|_{\rm s} =& \hspace{0.1cm} \frac{1}{12 \pi} \left(\frac{k}{a(\eta)H(\eta)}\right)^2 f_{\rm NL}^2 \int_0^{\infty}dt_1\int_{-1}^1 ds_1\int_0^{\infty} dt_2\int_{-1}^1 ds_2 \\
    &\hspace{2.5cm}\times\int_0^{2\pi}d\varphi_{12} \cos 2\varphi_{12} u_1v_1u_2v_2 \overline{\Tilde{J}(u_1,v_1,x)\Tilde{J}(u_2,v_2,x)}\\
    &\hspace{2.5cm}\times\frac{\Delta_g^2(v_1 k)}{v_1^3}\frac{\Delta_g^2(v_2 k)}{v_2^3}\Delta_g^2(k) \\
    =& \hspace{0.1cm} 0\,.
\end{aligned}
\end{equation}

\begin{equation}
\begin{aligned}
    \Omega_{\rm GW}(k,\eta)|_{\rm g_{\rm NL}} =&\hspace{0.1cm} \frac{1}{8 \pi} \left(\frac{k}{a(\eta)H(\eta)}\right)^2 g_{\rm NL}\int_0^{\infty}dt_1\int_{-1}^1 ds_1\int_0^{\infty} dt_2\int_{-1}^1 ds_2 \\
    &\hspace{2.5cm}\times\int_0^{2\pi}d\varphi_{12}\cos 2\varphi_{12}u_1v_1u_2v_2\overline{\Tilde{J}(u_1,v_1,x)\Tilde{J}(u_2,v_2,x)}\\
    &\hspace{2.5cm}\times\frac{\Delta_g^2(v_1 k)}{v_1^3}\frac{\Delta_g^2(v_2 k)}{v_2^3}\frac{\Delta_g^2(u_1 k)}{u_1^3}\\
    =& \hspace{0.1cm} 0 \,.
\end{aligned}
\end{equation}
The Feynman diagrams for the ``s'' and $g_{\rm NL}$ contributions are shown respectively on the right in the second row of Figure \ref{Fig::Feynman_fNL2} and on the left of Figure \ref{Fig::Feynman_gNL}. Further details are reported in Appendix \ref{SubSec::Feynman_fNL2} and \ref{SubSec::Feynman_gNL}.

\subsection{``new'' component of the spectrum}
\label{SubSec::new}
\begin{figure}[t!]
    \centering
    \includegraphics[width=0.8\textwidth]{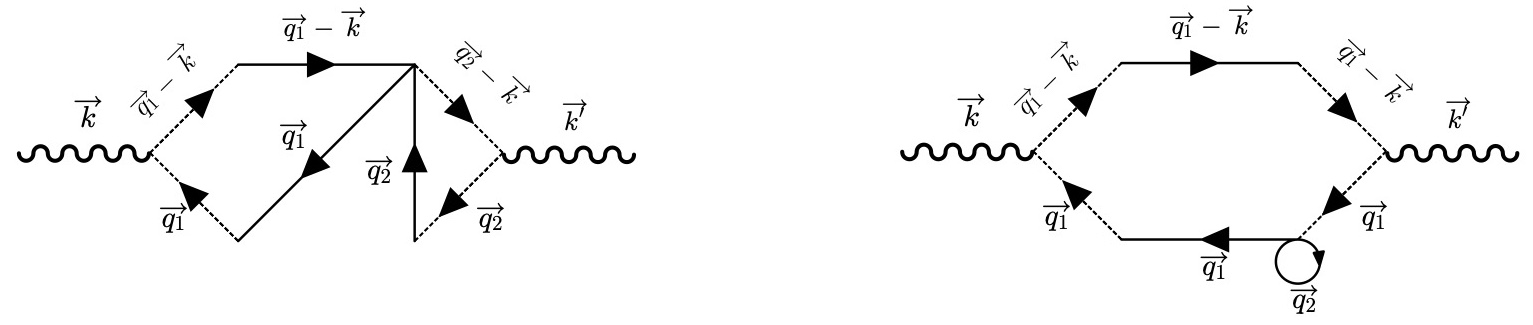}
    \caption{Feynman diagrams corresponding to the connected and disconnected $g_{\rm NL}$ contributions. The diagram on the left corresponds to the (vanishing) $g_{\rm NL}$ term, while the remaining one to the ``new'' contribution.}
    \label{Fig::Feynman_gNL}
\end{figure}
\begin{figure}
    \centering
    \includegraphics[width=0.6\textwidth]{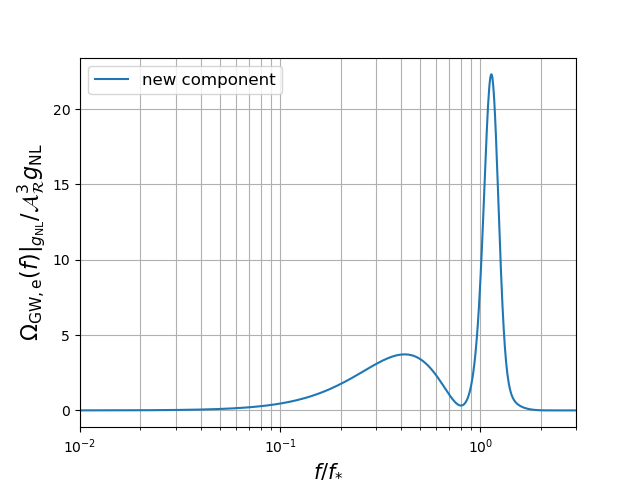}
    \caption{This plot shows the $\mathcal{O}(g_{\rm NL})$ component. We call name it ``new''  in the main text, since this contribution was not considered by \cite{2021JCAP...10..080A}.}
    \label{Fig::gNL}
\end{figure}

The ``new'' contribution coming from equation \eqref{Eq::new}, is studied in this paper for the first time. 
It arises naturally when the trispectrum for the primordial scalar perturbation, expanded at least up to third-order in a local model of nG, is computed via Wick's theorem. It is interesting to notice that, apart from numerical factors, it has the same shape of the ``Gaussian'' component multiplied by a further integration of the power spectrum. Hence
\begin{equation}
    \begin{aligned}
    P_{h,\lambda}(k,\eta)\big|_{\rm new} =& \hspace{0.1cm} 12 \hspace{0.1cm}g_{\rm NL} \langle\mathcal{R}_g^2\rangle P_{h,\lambda}(k,\eta)\big|_{\rm Gaussian}\,.
    \end{aligned}
\end{equation}
This form makes clear how this component can have a large impact on the total spectrum of GWs since, depending on the value of $g_{\rm NL}$, it could be of the same order of the ``Gaussian'' term or even become the dominant contribution.
The final expression reads\footnote{This result follows since the variance for a log-normal power spectrum corresponds simply to the amplitude $\mathcal{A}_{\mathcal{R}}$.}
\begin{equation}
\label{Eq::new}
\begin{aligned}
    \Omega_{\rm GW}(k,\eta)|_{\rm new} =& \hspace{0.1cm}12 \hspace{0.1cm}g_{\rm NL} \hspace{0.1cm}\mathcal{A}_{\mathcal{R}}\hspace{0.1cm}\Omega_{\rm GW}(k,\eta)|_{\rm Gaussian}\,, 
\end{aligned}
\end{equation}
and it does not need to be computed, once the Gaussian spectrum is known. The Feynman diagram for this contribution is shown on the right in Figure \ref{Fig::Feynman_gNL}. Further details are reported in Appendix \ref{SubSec::Feynman_gNL}. We emphasize that this component can be present also in the case where the other non-Gaussian components at the same order in perturbation theory ``t'', ``u'' and ``hybrid''  are vanishing: when $f_{\rm NL} = 0$, the $g_{\rm NL}$ parameter could still provide a non-vanishing contribution that would also depend on the sign of $g_{\rm NL}$.
The resulting spectrum is reported in Figure \ref{Fig::gNL}.

\subsection{Order \texorpdfstring{$\mathcal{O}(f^2_{\rm NL}g_{\rm NL})$}{fg} components}
\begin{figure}
    \centering
    \includegraphics[width=0.8\textwidth]{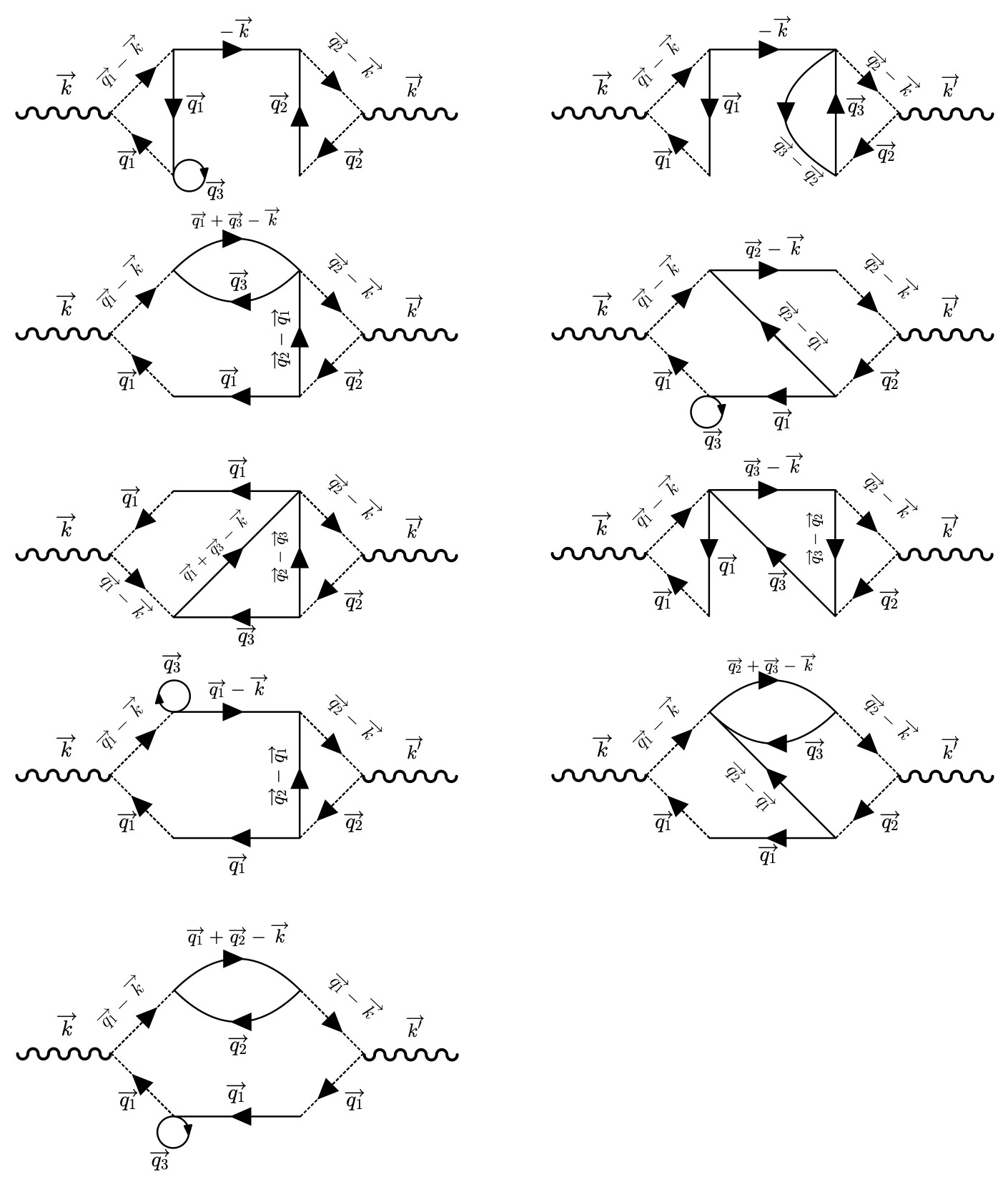}
    \caption{Feynman diagram corresponding to the connected and disconnected $f_{\rm NL}^2g_{\rm NL}$ contributions.}
    \label{Fig::Feynman_fNL2gNL}
\end{figure}
\begin{figure}
    \centering
    \includegraphics[width=0.6\textwidth]{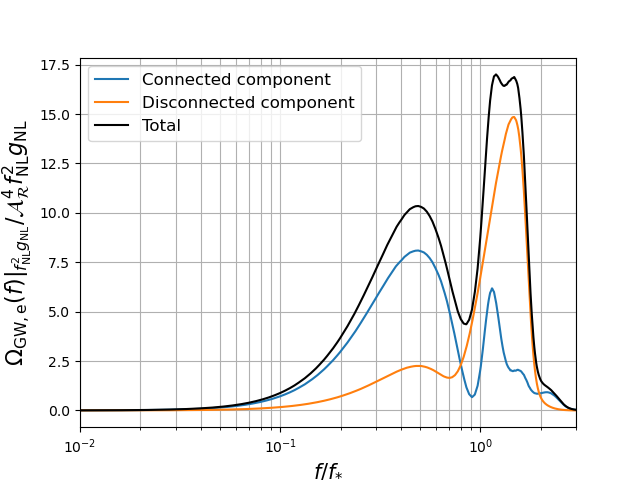}
    \caption{The plot shows the contributions at $\mathcal{O}(f_{\rm NL}^2g_{\rm NL})$. We report the connected component \eqref{Eq::fNL2gNL_Connected} in blue and the disconnected component \eqref{Eq::fNL2gNL_Disconnected} in orange. Their sum is reported in black.}
    \label{Fig::fNL2_gNL}
\end{figure}

The same procedure described above can be used to evaluate the terms contributing at higher orders, accounting for the additional integration over the internal momentum $\textbf{q}_3$ (see again Appendix \ref{manipulation}). In the case of the $\mathcal{O}(f^2_{\rm NL}g_{\rm NL})$ terms reported in equations \eqref{fnl2-gnl} and \eqref{fnl2-gnl-disc}, the two components can be treated separately. The first line of the connected component vanishes for the same reasoning done for the ``s'' and the ``$g_{\rm NL}$'' ones, since only the cosine coming from the polarization sum of the projection factors depends on the angle $\varphi_{12} = \phi_1-\phi_2$. The same happens in the second and the last lines when the change of variables $\phi_1 = \phi_2 + \varphi_{12}$ is performed, since $|\textbf{q}_2-\textbf{q}_3|$ is proportional to $\cos(\phi_2-\phi_3)$ and not affected by this change. All the other terms can be computed performing the same two changes of variables, with $\varphi_{ij} = \phi_i-\phi_j$. Finally one obtains
\begin{equation}
\label{Eq::fNL2gNL_Connected}
    \begin{aligned}
    \Omega_{\rm GW}(k,\eta)\big|_{\rm f_{\rm NL}^2g_{\rm NL}} =&\hspace{0.1cm} \frac{1}{16\pi^2}\left(\frac{k}{a(\eta)H(\eta)}\right)^2f_{\rm NL}^2g_{\rm NL}\prod_{i=1}^3\left[\int_0^\infty dt_i\int_{-1}^1 ds_i v_iu_i\right] \\
    & \hspace{1cm}\times\int_0^{2\pi}d\varphi_{12}\hspace{0.1cm}\overline{\Tilde{J}(u_1,v_1,x)\Tilde{J}(u_2,v_2,x)}\cos(2\varphi_{12})\\
    &\hspace{1cm}\times\Bigg[\hspace{0.1cm}\int_0^{2\pi}d\varphi_{23}\frac{\Delta^2_g(v_1k)}{v_1^3}\frac{\Delta^2_g(w_{a,12}k)}{w_{a,12}^3}\frac{\Delta^2_g(v_3k)}{v_3^3}\frac{\Delta^2_g(w_{b,13}k)}{w_{b,13}^3}\\
    &\hspace{1.7cm}+2\int_0^{2\pi}d\varphi_{23}\frac{\Delta^2_g(v_1k)}{v_1^3}\frac{\Delta^2_g(v_3k)}{v_3^3}\frac{\Delta^2_g(w_{b,13}k)}{w_{b,13}^3}\frac{\Delta^2_g(w_{a,23}k)}{w_{a,23}^3}\\
    &\hspace{1.7cm}+\int_0^{2\pi}d\varphi_{23}\frac{\Delta^2_g(v_1k)}{v_1^3}\frac{\Delta^2_g(w_{a,12}k)}{w_{a,12}^3}\frac{\Delta^2_g(v_3k)}{v_3^3}\frac{\Delta^2_g(w_{b,23}k)}{w_{b,23}^3}\hspace{0.1cm}\Bigg]\\
    &+6g_{\rm NL}\langle\mathcal{R}_g^2\rangle\left(\Omega_{\rm GW}(k,\eta)\big|_{\rm t} + \Omega_{\rm GW}(k,\eta)\big|_{\rm u}\right)\,.
    \end{aligned}  
\end{equation}
For the disconnected part we get
\begin{equation}
\label{Eq::fNL2gNL_Disconnected}
    \begin{aligned}
     \Omega_{\rm GW}(k,\eta)\big|_{\rm f_{\rm NL}^2g_{\rm NL},d} =&\hspace{0.1cm} 6 \hspace{0.1cm}g_{\rm NL}  \langle\mathcal{R}_g^2\rangle \Omega_{\rm GW}(k,\eta)\big|_{\rm hybrid}\,.
    \end{aligned}
\end{equation}
The resulting GW power spectra, normalized to the non-Gaussian parameters $f_{\rm NL}^2g_{\rm NL}$ and to the primordial amplitude $\mathcal{A}_{\mathcal{R}}^4$ are shown in Figure \ref{Fig::fNL2_gNL}. The ``double peak'' feature is still present but with the two peaks of almost the same height. This behaviour originates from the connected contribution that goes to 0 less sharply with respect to the $f_{\rm NL}^2$ case. The corresponding Feynman diagrams are reported in Figure \ref{Fig::Feynman_fNL2gNL} (see also Appendix \ref{SubSec::Feynman_Higher}).

\subsection{Order \texorpdfstring{$\mathcal{O}(g_{\rm NL}^2)$}{g} components}
\begin{figure}
    \centering
    \includegraphics[width=0.8\textwidth]{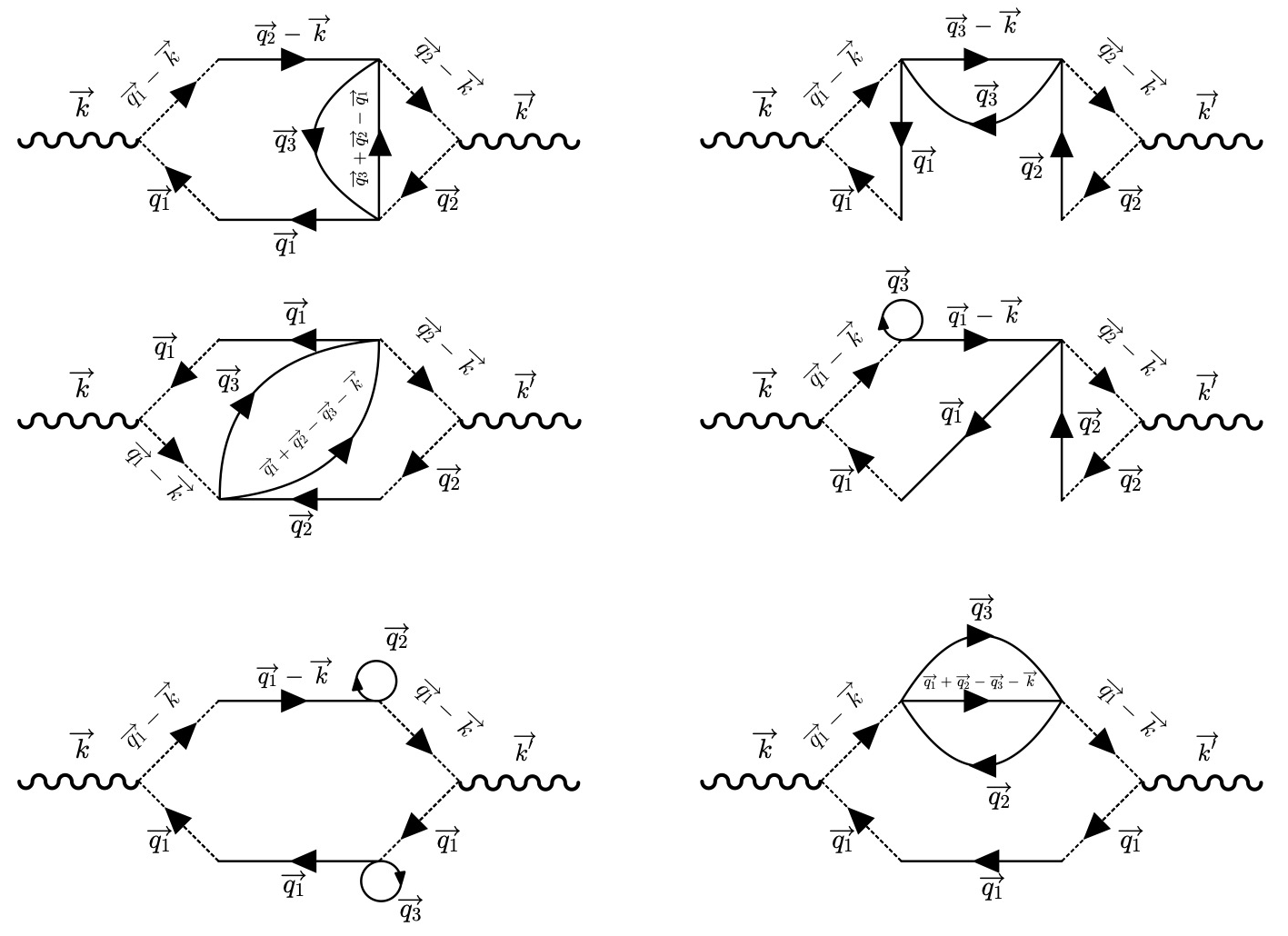}
    \caption{Feynman diagrams corresponding to the connected and disconnected $g_{\rm NL}^2$ contributions.}
    \label{Fig::Feynman_gNL2}
\end{figure}
For the components at order $\mathcal{O}(g_{\rm NL}^2)$, shown in equations \eqref{gnl2} and \eqref{gnl2-disc},  the second and the fourth lines of the connected part vanish again due to the integration of the cosine. The results for the other terms are
\begin{equation}
\label{Eq::gNL2_Connected}
    \begin{aligned}
    \Omega_{\rm GW}(k,\eta)\big|_{\rm g_{\rm NL}^2} =&\hspace{0.1cm} \frac{3}{64\pi^2}\left(\frac{k}{a(\eta)H(\eta)}\right)^2g^2_{\rm NL}\prod_{i=1}^3\left[\int_0^\infty dt_i\int_{-1}^1 ds_i v_iu_i\right]\\
    &\hspace{3cm}\times\int_0^{2\pi}d\varphi_{12}\int_0^{2\pi}d\varphi_{23}\overline{\Tilde{J}(u_1,v_1,x)\Tilde{J}(u_2,v_2,x)}\cos(2\varphi_{12})\\
    &\hspace{3cm}\times\Bigg[\hspace{0.1cm}\frac{\Delta^2_g(v_1k)}{v_1^3}\frac{\Delta^2_g(u_{1}k)}{u_{1}^3}\frac{\Delta^2_g(v_3k)}{v_3^3}\frac{\Delta^2_g(w_{q123}k)}{w_{q123}^3}\\
    &\hspace{3.4cm}+\frac{\Delta^2_g(v_1k)}{v_1^3}\frac{\Delta^2_g(v_{2}k)}{v_{2}^3}\frac{\Delta^2_g(v_3k)}{v_3^3}\frac{\Delta^2_g(w_{123}k)}{w_{123}^3}\hspace{0.1cm}\Bigg]
    \end{aligned}   
\end{equation}
\begin{equation}
\label{Eq::gNL2_Disconnected}
    \begin{aligned}
     \Omega_{\rm GW}(k,\eta)\big|_{\rm g^2_{\rm NL},disc} =& \hspace{0.1cm}54\hspace{0.1cm}g_{\rm NL}^2\langle\mathcal{R}_g^2\rangle^2 \Omega_{\rm GW}(k,\eta)\big|_{\rm Gaussian}\\
     &+\frac{1}{64\pi^2} \left(\frac{k}{a(\eta)H(\eta)}\right)^2 g^2_{\rm NL}\prod_{i=1}^3\left[\int_0^\infty dt_i\int_{-1}^1 ds_i  v_iu_i\right]\hspace{0.1cm}\overline{\Tilde{J}^2(u_1,v_1,x)}\\
     &\hspace{0.3cm}\times\int_0^{2\pi}d\varphi_{12}\int_0^{2\pi}d\varphi_{13}\frac{\Delta^2_g(v_1k)}{v_1^3}\frac{\Delta^2_g(v_{2}k)}{v_{2}^3}\frac{\Delta^2_g(v_3k)}{v_3^3}\frac{\Delta^2_g(w_{123}k)}{w_{123}^3}
    \end{aligned}
\end{equation}

\begin{figure}
    \centering
    \includegraphics[width=0.7\textwidth]{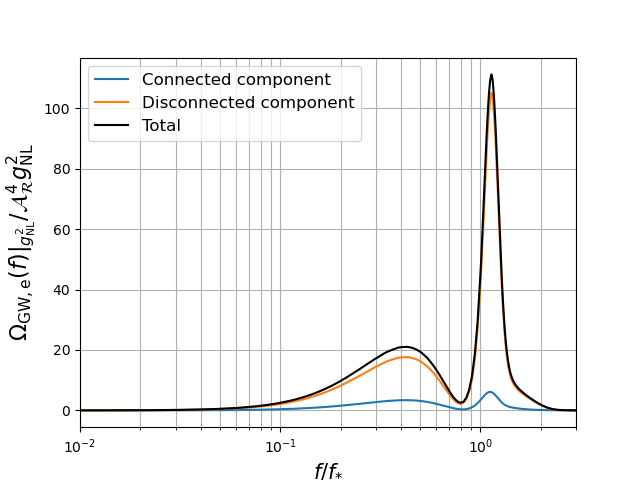}
    \caption{The plot shows the different contributions at $\mathcal{O}(g_{\rm NL}^2)$. We report the connected contribution, equation \eqref{Eq::gNL2_Connected}, in blue and the disconnected contribution, equation \eqref{Eq::gNL2_Disconnected}, in orange. Their sum is reported in black.}
    \label{Fig::gNL2}
\end{figure}

In Figure \ref{Fig::gNL2} we plot these contributions, normalized to the nG parameter $g_{\rm NL}^2$ and to the amplitude $\mathcal{A}_{\mathcal{R}}^4$. In this case we note that the first line of the disconnected component is proportional to the ``Gaussian'' spectrum. We also add that the disconnected contribution dominates the spectrum, being one order of magnitude greater than the connected one. The corresponding Feynman diagrams are reported in Figure \ref{Fig::Feynman_gNL2} (see also Appendix \ref{SubSec::Feynman_Higher}).

\subsection{Order \texorpdfstring{$\mathcal{O}(f_{\rm NL}^4)$}{f} components}
\begin{figure}
    \centering
    \includegraphics[width=0.8\textwidth]{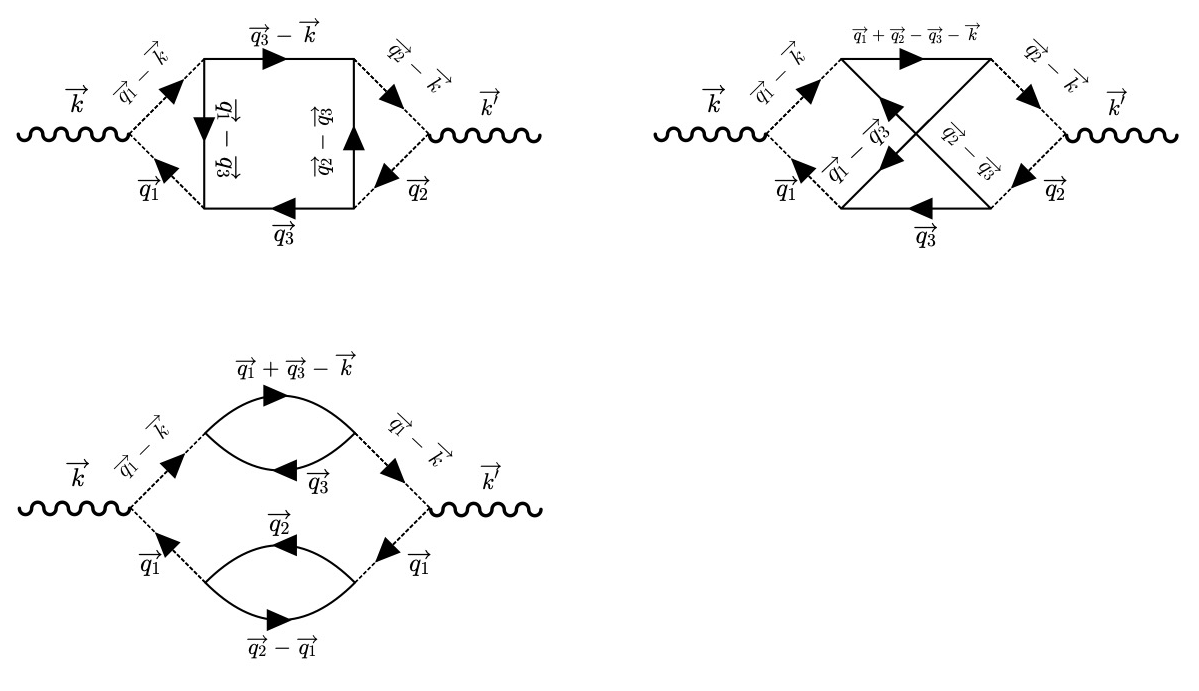}
    \caption{Feynman diagram corresponding to the connected and disconnected $f_{\rm NL}^4$ contributions.}
    \label{Fig::Feynman_fNL4}
\end{figure}

\begin{figure}
    \centering
    \includegraphics[width=0.6\textwidth]{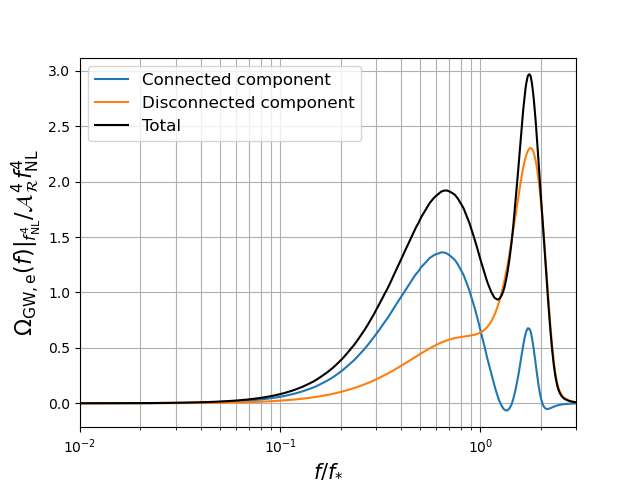}
    \caption{This plot shows the contributions at $\mathcal{O}(f_{\rm NL}^4)$. The connected component, eq. \eqref{Eq::fNL4_Connected}, contains both the ``planar'' and ``non-planar'' ones of \cite{2021JCAP...10..080A}, while the disconnected component, eq. \eqref{Eq::fNL4_Disconnected}, corresponds to the ``reducible'' one. Their sum is reported in black.}
    \label{Fig::fNL4}
\end{figure}

At order $\mathcal{O}(f_{\rm NL}^4)$, the terms coming from equations \eqref{fnl4} and \eqref{fnl4-disc} are
\begin{equation}
\label{Eq::fNL4_Connected}
    \begin{aligned}
    \Omega_{\rm GW}(k,\eta)\big|_{\rm f_{\rm NL}^4} =&\hspace{0.1cm} \frac{1}{96\pi^2}\left(\frac{k}{a(\eta)H(\eta)}\right)^2f^4_{\rm NL}\prod_{i=1}^3\left[\int_0^\infty dt_i\int_{-1}^1 ds_i v_iu_i\right]\\
    &\hspace{1.6cm}\times\int_0^{2\pi}d\varphi_{12}\int_0^{2\pi}d\varphi_{23}\overline{\Tilde{J}(u_1,v_1,x)\Tilde{J}(u_2,v_2,x)}\cos(2\varphi_{12})\\
    &\hspace{1.6cm}\times\Bigg[\hspace{0.1cm}2\frac{\Delta^2_g(v_3k)}{v_3^3}\frac{\Delta^2_g(u_{3}k)}{u_{3}^3}\frac{\Delta^2_g(w_{a,13}k)}{w_{a,13}^3}\frac{\Delta^2_g(w_{a,23}k)}{w_{a,23}^3}\\
    &\hspace{2cm}+\frac{\Delta^2_g(v_3k)}{v_3^3}\frac{\Delta^2_g(w_{a,13}k)}{w_{a,13}^3}\frac{\Delta^2_g(w_{a,23}k)}{w_{a,23}^3}\frac{\Delta^2_g(w_{123}k)}{w_{123}^3}\hspace{0.1cm}\Bigg]
    \end{aligned}   
\end{equation}
\begin{equation}
\label{Eq::fNL4_Disconnected}
    \begin{aligned}
     \Omega_{\rm GW}(k,\eta)\big|_{\rm f^4_{\rm NL},d} =&\hspace{0.1cm} \frac{1}{192\pi^2} \left(\frac{k}{a(\eta)H(\eta)}\right)^2 f^4_{\rm NL}\prod_{i=1}^3\left[\int_0^\infty dt_i\int_{-1}^1 ds_i  v_iu_i\right]\\
     &\hspace{1.6cm}\times\int_0^{2\pi}d\varphi_{12}\int_0^{2\pi}d\varphi_{23}\hspace{0.1cm}\overline{\Tilde{J}^2(u_1,v_1,x)}\\
     &\hspace{1.6cm}\times\frac{\Delta^2_g(v_2k)}{v_2^3}\frac{\Delta^2_g(v_3k)}{v_3^3}\frac{\Delta^2_g(w_{a,12}k)}{w_{a,12}^3}\frac{\Delta^2_g(w_{b,13}k)}{w_{b,13}^3}\,.
    \end{aligned}
\end{equation}
These contributions correspond to the ``planar'', ``non-planar'' and ``reducible'' components of \cite{2021JCAP...10..080A}, respectively. The resulting spectra, obtained through numerical integration, are shown in Figure \ref{Fig::fNL4}. The corresponding Feynman diagrams are reported in Figure \ref{Fig::Feynman_fNL4} (see also Appendix \ref{SubSec::Feynman_Higher}).

\subsection{Order \texorpdfstring{$\mathcal{O}(f_{\rm NL}h_{\rm NL})$}{fh} components}

\begin{figure}
    \centering
    \includegraphics[width=0.8\textwidth]{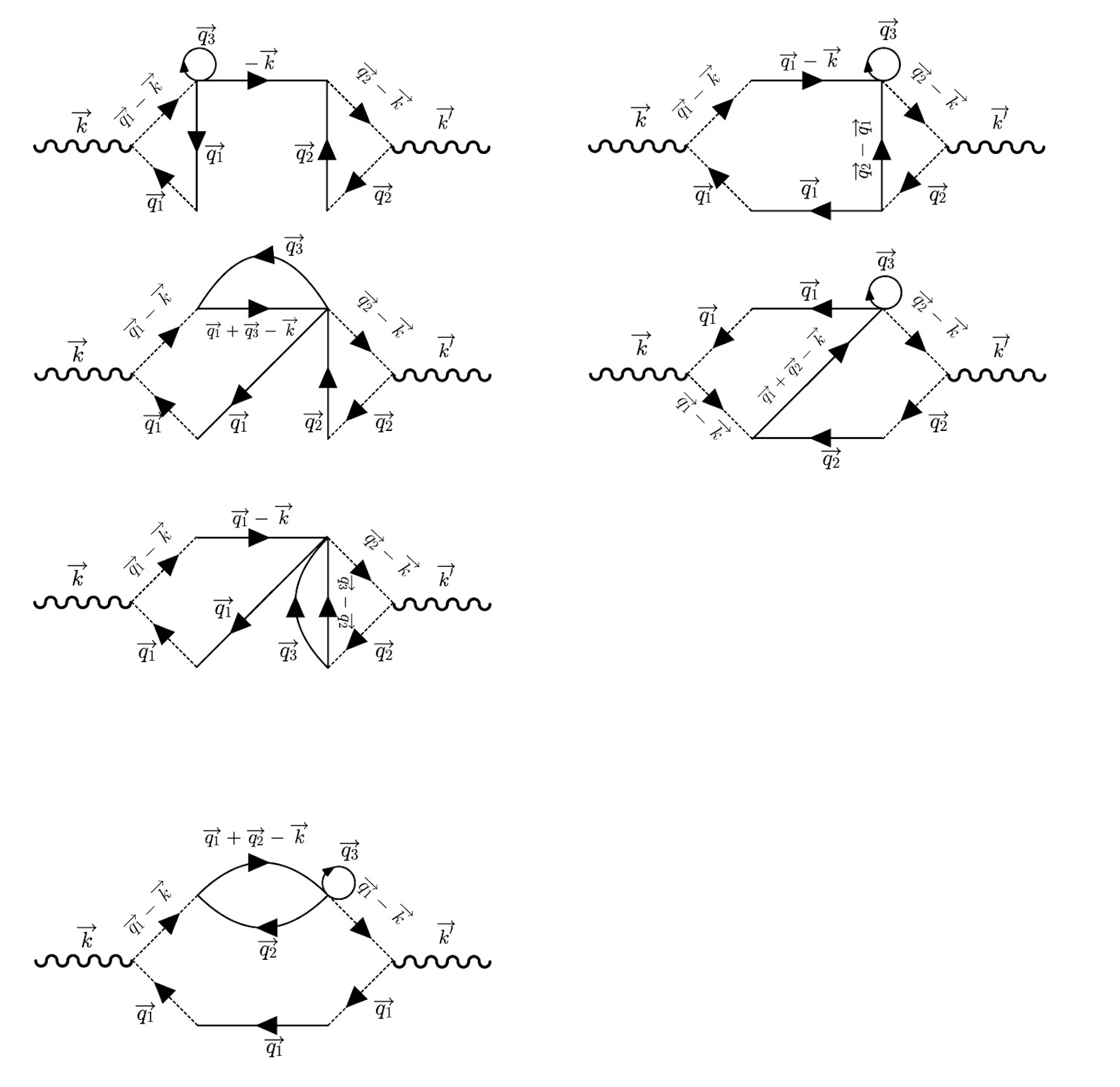}
    \caption{Feynman diagram corresponding to the connected and disconnected $f_{\rm NL}h_{\rm NL}$ contributions.}
    \label{Fig::Feynman_fNLhNL}
\end{figure}

\label{SubSec::fNLhNL}
At order $\mathcal{O}(f_{\rm NL}h_{\rm NL})$ the SIGW spectral density arising from equations \eqref{fnl-hnl} and \eqref{fnl-hnl-disc} is
\begin{equation}
\label{Eq::fNLhNL_Connected}
    \begin{aligned}
    \Omega_{\rm GW}(k,\eta)\big|_{\rm f_{\rm NL}h_{\rm NL}} =&\hspace{0.1cm}  12 \hspace{0.1cm}\frac{h_{\rm NL}}{f_{\rm NL}}  \langle\mathcal{R}_g^2\rangle \left(\Omega_{\rm GW}(k,\eta)\big|_{\rm t} + \Omega_{\rm GW}(k,\eta)\big|_{\rm u}\right)
    \end{aligned}   
\end{equation}
since the terms in square brackets in the second step of \eqref{fnl-hnl} vanish. The disconnected component reads
\begin{equation}
\label{Eq::fNLhNL_Disconnected}
    \begin{aligned}
     \Omega_{\rm GW}(k,\eta)\big|_{\rm f_{\rm NL}h_{\rm NL},d}
     =&\hspace{0.1cm}  12 \hspace{0.1cm}\frac{h_{\rm NL}}{f_{\rm NL}}  \langle\mathcal{R}_g^2\rangle \Omega_{\rm GW}(k,\eta)\big|_{\rm hybrid}
    \end{aligned}
\end{equation}    

\begin{figure}
    \centering
    \includegraphics[width=0.6\textwidth]{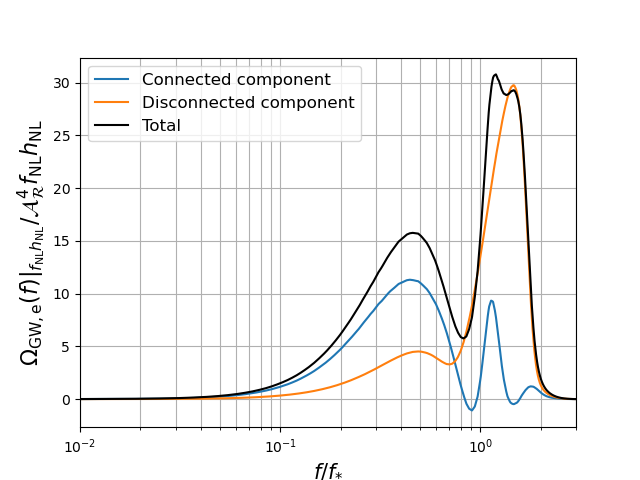}
    \caption{The plot shows the contributions at $\mathcal{O}(f_{\rm NL}h_{\rm NL})$. The connected component, eq. \eqref{Eq::fNLhNL_Connected}, is reported in blue, while the disconnected component, \eqref{Eq::fNLhNL_Disconnected} is reported in orange. Their sum is reported in black.}
    \label{Fig::fNL_hNL}
\end{figure}

The spectra for this component contributing to the scalar-induced GW spectrum are reported in Figure \ref{Fig::fNL_hNL}. Also in this case the ``double-peak'' feature is present, similarly to the $f_{\rm NL}^2$ case. We note that the final shape of these contribution is exactly the same of the $f_{NL}^2$ one at the next-to-leading order. This is because the non-vanishing components of the connected and disconnected terms, are actually proportional by the same factor to the u,t and hybrid terms. The corresponding Feynman diagrams are reported in Figure \ref{Fig::Feynman_fNLhNL} (see also Appendix \ref{SubSec::Feynman_Higher}).

\subsection{Order \texorpdfstring{$\mathcal{O}(i_{\rm NL})$}{i} components}
\begin{figure}
    \centering
    \includegraphics[width=0.4\textwidth]{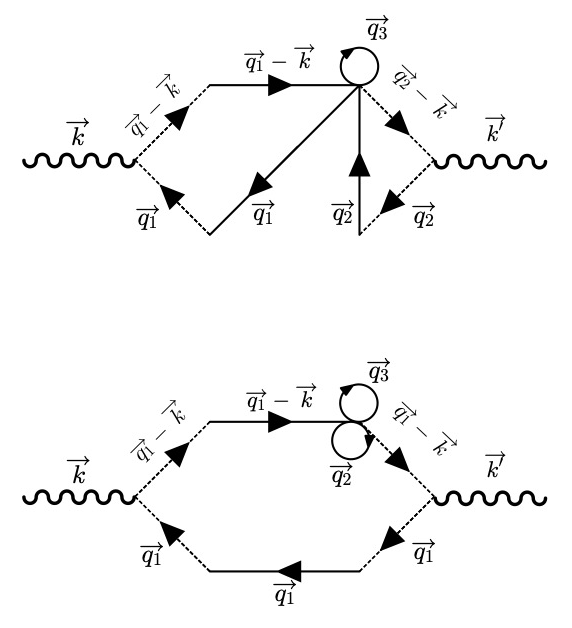}
    \caption{Feynman diagram corresponding to the connected and disconnected $i_{\rm NL}$ contributions.}
    \label{Fig::Feynman_iNL}
\end{figure}
\begin{figure}
    \centering
    \includegraphics[width=0.6\textwidth]{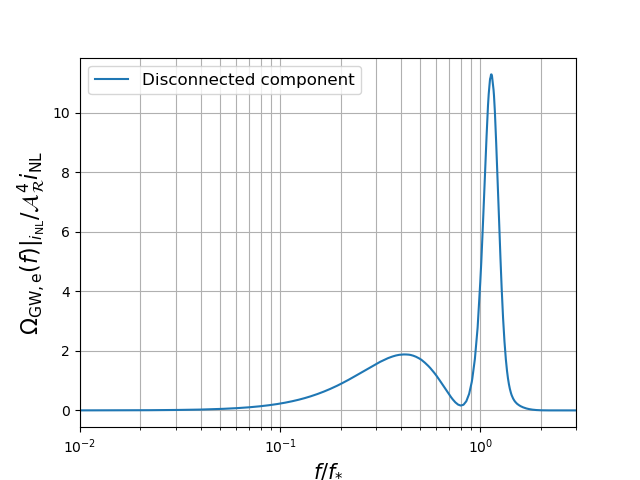}
    \caption{The plot shows the contributions at $\mathcal{O}(i_{\rm NL})$. We report only the disconnected contribution since the connected one vanishes.}
    \label{Fig::iNL}
\end{figure}

Lastly, the connected component at order $\mathcal{O}(i_{\rm NL})$ in equation \eqref{inl} vanishes for symmetry reasons. The only non-vanishing contribution is the disconnected one, \eqref{inl-disc},
\begin{equation}
\begin{aligned}
\label{Eq::iNL_Disconnected}
    \Omega_{\rm GW}(k,\eta)|_{\rm i_{\rm NL},disc} =& \hspace{0.1cm}60 \hspace{0.1cm}i_{\rm NL} \langle\mathcal{R}_g^2\rangle^2\hspace{0.1cm}\Omega_{\rm GW}(k,\eta)|_{\rm Gaussian}\,.
\end{aligned}
\end{equation}
We report the spectrum in Figure \ref{Fig::iNL}, obtained by simply multiplying the ``Gaussian'' one by a proper factor. The corresponding Feynman diagrams are reported in Figure \ref{Fig::Feynman_iNL} (see also Appendix \ref{SubSec::Feynman_Higher}). 
We conclude noting that a similar contribution is expected at each order in the expansion. When considering the $N$-point correlation function (with $N$ even) it will always be possible to build a correlator with three simple Gaussian fields and a term proportional to $m_{\rm NL}\, R_g^m$, with $m=N-3$ (e.g. for the 6-point function we had $m = 3$ and we had the ``new’’ contribution, while for the 8-point function we have $m = 5$ and we originate the $i_{\rm NL}$ term). The number $m$ defines the highest possible power in the local expansion at the order considered (accounting for a term proportional to $\mathcal{R}_g^{m+1}$ would lead to a $(N+1)$-point correlation function that, since $N$ is even, will be vanishing for a Gaussian field). The corresponding Feynman diagram would have the same topology of the ``new'' or $i_{\rm NL}$ ones, but with $(m-1)/2$ factorizable loops\footnote{This can be understood as follows: among the $m$ possible lines arising from the $m_{\rm NL}$ vertex, only one will correlate with a Gaussian field. The remaining $m-1$ lines will form a closed loop in pairs.}.

\subsection{SIGWB signal vs LISA Sensitivity}

In this subsection we compare the spectral density of GWs with the sensitivity of the LISA detector, once all the different contributions are summed together. Since all the spectra discussed until now are related to the \textit{emitted} spectrum of GW, namely $\Omega_{\rm GW,e}(k)$, we need to redshift them to present times via equation \eqref{altempoattuale}.
As mentioned in Appendix \ref{manipulation}, the range of frequencies probed by LISA are sufficiently high to consider that the corresponding scales re-enter the horizon before the epoch of equivalence. With this assumption $\Omega_{\rm GW,e}(k) = \Omega_{\rm GW,eq}(k)$ and $g_{*,e} = g_{*,eq}$. As it is possible to verify in Figure 2 of \cite{2006PhRvD..73l3515W}, the effective number of relativistic degrees of freedom is fixed to $g_{*} \sim 3.36$ for temperatures of the universe below $T \sim 0.1\, \si{\mega\electronvolt}$. This allows to set $\left(g_{*,0}/g_{*,e}\right)^{1/3} = 1$ in the above equation, since the temperature at equivalence was $T_{eq} \sim 0.8\, \si{\electronvolt}$. Recalling that $\Omega_{rad,0}h^2 \sim 4.2 \times 10^{-5}$ it is straightforward to obtain the spectrum of scalar-induced GW observed today.\par

In the following we report the plots obtained varying the nG parameters in order to observe their effect on the spectrum. The choice of the parameters is to highlight the effects of the different contributions. We input $\mathcal{A}_{\mathcal{R}} = 10^{-2}$, $\sigma = 1/10$ and $f_* = 0.005$ Hz. 

\begin{figure}
    \centering
    \includegraphics[width=1\textwidth]{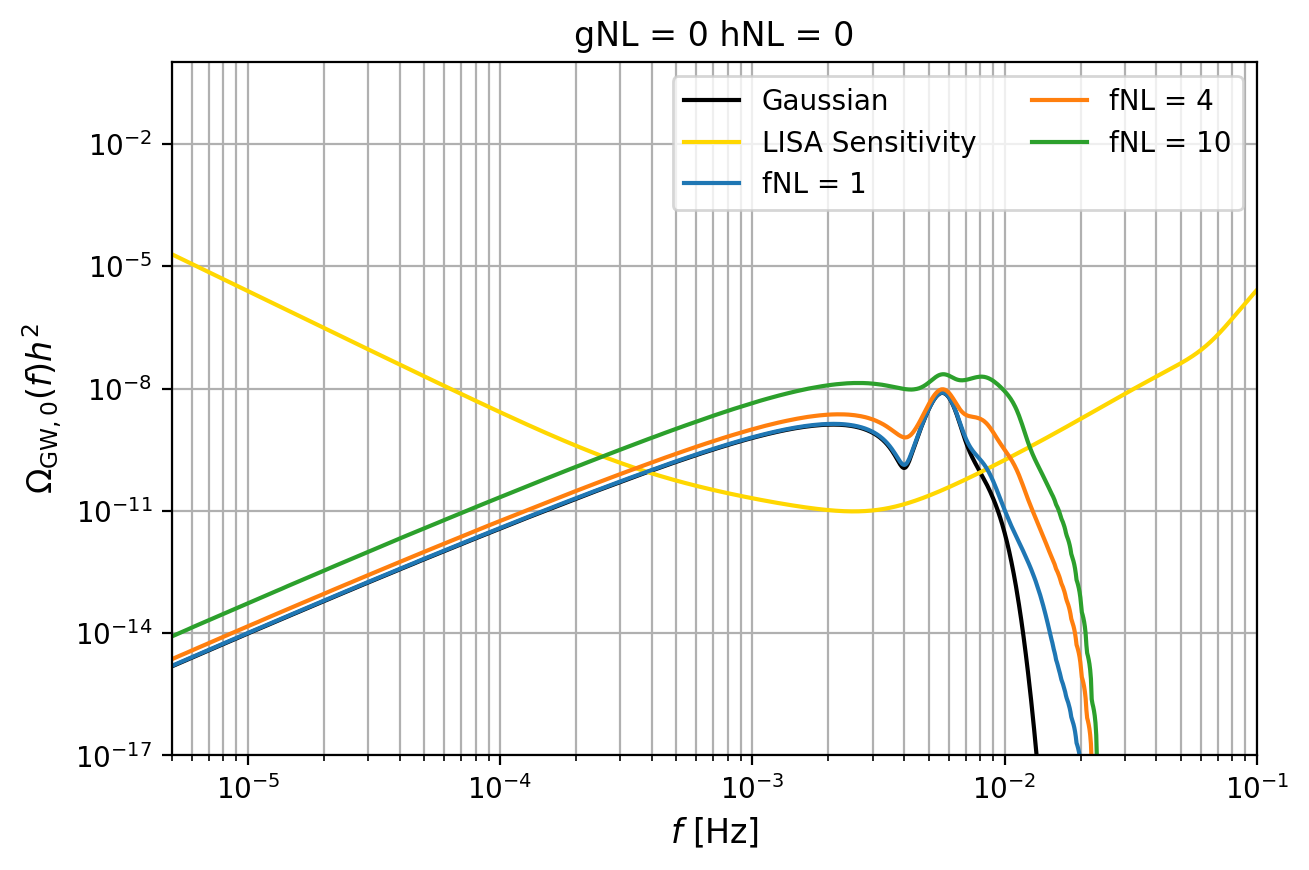}
    \caption{The figure shows the GW spectrum obtained for different values of $f_{\rm NL}$, fixing $g_{\rm NL}$ and $h_{\rm NL}$ to 0.}
    \label{Fig::Varying_fNL}
\end{figure}

We report in Figure \ref{Fig::Varying_fNL} the resulting total spectra obtained considering only those contributions proportional to any power of $f_{\rm NL}$, up to order $\mathcal{A}_{\mathcal{R}}^4$. We consider three different values of $f_{\rm NL}$ to show how the corresponding contributions affect the final spectrum. Firstly, we remark that such corrections will always be positive, being proportional to even powers of the nG parameter. For sufficiently high values of $f_{\rm NL}$, these terms could become high enough to be comparable to the Gaussian one, as it can be appreciated from the shift upwards of the resulting spectrum. The main imprint arising from these terms can be observed in the UV tail. The higher the value of $f_{\rm NL}$, in fact, the more the ``double peak'' feature is important, leading to a large enhancement of the spectrum at those frequencies. A further bump in the far UV tail arising from the $f_{\rm NL}^4$ spectrum could also be observed. We note that even in the case of $f_{\rm NL} = 1$ or 4, the variations of the spectrum in the UV range can still be appreciated and are mainly due to the $f_{\rm NL}^2$ corrections. 

\begin{figure}
    \centering
    \includegraphics[width=1\textwidth]{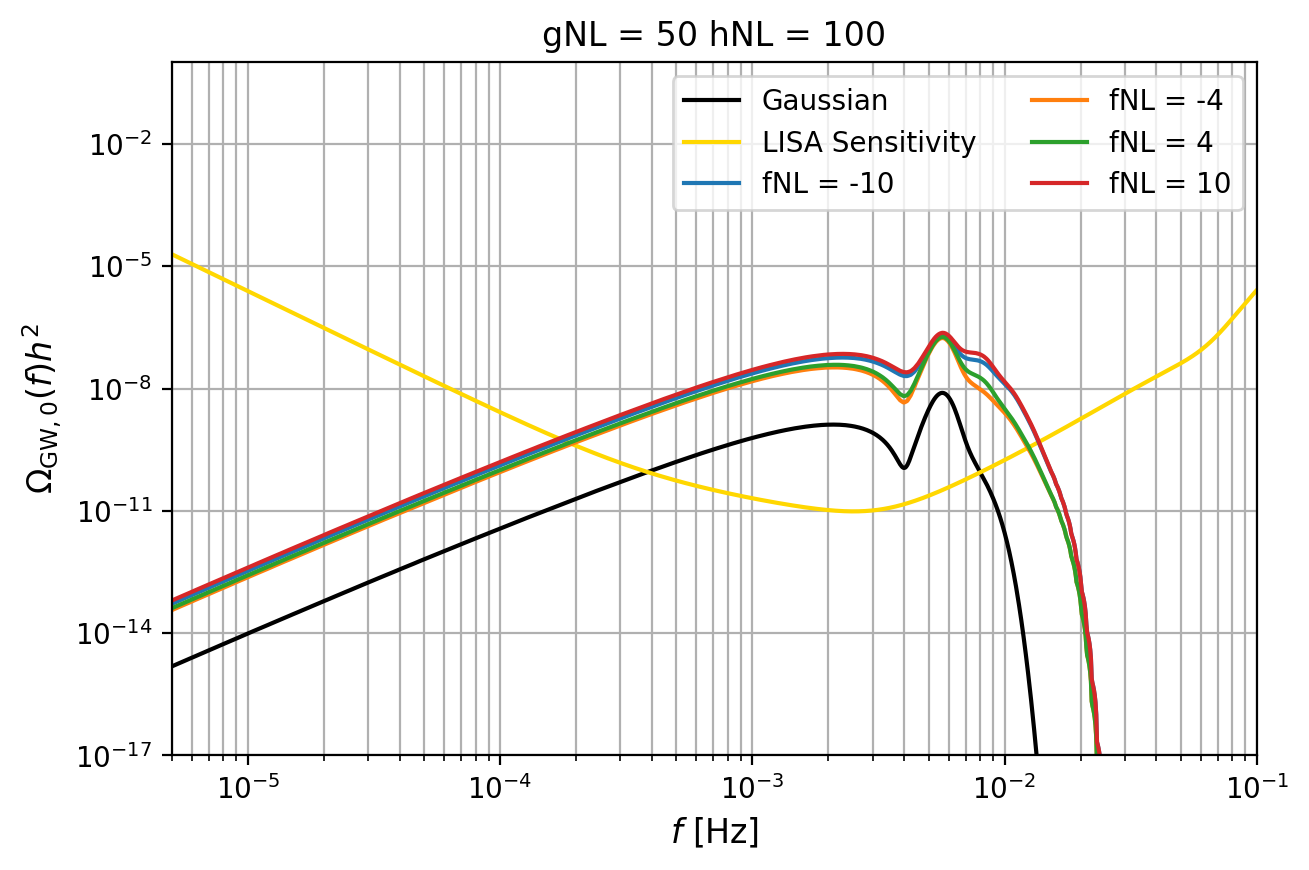}
    \caption{The figure shows the GW spectrum obtained for different values of $f_{\rm NL}$. In this case we consider non-vanishing values for $g_{\rm NL}$ and $h_{\rm NL}$ in order to account for their contributions to the GW spectrum.}
    \label{Fig::Varying_fNL_2}
\end{figure}

Instead, in Figure \ref{Fig::Varying_fNL_2} we report the spectra obtained accounting for different values of $f_{\rm NL}$ but also considering a non-vanishing $g_{\rm NL}$ and $h_{\rm NL}$. We do not include $i_{\rm NL}$ in the analysis since it provides a subdominant contribution, whose effect would just result in a shift upward or downward of the spectrum (in order to provide an appreciable change in the spectrum, $i_{\rm NL}$ has to be at least of order $10^4$, when $\mathcal{A}_{\mathcal{R}}=10^{-2}$ and even higher for lower values of the amplitude). Due to the non-vanishing values of $g_{\rm NL}$ and $h_{\rm NL}$, in this case more terms are expected to contribute. In the plot, the effect of $g_{\rm NL}$ consists mainly in a shift of the spectrum upward (coming also from the ``new'' term) and in a modification of the UV tail coming both from the $g_{\rm NL}^2$ and the $f_{\rm NL}^2 g_{\rm NL}$ terms (whose main effect consists in enhancing the resonance peak). A further modification of the tail comes from the $f_{\rm NL}h_{\rm NL}$ contribution that, as shown in Figure \ref{Fig::fNL_hNL}, mainly affects this part of the spectrum. More specifically, the effect of the $f_{\rm NL}h_{\rm NL}$ terms, even if in principle subdominant, depends on the sign of both $h_{\rm NL}$ and $f_{\rm NL}$. As Figure \ref{Fig::Varying_fNL_2} shows, in fact, for positive values of $h_{\rm NL}$, when $f_{\rm NL}$ is negative, the UV part after the peak results damped in the range corresponding to the peak shown in Figure \ref{Fig::fNL_hNL}.\par

\begin{figure}
    \centering
    \includegraphics[width=\textwidth]{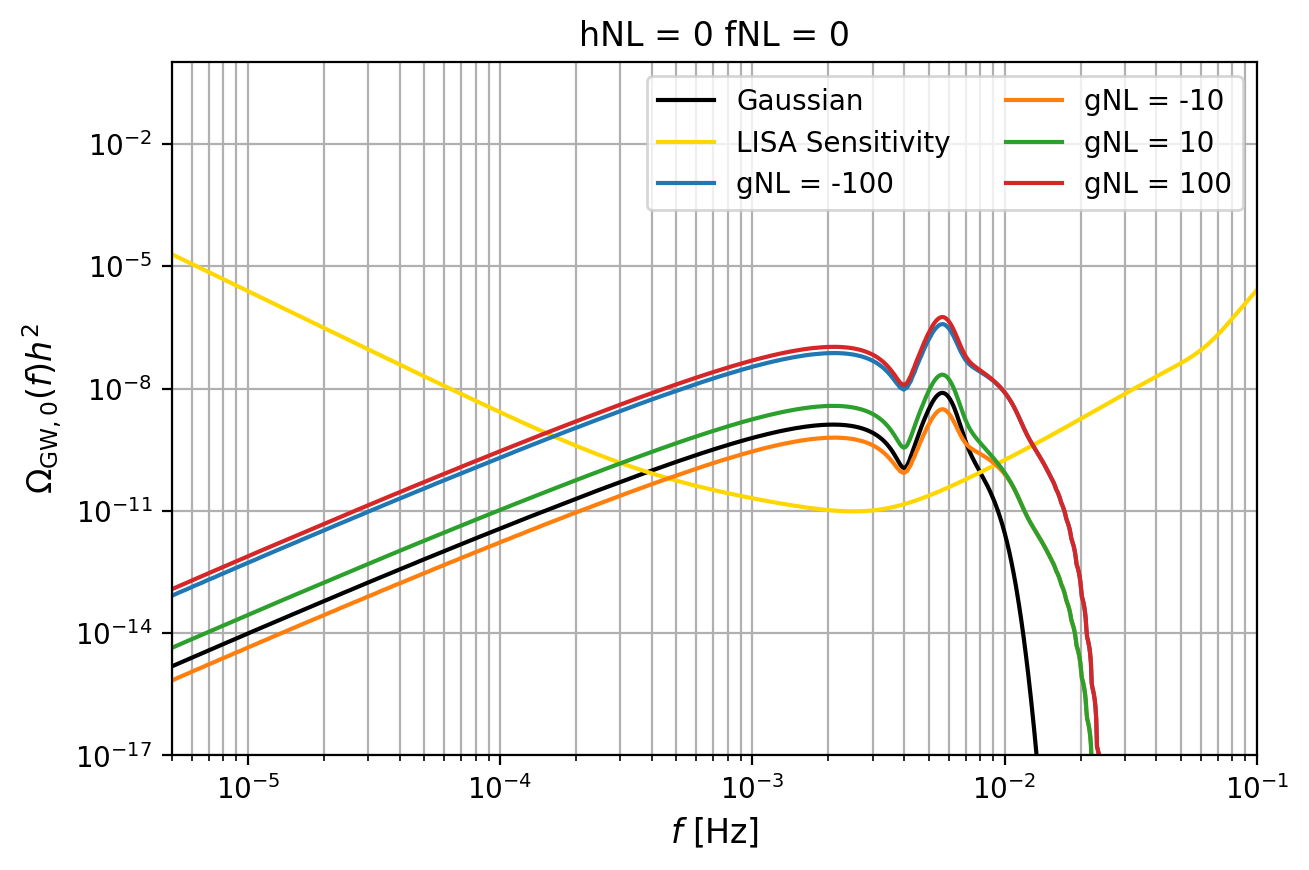}
    \caption{The figure shows the effects of the $g_{\rm NL}$ contributions to the spectrum (in absolute value): we fix all the other nG parameter to 0. The main effect of this contribution consist in a shift of the spectrum coming from the ``new'' contribution and a modification of the UV tail coming from the $g_{\rm NL}^2$ one.}
    \label{Fig::Varying_gNL}
\end{figure}

We report in Figure \ref{Fig::Varying_gNL} the spectra obtained by varying $g_{\rm NL}$ to different values. Since we now consider only $g_{\rm NL}$ to be non-vanishing, the corrections arise only from the $g_{\rm NL}^2$ terms and the ``new'' term. As expected, the main effect amounts in a shift of the spectrum upward or downward, depending on the value of the nG parameter. A further enhancement can be appreciated in the UV tail and it comes mainly from the $g_{\rm NL}^2$ term, that decreases more slowly for high values of the frequency. Furthermore, even if in principle subdominant, this latter term (when normalized) reaches values of order 100 around the peak as shown in Figure \ref{Fig::gNL2}. So when $\mathcal{A}_{\mathcal{R}} = 10^{-2}$, around the peak the spectrum is able to compensate a $\mathcal{A}_{\mathcal{R}}$ factor and this contribution becomes comparable to the ``new'' one. For lower values of the amplitude such effects can still provide a non-negligible contribution to the spectrum, depending on the value of $g_{\rm NL}$. Of course the shift upward and downward is degenerate with the amplitude $\mathcal{A}_{\mathcal{R}}$ and consequently with the Gaussian term, but the further UV bump comes mainly from the $g_{\rm NL}^2$ term, that contributes with a different frequency shape and in principle can be used to break such a degeneracy.

\begin{figure}
    \centering
    \includegraphics[width=1\textwidth]{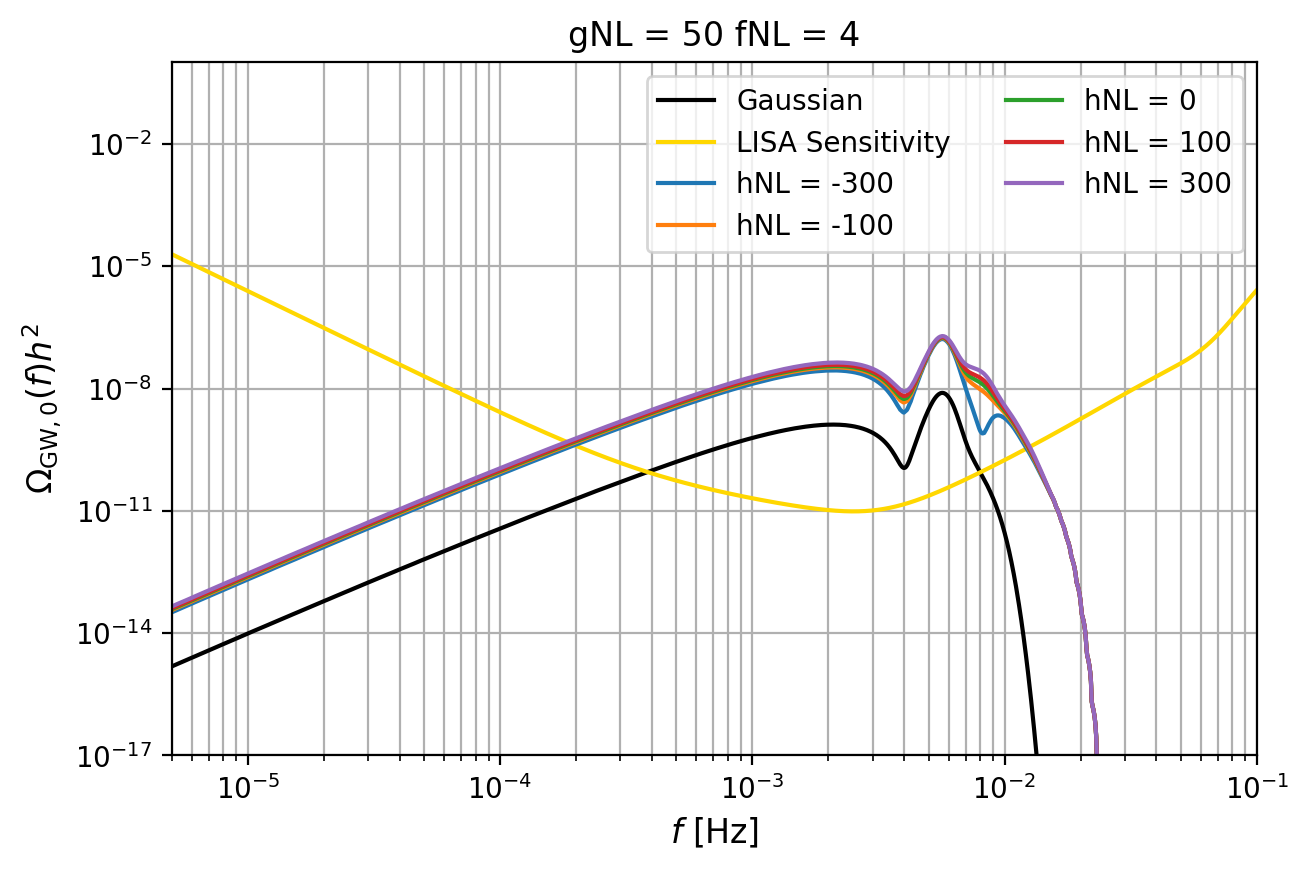}
    \caption{The figure shows the effect of $h_{\rm NL}$ one the spectrum. Since these terms are present only as products of $f_{\rm NL}h_{\rm NL}$, a non-vanishing $f_{\rm NL}$ is required. We fix $g_{\rm NL} = 50$ and $f_{\rm NL} = 4$.}
    \label{Fig::Varying_hNL}
\end{figure}

Then we report in Figure \ref{Fig::Varying_hNL} the spectra obtained varying $h_{\rm NL}$ after fixing $g_{\rm NL}= 50$ and $f_{\rm NL} = 4$. In order to appreciate the effect of the $h_{\rm NL}$ contribution it is necessary to have a sufficiently high and non-vanishing value of $f_{\rm NL}$. Furthermore, if on one hand this contribution dominates with respect to the $i_{\rm NL}$ and $f_{\rm NL}^4$ ones, on the other hand it is comparable to the $f_{\rm NL}^2 g_{\rm NL}$ one and clearly subdominant with respect to the $g_{\rm NL}$ ones, as it can be appreciated from the Figures reported before. For the chosen values of $g_{\rm NL}$ and $f_{\rm NL}$, when $h_{\rm NL} = \pm 100$ the corresponding contribution will be still subdominant with respect to the $g_{\rm NL}$ one: the orange, green and red curve are almost superimposed. When considering a higher value for $h_{\rm NL}$\footnote{The order of magnitude needed to make this contribution appreciable can be understood easily, considering that $f_{\rm NL}h_{\rm NL}$ has to be at least of order $\sim 10 \mathcal{A}_{\mathcal{R}}^{-2}$ to overcome the $\mathcal{A}_{\mathcal{R}}^2$ factor and then it has to be high enough to be comparable to the eventual $g_{\rm NL}$ contribution.}, keeping $f_{\rm NL}\sim\mathcal{O}(1)$ we can easily observe that the corresponding contribution modifies the spectrum accordingly. For positive values of the parameter, the ``double peak'' feature that characterizes the $f_{\rm NL}h_{\rm NL}$ contribution becomes important and generates a small bump in the purple spectrum reported in the Figure. This effect then rapidly decreases (as the $f_{\rm NL}h_{\rm NL}$ spectrum goes rapidly to 0 after the ``double peak'') and the $g_{\rm NL}$ behaviour dominates. For negative values of the parameter $h_{\rm NL}$, such contributions cause a suppression of the spectrum leading to troughs where in the positive case the spectrum had an enhancement (e.g. after the peak where the effect of the $h_{\rm NL}$ term becomes more important). We recall that the term generating this contributions are exactly the same that give rise to the $f_{\rm NL}^2$ one, multiplied by a proper factor, as shown before. This in principle makes the $h_{\rm NL}$ and $f_{\rm NL}$ parameters degenerate at this order since they leave the same imprint on the spectrum. Due to the presence of further terms depending on the latter parameter at the next-to-next-to-leading order, such a degeneracy could in principle be broken.

\begin{figure}
    \centering
    \includegraphics[width=1\textwidth]{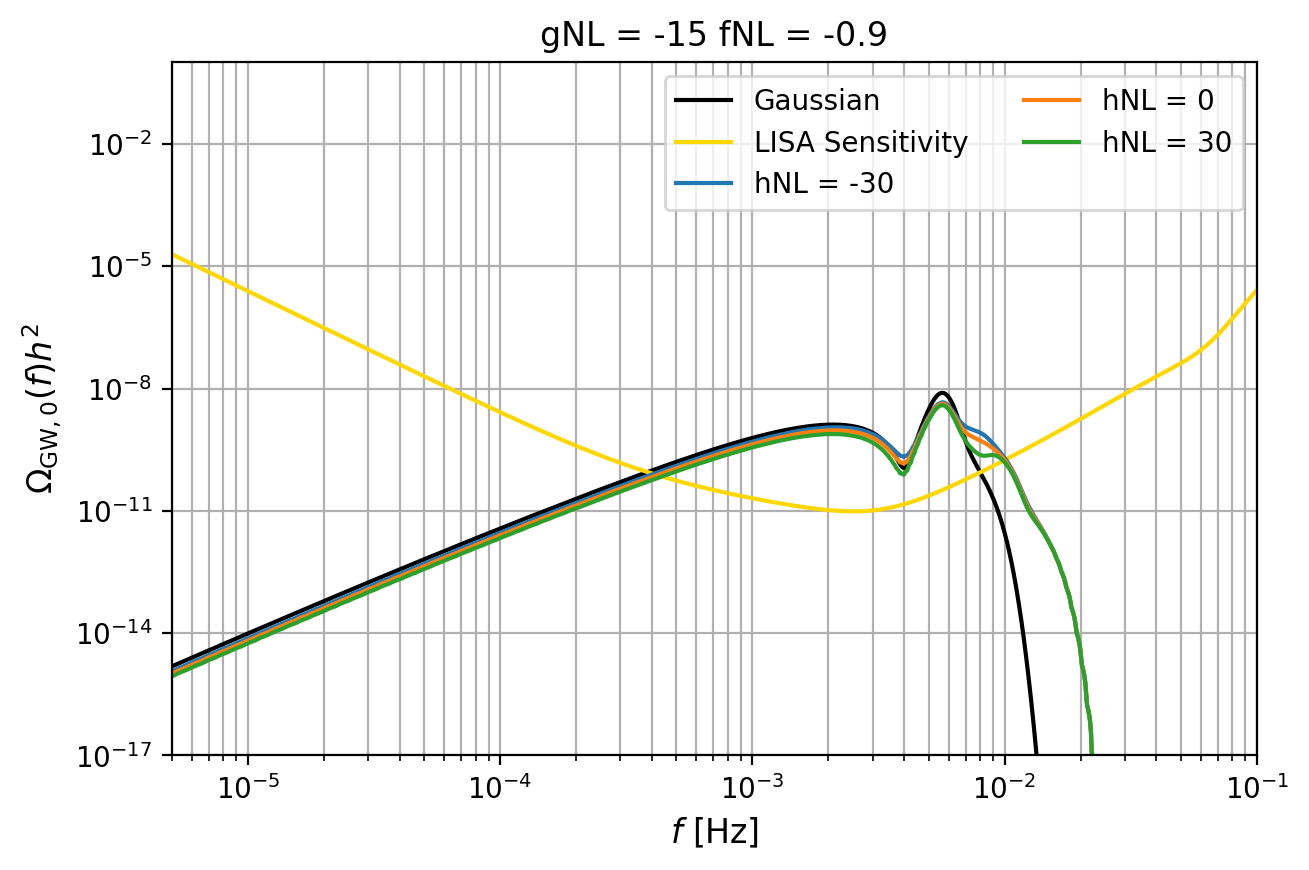}
    \caption{The plot shows the effect of $h_{\rm NL}$ for different values of the parameter. We show this plot to underline that without assuming a hierarchy between the nG parameters, it is still possible to observe the effects of nG without any strong deviation from the Gaussian spectrum.}
    \label{Fig::Varying_hNL_2}
\end{figure}

Finally we report in Figure \ref{Fig::Varying_hNL_2} different spectra obtained fixing the values of $g_{\rm NL}$ and $f_{\rm NL}$ to $-15$ and $-0.9$ respectively and varying $h_{\rm NL}$. We recall that we fix $i_{\rm NL} = 0$ just because its effect consist in enhancing the Gaussian spectrum, but such effect becomes relevant only when $i_{\rm NL}\sim\mathcal{A}_\mathcal{R}^{-2}$. We observe that even without assuming a hierarchical behaviour between the nG parameters, it is possible to obtain a spectrum of the same order of the Gaussian one (corresponding to the black curve in the figure), but with some nG features that still can be observable in the spectrum. The most relevant effects due to primordial nG can be appreciated in the UV tail where the $f_{\rm NL}h_{\rm NL}$ and the $g_{\rm NL}$ contributions originate a bump after the resonance peak. The $h_{\rm NL}$ terms, depending on the sign of the nG parameter, can enhance or suppress the spectrum, as it can be observed comparing the blue or the green one with the orange line (corresponding to $h_{\rm NL}=0$). In this latter case, being the $h_{\rm NL}$ contributions vanishing and the $f_{\rm NL}$ ones always positive, the suppression in the IR tail comes mainly from $g_{\rm NL}$ contributions. 

We also verified that for the amplitude chosen in this work, when considering Planck best fit values for $f_{\rm NL}$ and $g_{\rm NL}$, the nG corrections lead to a strong enhancement of the spectrum that overcomes the bounds on the energy density of GWs coming from Big Bang Nucleosynthesis \cite{Caprini:2018mtu}. 

\section{Fisher forecast analysis}
\label{Fishersection}

In this section we report the theoretical steps at the base of the Fisher forecast that we perform both on the shape parameters (i.e $A_{\mathcal{R}}$, $\sigma$, $f_*$) and on the nG ones (i.e. $f_{\rm NL}$, $g_{\rm NL}$, $h_{\rm NL}$, $i_{\rm NL}$). This method allows to predict the uncertainty we could obtain with future experiments on the parameters of interest of the model. The Fisher information matrix is defined as

\begin{equation}
     F_{\alpha\beta} = \left\langle-\frac{\partial^2 \ln \mathcal{L}}{\partial\lambda_{\alpha}\partial\lambda_{\beta}} \right\rangle \bigg|_{\vec{\lambda} = \vec{\lambda}_0},
\end{equation}
with $\lambda_\alpha$ the parameters considered, $\vec{\lambda}_0$ their best fit values and $\mathcal{L}$ the likelihood. We consider a Gaussian likelihood 
\begin{equation}
    \ln \mathcal{L} =-\frac{N_c}{2}\sum_{i}\sum_{k}\frac{\left(\mathcal{D}_{i}^{(k)}-\mathcal{D}_{i}^{(k), \rm{th}}\right)^2}{\sigma_{i}^{(k) 2}}
\end{equation}
where $i$ runs over the three LISA (TDI) channels and $k$ runs over frequency bins $f_k$.  $N_c$ is the number of data segments in the analysis. Notice that since we are using the  AET basis the likelihood reduces to the sum of the diagonal elements (in channel space). The variance can be expressed in terms of the theoretical ansatz as $\sigma_{i}^{(k) 2} = \left(\mathcal{D}_{i}^{(k), \rm{th}}\right)^2$ and $\mathcal{D}_{i}^{(k), \rm{th}} (f_k, \vec{\lambda}) = \mathcal{R}_{ii}(f_k)h^2 \Omega_{\rm GW}(f_k, \vec{\lambda}) + N^{\Omega}_{ii}(f_k)$
with $\mathcal{R}_{ij}(f_k)$ the response function and $N^{\Omega}_{ij}(f_k)$ the noise associated to the TDI channels. 
Hence we obtain

\begin{equation}
    F_{\alpha\beta} = T_d \sum_{i \in \{A,E,T\}} \int_{f_{\rm min}}^{f_{\rm max}} \frac{\partial \mathcal{D}_{i}^{\rm{th}}}{\partial \lambda_\alpha} \frac{\partial \mathcal{D}_{i}^{\rm{th}}}{\partial \lambda_\beta} \frac{1}{\sigma_i^{2}}df
\end{equation}
where $T_d$ is the total observation time and $\sigma$ accounts for the noise in all the three TDI channels. We assume the signal to be centered in the LISA band, a mission duration $T_d = 4$ years, a fixed LISA noise model and different values for the nG parameters (taking into account the Planck bounds). In the following we report the results of the forecast obtained firstly accounting only for contributions at order $\mathcal{A}_{\mathcal{R}}^3$ and then adding the ones at order $\mathcal{A}_{\mathcal{R}}^4$, to observe the effect of the nG corrections on the forecast.

\subsection{Forecast up to \texorpdfstring{$\mathcal{O}$}{O}\texorpdfstring{$(\mathcal{A}_{\mathcal{R}}^3)$}{A}}
\label{SubSec::FisherA3}
\begin{figure}
    \centering
    \includegraphics[width=0.6\textwidth]{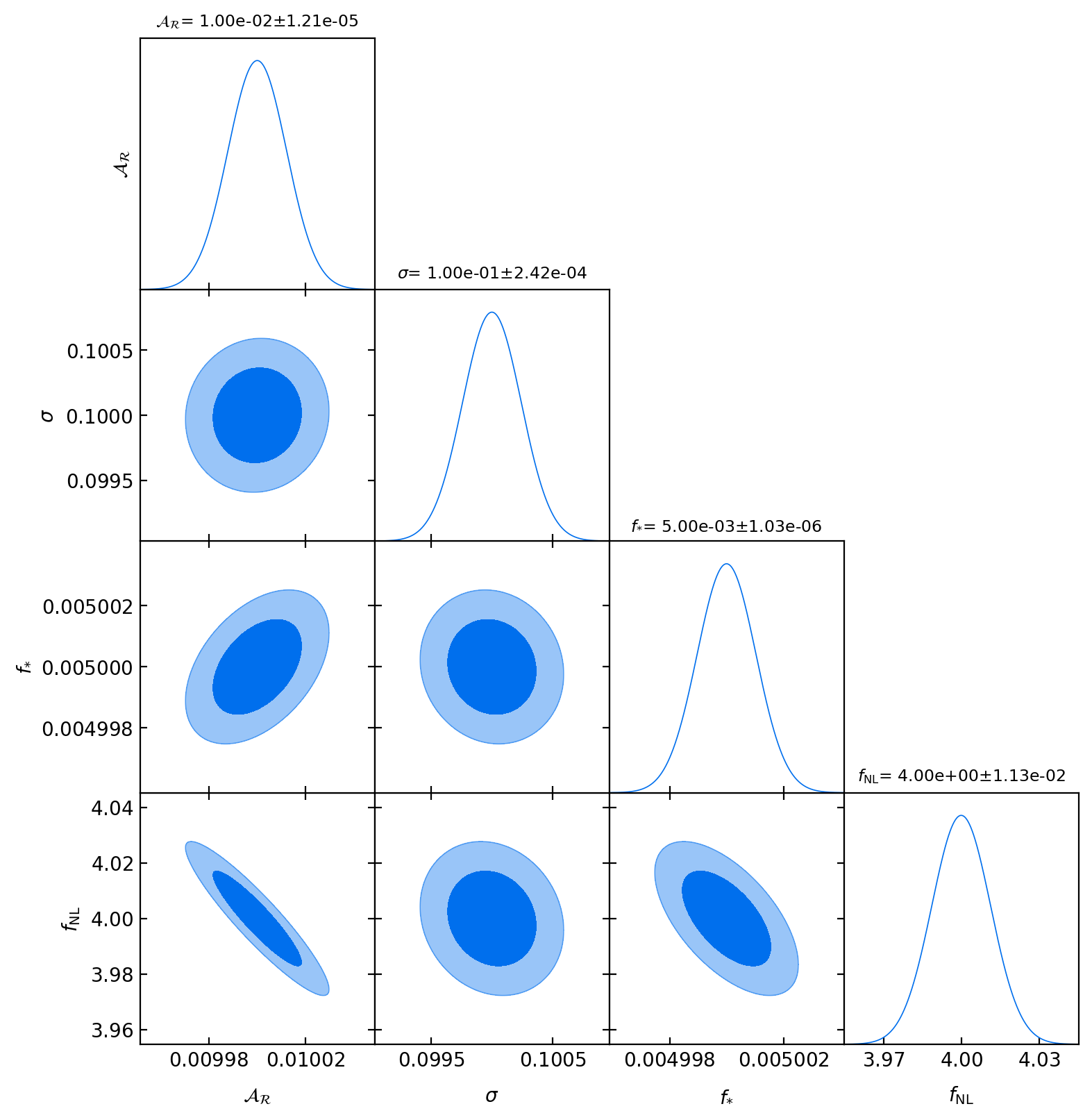}
    \caption{We report the triangular plot obtained from the Fisher forecast for $\mathcal{A}_{\mathcal{R}} = 10^{-2}$ and $f_{\rm NL} = 4$, $\sigma = 0.1$ and $f_* = 0.005$ Hz. The other nG parameters are fixed to 0.}
    \label{Fig::Fisher_Ar3_v4}
\end{figure}

We perform a Fisher forecast firstly accounting for all the contributions up to order $\mathcal{A}_{\mathcal{R}}^3$ (hence only the ones $\propto f_{\rm NL}^2$ and in principle the ``new'' term). The shape parameters are specific of the chosen seed (log-normal in this work) and leave very specific imprints on the spectrum itself. The nG parameters, on the other hand, leave different features, especially on the tails of $\Omega_{\rm GW}$, as explained in the previous section. Furthermore, the nG parameters themselves are multiplicative parameters, hence regulating the amplitude of the contributions they generate, while the different features they generate in the spectrum depend on the interaction of the corresponding scalar perturbations in the scalar trispectrum (i.e. on the various combinations giving rise to the different topologies in the Feynman diagrams at each order). We consider $\mathcal{A}_{\mathcal{R}} = 10^{-2}$, $\sigma = 0.1$ and $f_* = 5 \cdot 10^{-3}$ Hz for the shape parameters and $f_{\rm NL} = 4$, compatible with Planck best fit values (at 68\% CL). For what regards $g_{\rm NL}$, due to its degeneracy with the Gaussian contribution (see subsection \ref{SubSec::new} and the corresponding Feynman diagram), its inclusion in the forecast would generate a singular matrix or could completely spoil any constraint on $f_{\rm NL}$ and $\mathcal{A}_{\mathcal{R}}$. We consider $g_{\rm NL}$ to be vanishing, but we specify that due to its relative importance, fixing it to a different value could affect (worsening or improving) the constraints on $\mathcal{A}_{\mathcal{R}}$ and $f_{\rm NL}$. The results obtained are reported in Figure \ref{Fig::Fisher_Ar3_v4}. We observe that it is possible to obtain tight bounds, on all the parameters considered. The shape parameters can be bound with a relative error up to $10^{-4}$, while $f_{\rm NL}$, can be constrained up to $\mathcal{O}(10^{-3})$ level. We notice the presence of a degeneracy between $\mathcal{A}_{\mathcal{R}}$ and $f_{\rm NL}$, both of them being parameters that affect the magnitude of the spectrum itself (this behaviour is thus expected with all the nG parameters considered). Such a degeneracy is broken by the different shape (e.g., the ``double peak'' feature) that the different nG contributions leave on the spectrum with respect to the Gaussian case (proportional only to the amplitude).

\subsection{Forecast up to \texorpdfstring{$\mathcal{O}$}{O}\texorpdfstring{$(\mathcal{A}_{\mathcal{R}}^4)$}{A}}

\begin{figure}[t!]
    \centering
    \includegraphics[width=\textwidth]{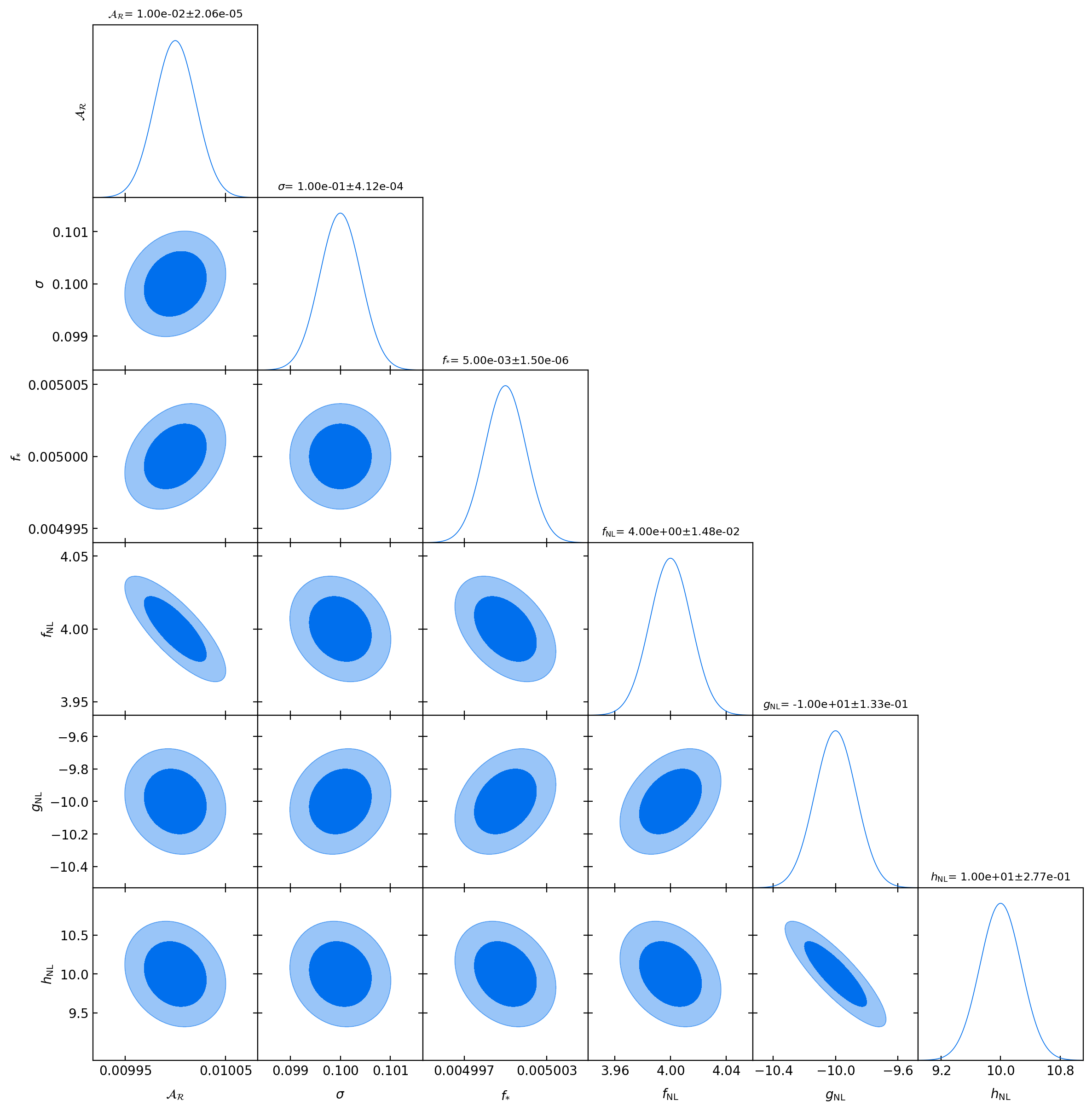}
    \caption{We report the triangular plot obtained from the Fisher forecast for all the parameters included in this work. We just fix $i_{\rm NL} = 0$, without including it in the forecast.}
    \label{Fig::Fisher_Ar4_v1}
\end{figure}

In this subsection, instead, we report the outcome of the Fisher forecast obtained including all the contributions up to order $\mathcal{A}_{\mathcal{R}}^4$. We keep the same values considered in the previous subsection for all the parameters and we add $g_{\rm NL} = -10$, compatible with Planck best fit values (at 68\% CL) and we input $h_{\rm NL} = 10$ as an example. Similarly to the $g_{\rm NL}$ case, at this order the $i_{\rm NL}$ correction is again degenerate with the Gaussian one. For this reason, we consider it to be vanishing. The results are reported in Figure \ref{Fig::Fisher_Ar4_v1}.

Even if on one hand we expect that all the bounds to worsen due to the larger parameter space, on the other hand we find that, due to the higher order contributions, it is still possible to place quite tight constraints (ranging from almost $\mathcal{O}(10^{-3})$ accuracy in the worst case, up to $10^{-4}$ in the best case) on $\mathcal{A}_{\mathcal{R}}$, $\sigma$ and $f_*$. This can be understood, again, recalling that they provide very specific features to the spectrum (like the position of the resonance peak). Furthermore we observe that it is still possible to measure $f_{\rm NL}$ at $\mathcal{O}(10^{-3})$, while $g_{\rm NL}$ and $h_{\rm NL}$ up to $\mathcal{O}(10^{-2})$ . The presence of terms $\propto h_{\rm NL} f_{\rm NL}$ and $\propto g_{\rm NL}$, in fact, induces further peaks or bumps in the spectrum with respect to the case with $h_{\rm NL}=0$, whose shape depends on the features of the scalar spectrum considered. Thus they add more information that helps in improving the constraints on both the shape parameters and the nG ones. 

We observe a slight degeneracy of $h_{\rm NL}$ with $f_{\rm NL}$ originating from the full degeneracy among the $h_{\rm NL}f_{\rm NL}$ terms and the $f_{\rm NL}^2$ ones. As reported in Subsection \ref{SubSec::fNLhNL} and as it can be observed from the Feynman diagrams, these contributions arise exactly from the same combination of the u, t and hybrid terms hence, modulo a different multiplicative constant, the frequency shape of these contributions is completely identical, making it difficult to distinguish the two. The degeneracy is possibly broken by the presence of further different $f_{\rm NL}$-dependent terms like the $f_{\rm NL}^4$ and $f_{\rm NL}g_{\rm NL}^2$ ones (this can be understood with the presence of new topologies in the Feynman diagrams). The importance of these contributions, of course, depends on the values chosen for the nG parameters. We conclude underlying that the presence of $g_{\rm NL}^2$ and $g_{\rm NL} f_{\rm NL}^2$ terms at this order is able to break the degeneracy on $g_{\rm NL}$ found at the previous order, allowing to impose very tight constraints on such a parameter without spoiling the whole forecast. 

\section{Primordial Black Holes Implications}
\label{pbh section}
We conclude discussing the PBH implications related to the production of the SIGW spectra discussed in this work. The presence of large initial fluctuations can lead to the formation of overdense region as the former enter the horizon and their value exceeds the collapse threshold. Hence, after the collapse, these regions can form PBHs and since this process occurs at the horizon re-entry of the perturbed region \cite{2017PhRvD..95l3510I}, this explains why PBHs constitute a counterpart of the SIGWB \cite{Saito:2008jc, Bugaev:2010bb, Inomata:2018epa,  Lu:2019sti}. Their formation depends on the peak value of the primordial scalar perturbation and the masses generated are determined by the corresponding frequency \cite{Saito:2008jc}. It is possible to evaluate the fraction of the Universe that turns into PBH at the epoch of formation following \cite{Young:2022phe, Ferrante:2022mui}
\begin{equation}
\label{Eq::Beta_PBH}
    \beta_{\rm NG} = \frac{\rho_{\rm PBH}}{\rho_{\rm tot}}\Big|_{\rm form} = \int_{\mathcal{D}} \mathcal{K}(\mathcal{C}-\mathcal{C}_{\rm th})^{\gamma} {\rm{P}}_{\rm G}(\mathcal{C}_{\rm G},\zeta_{\rm G}) d\mathcal{C}_{\rm G} d\zeta_{\rm G}
\end{equation}
accounting for both the non-linearity, arising from the non-linear relation between the curvature perturbation and the density contrast, and the primordial nG (in this work accounted with the local expansion). The joint probability density function  for the Gaussian curvature perturbation $\zeta_{\rm G}$ and the Gaussian compaction function $\mathcal{C}_{\rm G}$ in equation \eqref{Eq::Beta_PBH} yields 
\begin{equation}
    {\rm P}_{\rm G} (\mathcal{C}_{\rm G}, \zeta_{\rm G}) = \frac{1}{(2 \pi) \sigma_c \sigma_r \sqrt{1-\gamma_{cr}^2}}\exp{\left(\frac{\zeta_{\rm G}}{2 \sigma^2_{r}}\right)} \exp{\left[ - \frac{1}{2(1-\gamma_{cr}^2)} \left( \frac{\mathcal{C}_{\rm G}}{\sigma_c} - \frac{\gamma_{cr} \zeta_{\rm G}}{\sigma_r}\right)^2\right]}\,,
\end{equation}
where in the last expression $\sigma_c$, $\sigma_r$ are the variances respectively of $\mathcal{C}_{\rm G}$ and $\zeta_{\rm G}$, while $\gamma_{cr}$ is defined as
\begin{equation}
    \gamma_{cr} = \frac{\sigma^2_{cr}}{\sigma_c \sigma_r}\,.
\end{equation}
The compaction function $\mathcal{C}$, instead, results \cite{Ferrante:2022mui}
\begin{equation}
    \mathcal{C} = \mathcal{C}_{\rm G} \frac{d \zeta^{\rm NG}}{d \zeta_{\rm G}} - \frac{1}{4 \Phi} \mathcal{C}_{\rm G}^2 \left(\frac{d \zeta^{\rm NG}}{d \zeta_{\rm G}}\right)^2
\end{equation}
Following \cite{Ferrante:2022mui}, we take $\mathcal{K} = 3.3$ for a log-normal power spectrum and $\gamma = 0.36$. Regarding the threshold $\mathcal{C}_{\rm th}$, that depends on the width $\sigma$ of the power spectrum, we follow the prescription of \cite{2021PhRvD.103f3538M}, while the domain of integration yields $\mathcal{D} = \{ \mathcal{C}_{\rm G}, \zeta_{\rm G} \in \mathbb{R} \,: \mathcal{C}(\mathcal{C}_{\rm G},\zeta_{\rm G}) > \mathcal{C}_{\rm th}\, \land\, \mathcal{C}_1(\mathcal{C}_{\rm G},\zeta_{\rm G}) < 2 \Phi \}$. More details about the quantities and the constants needed to evaluate the domain and the functions to evaluate the integral can be found in the mentioned papers. Substituting the same nG parameters considered in the previous Section, i.e. $f_{\rm NL} = 4$, $g_{\rm NL} = -10$, $h_{\rm NL} = 10$ and fixing $i_{\rm NL} = 10$ as an example, we find $\beta_{\rm NG} \simeq 4.4 \cdot 10^{-9}$ at the formation time. In addition we verified that when considering the case $r_{\rm dec} = 0.4$ as in \cite{Ferrante:2022mui} for the curvaton model, chosen as an example, and stopping up to the $g_{\rm NL}$ contribution (hence $N = 3$ in the reference paper, corresponding to impose vanishing coefficients for the $\zeta_{\rm G}^4$, $\zeta_{\rm G}^5$, $\dots$ terms) we get $\beta_{\rm NG}\sim10^{-15}$ in accordance with their results\footnote{We add that this value, even when adding the further terms of the expansion, will not change substantially, since as the authors show for the value of $\sigma$ chosen, the expansion itself has already converged.}. When keeping the same coefficients of the $\zeta_{\rm G}^2$ and $\zeta_{\rm G}^3$ terms, but considering $-10$ and 50 for the $\zeta_{\rm G}^4$, $\zeta_{\rm G}^5$ terms\footnote{Note that in this case we are not specifically considering the curvaton model, as done by the authors of \cite{Ferrante:2022mui}, since in such a case the value of the coefficients of the expansion is fixed depending on $r_{dec}$. We just aim to remark how variations of the model can impact on the amount of PBH and possibly help in excluding (or validating) some models of inflation.} we get $\beta_{\rm NG} \simeq 2 \cdot 10^{-16}$, almost an order of magnitude less than the corresponding value in the curvaton model. This result further underlines the impact of primordial non-Gaussianity on the formation of PBHs.

\section{Conclusions and Discussion}\label{conclusionsection}

In this work we studied the signatures and the impact of primordial nG on the SIGW signal, expected to be among the sources present in the LISA band. 
We considered the nG local expansion up to $\mathcal{R}^5_g$. In this way we accounted for all the contributions to the tensor power spectrum up to the next-to-next-to leading order. 

We found additional components with respect to  \cite{2021JCAP...10..080A} that did not account for any $g_{\rm NL}$ contribution, like the ``new'' term at Next-To-Leading order.
This latter is of particular importance since it depends linearly on $g_{\rm NL}$ and thus on its sign. We specify that some of the contributions we found (i.e., the disconnected ones) are related to corrections to the Gaussian statistics encoded in the power spectrum and should be distinguished from the genuine nG ones (i.e., coming from the connected trispectrum) \cite{2023JCAP...03..057G}.\\
Often in literature a hierarchy among the nG parameters is assumed, leading to neglect the corresponding contributions. However, even assuming a hierarchy among the nG parameters (e.g., $g_{\rm NL}\ll f_{\rm NL}$), we show that, especially around the peak of the spectra, the $g_{\rm NL}$ contribution is an order of magnitude greater than the $f_{\rm NL}^2$ one when all the parameters are normalized to 1. So, these terms can provide a valuable contribution. Furthermore, we find that at $\mathcal{O}(\mathcal{A}_{\mathcal{R}}^3$) the only non-vanishing $g_{\rm NL}$ contribution arises from a disconnected term in the scalar trispectrum and can be rewritten in terms of the (tensorial) Gaussian contribution. On the one hand, this simplifies the numerical analysis,  but, on the other hand, it provides a possible source of degeneracy between the amplitude and the nG parameters. The same line of reasoning does not apply for the $f_{\rm NL}$ terms at this order, since they show some particular features (the ``double peak'' in the UV tail) that enhances the spectrum at high frequencies.

 Going to the next order in the expansion $\mathcal{O}(\mathcal{A}_{\mathcal{R}}^4$), we observe the presence of further additional terms that in principle could help in breaking the degeneracies. The spectra that we find are all the possible ones at this order, so no further corrections can arise. Any higher power in $\mathcal{R}$ in the local expansion, when building the tensor power spectrum, leads to contributions at order $\mathcal{A}_{\mathcal{{R}}}^5$ (or 10-point correlation function in $\mathcal{R}$). 
These new terms, for high enough values of the nG parameters and for the parameters we consider in this work, can still produce valuable contributions to the final spectrum. The $g_{\rm NL}^2$ term, for instance, for the values allowed by Planck, can provide a dominant contribution comparable to the leading order. We note the presence of terms depending linearly on the various nG parameters and so in principle sensitive to their sign. We studied the effects of each non-Gaussian parameters and we reported some examples of the spectra and of the different features that could arise if any of the various contributions dominate.

Finally, we performed a Fisher forecast to estimate the capability of the future space-based interferometer LISA to constrain the various parameters. We consider values of $f_{\rm NL}=4$ and $g_{\rm NL}=-10$ compatible with Planck best fit values (at 68\% CL) and we chose $h_{\rm NL}=10$ as an example. We also considered three further parameters in the forecast, i.e. $\mathcal{A}_{\mathcal{R}}$, $\sigma$ and $f_*$, that we input respectively to the reference values of $10^{-2}$, 0.1 and $0.005$ Hz. We firstly perform the analysis considering only the contributions up to the next-to leading order, neglecting the $g_{\rm NL}$ contribution since it is degenerate with the amplitude (being exactly proportional to the Gaussian spectrum). We find that LISA would be able to constrain the shape parameters up to order $\mathcal{O}(10^{-4})$, while $f_{\rm NL}$ can be measured with precision of $\mathcal{O}(10^{-3})$. Even with the addition of further nG parameters we find that LISA will still be able to constrain the shape parameters up to $\mathcal{O}(10^{-4})$, due to the different effects they provide on the spectrum, that are strictly dependent on the position of the peak, on the amplitude and on the width. Also the other nG parameters, instead, can be constrained up to $\mathcal{O}(10^{-3})$. The inclusion of astrophysical foregrounds, expected to be present in the LISA band, could also impact our estimates \cite{Babak:2023lro}.

We also verified the implications of the nG parameters on the amount of PBHs, following \cite{Young:2022phe, Ferrante:2022mui}. For the values of the nG parameters considered in the Fisher forecast we find $\beta_{\rm NG} \simeq 4.4 \cdot 10^{-9}$. In addition we reproduced the results for the curvaton model obtained by the authors of \cite{Ferrante:2022mui}. With the purpose of underlying the effects of primordial nG in the abundance of PBH, we consider different nG coefficients with respect to the reference paper obtaining a variation in the amount of PBH of almost one order of magnitude, underlying how PBH and correspondingly SIGW could be important in validating (or excluding) models of primordial inflation.

We conclude mentioning that in principle $\langle h^{\lambda,(2)}_{ij}h^{\lambda,(3)}_{ij}\rangle$ or higher order terms in the tensor power spectrum can be responsible for further non-negligible contributions, as shown in \cite{chang2023new}. These additional terms could possibly affect the constraints on the nG parameters and have an impact on our results. Being interested mainly in the contributions coming from the leading order in the tensor spectrum, we leave the analysis of these additional terms for a future work.

\acknowledgments

G.P. thanks Alessandro Greco for useful discussions. C.T. thanks Ottavia Truttero and Matteo Peron for their valuable advice. The authors thank Gabriele Franciolini, Juan Garcia-Bellido, Mauro Pieroni, Sam Young, Sébastien Renaux-Petel and Antonio Riotto for their valuable comments and feedbacks. A.R. acknowledges financial support from the Supporting Talent in ReSearch@University of Padova (STARS@UNIPD) for the project ``Constraining Cosmology and Astrophysics with Gravitational Waves, Cosmic Microwave Background and Large-Scale Structure cross-correlations''. SM acknowledges support from the COSMOS network (www.cosmosnet.it) through the ASI (Italian Space Agency) Grants 2016-24-H.0, 2016-24-H.1-2018 and 2020-9-HH.0.

\newpage
\appendix

\section{Manipulation of integrals for the numerical computation}\label{manipulation}

This appendix retrieves the main steps to rewrite the integrals of the components of the GW spectrum following e.g. \cite{2021JCAP...10..080A}. We take the ``t'' component, equation \eqref{t}, as an example, underlying that the same reasoning can be applied to each of the other terms.
We start from 
\begin{equation}
    \begin{aligned}
    \sum_{\lambda=+,\times}\overline{P_{h,\lambda}(k,\eta)}\big|_{t} =& \hspace{0.1cm} 2^8 f_{\rm NL}^2 \int \frac{d^3 q_1}{(2\pi)^3} \int \frac{d^3 q_2}{(2\pi)^3}\sum_{\lambda=+,\times}\left[Q_{\lambda}(\textbf{k},\textbf{q}_1)Q_{\lambda}(\textbf{k},\textbf{q}_2)\right]\\
    &\hspace{0.1cm}\times\overline{I(|\textbf{k}-\textbf{q}_1|,q_1,\eta)I(|\textbf{k}-\textbf{q}_2|,q_2,\eta)}  P_{\mathcal{R}_g}(q_2)P_{\mathcal{R}_g}(|\textbf{k}-\textbf{q}_2|)P_{\mathcal{R}_g}(|\textbf{q}_1-\textbf{q}_2|)\,.
    \end{aligned}
    \label{inizio}
\end{equation}
In order to evaluate the final spectrum of SIGW we sum over polarizations and we perform the oscillation-average of each term contributing to the power spectrum, i.e. equations \eqref{g}-\eqref{inl-disc}. Then we insert the results in the expression of the spectral density of GW, equation \eqref{omega}, and sum over all the contributions to get the corrections to the ``Gaussian'' term at different orders.

\subsection*{First change of variables}
We start setting
\begin{equation}
\begin{aligned}
    \textbf{k} &= k(0,0,1)\,,\\
    \textbf{q}_1 &= q_1(\cos\phi_1 \sin\theta_1,  \sin\phi_1 \sin\theta_1, \cos\theta_1)\hspace{0.3cm} \\
    \textbf{q}_2 &= q_2(\cos\phi_2 \sin\theta_2,  \sin\phi_2 \sin\theta_2, \cos\theta_2)\,,
\end{aligned}
\end{equation}
where $\textbf{k}$ is aligned along $\hat{z}$ and all the angles are referred to it. Firstly we rewrite the 3D integral in $d^3q_i$ in spherical coordinates as
\begin{equation}
\begin{aligned}
    \int d^3q_i =& \int_0^{\infty}q_i^2dq_i\int_0^{\pi}\sin \theta_i d\theta_i\int_0^{2\pi}d\phi_i =  \int_0^{\infty}q_i^2dq_i\int_{-1}^{1}d\cos\theta_i\int_0^{2\pi}d\phi_i \,.
\end{aligned}
\end{equation}
and then we introduce a first suitable change of variables
\begin{align}
   &v_i = \frac{q_i}{k}\,, \\
   &u_i = \frac{|\textbf{k}-\textbf{q}_i|}{k} = (1+v_i^2-2v_i\cos\theta_i)^{1/2} = \left(1+v_i^2-2\hspace{0.1cm}\frac{\textbf{k}\cdot\textbf{q}_i}{k^2}\right)^{1/2}\,,
\end{align}
whose determinant for Jacobian is det$J_i = -ku_i/v_i$. The integrals than become
\begin{equation}
    \int d^3q_i = k^3\int_0^{\infty}dv_i\int_{|1-v_i|}^{1+v_i} du_i\hspace{0.1cm} u_i v_i\int_0^{2\pi}d\phi_i\,.
\end{equation}

\subsection*{Polarization-summed projection factor}
After introducing these new variables, it is useful to rewrite the projection factor, defined as
\begin{equation}
    Q_{\lambda}(\textbf{k},\textbf{q}) = \frac{q^2}{\sqrt{2}}\sin^2\theta\times
    \begin{cases}
    \cos2\phi, \hspace{0.2cm} \lambda = + \\
    \sin2\phi, \hspace{0.2cm}\lambda = \times
    \end{cases}\,.
\end{equation}
Summing over the polarizations one obtains
\begin{equation}
    \begin{aligned}
    \sum_{\lambda=+,\times}\Big[Q^2_{\lambda}(\textbf{k},\textbf{q})\Big] = \frac{q^4}{2}\sin^4\theta\cos^2 2\phi+\frac{q^4}{2}\sin^4\theta\sin^2 2\phi = \frac{q^4}{2}\sin^4\theta\,.
    \end{aligned}
\end{equation}
With the new coordinates $u_i$ and $v_i$, we get
\begin{align}
    \cos\theta_i =& \frac{1+v_i^2-u_i^2}{2v_i}\hspace{0.3cm}\\
    \sin^2\theta_i =& \frac{4v_i^2-(1+v_i^2-u_i^2)^2}{4v_i^2}\,,
\end{align}
and the sum becomes
\begin{equation}
    \begin{aligned}
    \sum_{\lambda=+,\times}\Big[Q^2_{\lambda}(\textbf{k},\textbf{q})\Big] = \frac{v^4k^4}{2}\left(\frac{4v^2-(1+v^2-u^2)^2}{4v^2}\right)^2\,.
    \end{aligned}
\end{equation}
The term in equation \eqref{inizio}, instead, reads
\begin{equation}
    \begin{aligned}
    &\sum_{\lambda=+,\times}\left[Q_{\lambda}(\textbf{k},\textbf{q}_1)Q_{\lambda}(\textbf{k},\textbf{q}_2)\right] = \hspace{0.1cm}\frac{q_1^2q_2^2}{2}\sin^2\theta_1\sin^2\theta_2(\cos2\phi_1\cos2\phi_2 + \sin2\phi_1\sin2\phi_2)\\
    &\hspace{4.5cm}= \hspace{0.1cm} \frac{q_1^2q_2^2}{2}\sin^2\theta_1\sin^2\theta_2\cos(2(\phi_1-\phi_2))\\
    &\hspace{1cm}= \hspace{0.1cm} \frac{k^4}{2}v_1^2v_2^2\left(\frac{4v_1^2-(1+v_1^2-u_1^2)^2}{4v_1^2}\right)\left(\frac{4v_2^2-(1+v_2^2-u_2^2)^2}{4v_2^2}\right)\cos(2(\phi_1-\phi_2))\,.
    \end{aligned}
\end{equation}

\subsection*{Kernel}

For what regards the kernel $I(|\textbf{k}-\textbf{q}|,q,\eta)$ we refer to \cite{2018PhRvD..97l3532K} and the Appendix A of \cite{2021JCAP...10..080A}. The kernel is defined as
\begin{equation}
    I(|\textbf{k}-\textbf{q}|,q,\eta) = I(p,q,\eta) = \int_{\eta_i}^{\eta}d\Bar{\eta}G_{\textbf{k}}(\eta,\Bar{\eta})\frac{a(\Bar{\eta})}{a(\eta)}f(p,q,\Bar{\eta})\,.
    \label{ker}
\end{equation}
To build this quantity it is necessary to get the solution to the equation of motion of the transfer function $\phi(k\eta)$ which appears inside the growing mode $f(p,q,\eta)$.

In absence of isocurvature perturbations, the Newtonian potential $\Phi(\textbf{k},\eta)$ in equation (\ref{primordial}) evolves according to
\begin{equation}
    \Phi''(\textbf{k},\eta) + 3(1+w)\mathcal{H}\Phi'(\textbf{k},\eta)+wk^2\Phi(\textbf{k},\eta)=0\,.
\end{equation}
which, exploiting the relation $\mathcal{H}(\eta) = 2/(\eta(1+3w))$ and a change of variable $y=\sqrt{w}k\eta$, can be rewritten as
\begin{equation}
    \frac{d^2\Phi}{dy^2}+\frac{6(1+w)}{1+3w}\frac{1}{y}\frac{d\Phi}{dy} + \Phi=0 \hspace{0.3cm} \rightarrow \hspace{0.3cm} \frac{d^2\Phi}{dy^2}+2\gamma\frac{1}{y}\frac{d\Phi}{dy} + \Phi=0 \,.
\end{equation}
In the last step we defined $\gamma=\frac{3(1+w)}{1+3w}$. When $w\neq 0$, the solutions are given in terms of the Spherical Bessel functions of the first and second kind $j_{\nu}$ and $y_{\nu}$
\begin{equation}
    \phi(k\eta) =\frac{C_1j_{\gamma-1}(y)+C_2y_{\gamma-1}(y)}{y^{\gamma-1}}\,.
\end{equation}
Imposing the super-horizon initial conditions $\phi(k\eta \rightarrow 0) = 1$ and $\partial_x\phi(k\eta \rightarrow 0)=0$, it is possible to fix the two constants $C_1$ and $C_2$ exploiting the limiting forms of the Spherical Bessel functions
\begin{equation}
    \phi(k\eta \rightarrow 0) = 1 = \frac{1}{y^{\gamma-1}}(C_1j_{\gamma-1}(y\rightarrow0)+C_2y_{\gamma-1}(y\rightarrow0) \sim \frac{1}{y^{\gamma-1}}\left(C_1y^{\gamma-1}+\frac{C_2}{y^{\gamma}}\right)\,,
    \end{equation}
and hence $C_1 =1$ and $C_2=0$.\par
The frequencies at play are sufficiently high to consider that the corresponding scales re-enter the horizon before the epoch of equivalence. This allows to fix $w=1/3$, corresponding to scalar-induced GW produced in radiation domination. Hence $\gamma=2$ and the relevant Spherical Bessel function is $j_1(z) = \frac{\sin z}{z^2} - \frac{\cos z}{z}$, resulting in
\begin{equation}
     \phi(k\eta) =\frac{1}{y^3}(\sin y - y\cos y) = \left(\frac{\sqrt{3}}{k\eta}\right)^3\left[\sin\Big(\frac{k\eta}{\sqrt{3}}\Big)-\frac{k\eta}{\sqrt{3}}\cos\Big(\frac{k\eta}{\sqrt{3}}\Big)\right]\,.
\end{equation}
The growing mode then becomes
\begin{equation}
    \begin{aligned}
    f_{RD}(p,q,\eta) = &\hspace{0.1cm}\frac{4}{3}\frac{1}{p^3q^3\eta^6}\Big[18pq\eta^2\cos\Big(\frac{p\eta}{\sqrt{3}}\Big)\cos\Big(\frac{q\eta}{\sqrt{3}}\Big)+ 2\sqrt{3}p\eta(q^2\eta^2-9)\cos\Big(\frac{p\eta}{\sqrt{3}}\Big)\sin\Big(\frac{q\eta}{\sqrt{3}}\Big)\\
    &\hspace{1.7cm}+2\sqrt{3}q\eta(p^2\eta^2-9)\sin\Big(\frac{p\eta}{\sqrt{3}}\Big)\cos\Big(\frac{q\eta}{\sqrt{3}}\Big)\\
    &\hspace{1.7cm}+\big(54-6(q^2+p^2)\eta^2+p^2q^2\eta^4\big)\sin\Big(\frac{p\eta}{\sqrt{3}}\Big)\sin \Big(\frac{q\eta}{\sqrt{3}}\Big)\Big]\,.
    \end{aligned}
\end{equation}
Finally we evaluate the Green's function $G_{\textbf{k}}(\eta,\Bar{\eta})$. The general expression for it, taken from Appendix A in \cite{2021JCAP...10..080A}, results
\begin{equation}
    G_{\textbf{k}}(\eta,\Bar{\eta}) = k\eta\Bar{\eta}\hspace{0.1cm}[j_{\alpha-1}(k\Bar{\eta})y_{\alpha-1}(k\eta)-j_{\alpha-1}(k\eta)y_{\alpha-1}(k\Bar{\eta})]\,.
\end{equation}
Again, in radiation domination $\alpha = 2/(1+3w) = 1$, and hence the only relevant Spherical Bessel functions are
$j_0(z) = \sin z/z$ and $y_0(z) = -\cos z/z$. Thus
\begin{equation}
    G_{\textbf{k},RD}(\eta,\Bar{\eta}) = \frac{\sin k(\eta-\Bar{\eta})}{k}\,.
\end{equation}
The last factor in (\ref{ker}), $a(\Bar{\eta})/a(\eta)$, can be computed exploiting the relation $a(\eta)=a_0\Big(\frac{\eta}{\eta_0}\Big)^{\alpha}$, with $\alpha = 2/(1+3w)$
\begin{equation}
    \frac{a(\Bar{\eta})}{a(\eta)} = \frac{a_0\Big(\frac{\Bar{\eta}}{\eta_0}\Big)^{\alpha}}{a_0\Big(\frac{\eta}{\eta_0}\Big)^{\alpha}} = \Big(\frac{\Bar{\eta}}{\eta}\Big)^{\alpha}\,,
\end{equation}
hence $a(\Bar{\eta})/a(\eta) = \Bar{\eta}/\eta$ in radiation domination. Putting everything together, the kernel becomes
\begin{equation}
    I_{RD}(p,q,\eta) = \int_{\eta_i}^{\eta}d\Bar{\eta}\hspace{0.1cm}\frac{\sin k(\eta-\Bar{\eta})}{k}\frac{\Bar{\eta}}{\eta}f_{RD}(p,q,\Bar{\eta})\,.
\end{equation}
Performing now the change of variables $u = p/k = |\textbf{k}-\textbf{q}|/k$, $v=q/k$, $x=k\eta$ and $\Bar{x}=k\Bar{\eta}$\,,
\begin{equation}
    I_{RD}(u,v,x) = \frac{1}{k^2} \int_{x_i}^{x}d\Bar{x}\hspace{0.1cm}\sin(x-\Bar{x})\frac{\Bar{x}}{x}f_{RD}(u,v,\Bar{x})\,.
\end{equation}
We then introduce, as in \cite{2021JCAP...10..080A}, $\Tilde{I}(u,v,x)\equiv k^2 I(vk,uk,x/k) = k^2 I(p,q,\eta)$, which is required to compute the observable GW spectrum, since its oscillation average has a known asymptotic limit for $x\gg1$, corresponding to the ``late-time'' limit or, equivalently, to consider scales which are significantly inside the horizon. It results
\begin{equation}
\begin{aligned}
    \overline{\Tilde{I}_{RD}(u_1,v_1,x\rightarrow\infty)\Tilde{I}_{RD}(u_2,v_2,x\rightarrow\infty)} = \frac{1}{2x^2}\Tilde{I}_A(u_1,v_1)\Tilde{I}_A(u_2,v_2)&\Big[\Tilde{I}_B(u_1,v_1)\Tilde{I}_B(u_2,v_2)\\
    &+\pi^2\Tilde{I}_C(u_1,v_1)\Tilde{I}_C(u_2,v_2)\Big]    \,,
\end{aligned}
\end{equation}
where
\begin{align}
    &\Tilde{I}_A(u,v)=\frac{3(u^2+v^2-3)}{4u^3v^3}\,,\\
    &\Tilde{I}_B(u,v)=-4uv+(u^2+v^2-3)\ln\Big|\frac{3-(u+v)^2}{3-(u-v)^2}\Big|\\
    &\Tilde{I}_C(u,v)=(u^2+v^2-3)\Theta(u+v-\sqrt{3})\,.
\end{align}
With this in mind, the oscillation average of the square of the kernel, in radiation domination and in the sub-horizon limit, is
\begin{equation}
    \begin{aligned}
    \overline{I_{RD}^2(p,q,\eta)} =&\hspace{0.1cm} \frac{1}{k^4}\overline{\Tilde{I}_{RD}(u,v,x\rightarrow\infty)\Tilde{I}_{RD}(u,v,x\rightarrow\infty)} = \frac{1}{2x^2k^4}\Big(\frac{3(u^2+v^2-3)}{4u^3v^3}\Big)^2
    \\
    \times\Big[&\Big(-4uv+(u^2+v^2-3)\ln\Big|\frac{3-(u+v)^2}{3-(u-v)^2}\Big|\Big)^2+\pi^2(u^2+v^2-3)^2\Theta(u+v-\sqrt{3})\Big]    \,,
    \end{aligned}
\end{equation}
while in the general case in which there are $u_1, v_1, u_2$ and $v_2$, as in equation \eqref{inizio}, we get
\begin{equation}
    \begin{aligned}
    &\overline{I_{RD}(|\textbf{k}-\textbf{q}_1|,q_1,\eta)I_{RD}(|\textbf{k}-\textbf{q}_2|,q_2,\eta)} = \hspace{0.1cm} \frac{1}{k^4}\overline{\Tilde{I}_{RD}(u_1,v_1,x\rightarrow\infty)\Tilde{I}_{RD}(u_2,v_2,x\rightarrow\infty)} \\
    &\hspace{0.5cm}=\hspace{0.1cm}\frac{1}{2x^2k^4}\frac{3(u_1^2+v_1^2-3)}{4u_1^3v_1^3}\frac{3(u_2^2+v_2^2-3)}{4u_2^3v_2^3}\\
    &\hspace{1cm}\times\Big[\Big(-4u_1v_1+(u_1^2+v_1^2-3)\ln\Big|\frac{3-(u_1+v_1)^2}{3-(u_1-v_1)^2}\Big|\Big)\\
    &\hspace{2cm}\times\Big(-4u_2v_2+(u_2^2+v_2^2-3)\ln\Big|\frac{3-(u_2+v_2)^2}{3-(u_2-v_2)^2}\Big|\Big)\\
    &\hspace{2.5cm}+\pi^2(u_1^2+v_1^2-3)(u_2^2+v_2^2-3)\Theta(u_1+v_1-\sqrt{3})\Theta(u_2+v_2-\sqrt{3})\Big]\,.
    \end{aligned}
\end{equation}
From this point on, the notation $RD$ will be dropped for simplicity, but radiation domination is always implied in this work.

\subsection*{Resulting Integral}
We now conclude introducing $\varphi_{12} = \phi_1-\phi_2$ and defining
\begin{equation}
\begin{aligned}
    w_{a,12} =& \frac{|\textbf{q}_1-\textbf{q}_2|}{k} = (v_1^2+v_2^2-2v_1v_2(\cos\theta_1\cos\theta_2+\sin\theta_1\sin\theta_2\cos\varphi_{12}))^{1/2} \\
    =& \left(v_1^2+v_2^2-2\frac{\textbf{q}_1\cdot\textbf{q}_2}{k^2}\right)^{1/2}   
\end{aligned}
\end{equation}
and
\begin{equation}
    \Tilde{J}(u_i,v_i,x) = v_i^2\sin^2\theta_i\Tilde{I}(u_i,v_i,x) = \frac{4v_i^2-(1+v_i^2-u_i^2)^2}{4}\Tilde{I}(u_i,v_i,x)       \,. 
\end{equation}
Equation (\ref{inizio}) can thus be rewritten as
\begin{equation}
    \begin{aligned}
    &\sum_{\lambda=+,\times}\overline{P_{h,\lambda}(k,\eta)}\big|_{t} = \hspace{0.1cm} \frac{2^5\pi}{k^3}  f_{\rm NL}^2 \int_{0}^{\infty} d v_1 \int_{|1-v_1|}^{1+v_1} du_1 \int_{0}^{\infty} d v_2 \int_{|1-v_2|}^{1+v_2} du_2\int_{0}^{2\pi}d\varphi_{12} \\
    &\hspace{3.5cm}\times u_1v_1u_2v_2\left[v_1^2v_2^2\sin^2\theta_1\sin^2\theta_2\cos(2\varphi_{12})\right]\overline{\Tilde{I}(u_1,v_1,x)\Tilde{I}(u_2,v_2,x)}\hspace{0.1cm}\\
    &\hspace{3.5cm}\times  \frac{\Delta^2_{g}(v_2k)}{v_2^3}\frac{\Delta_g^2(u_2k)}{u_2^3}\frac{\Delta^2_g(w_{a,12} k)}{w_{a,12}^3}\\
    &= \hspace{0.1cm} \frac{2^5\pi}{k^3} f_{\rm NL}^2 \int_{0}^{\infty} d v_1 \int_{|1-v_1|}^{1+v_1} du_1 \int_{0}^{\infty} d v_2 \int_{|1-v_2|}^{1+v_2} du_2\int_{0}^{2\pi}d\varphi_{12} \hspace{0.1cm}u_1v_1u_2v_2\cos(2\varphi_{12})\hspace{0.1cm} \\
    &\hspace{3.5cm}\times\overline{\Tilde{J}(u_1,v_1,x)\Tilde{J}(u_2,v_2,x)}\hspace{0.1cm} \frac{\Delta^2_{g}(v_2k)}{v_2^3}\frac{\Delta_g^2(u_2k)}{u_2^3}\frac{\Delta^2_g(w_{a,12} k)}{w_{a,12}^3}\,,
    \end{aligned}
\end{equation}
where we used the definition of dimensionless power spectrum and integrated over the internal angle (that leads to a further $2\pi$ factor).

\subsection*{Second change of variables}
A useful second change of variable can be performed introducing
\begin{equation}
    \begin{aligned}
    s_i =&\hspace{0.1cm} u_i-v_i\,,\\
    t_i =&\hspace{0.1cm} u_i+v_i-1\,.
    \end{aligned}
\end{equation}
This variables are chosen since the corresponding integration domain is rectangular, allowing to simplify the numerical evaluation of the integrals, especially when multidimensional integrals are considered. The determinant of the Jacobian results -1/2.

Finally the integrals can be rewritten as
\begin{equation}
    \int_0^{\infty}dv_i\int_{|1-v_i|}^{1+v_i} du_i = \frac{1}{2}\int_0^{\infty}dt_i\int_{-1}^{1}ds_i\,.
\end{equation}
Hence
\begin{align}
    \cos\theta_i =& \hspace{0.1cm}\frac{1-s_i(1+t_i)}{t_i-s_i+1}\,,\\
    \sin^2\theta_i =&\hspace{0.1cm} \frac{(1-s_i^2)t_i(2+t_i)}{(t_i-s_i+1)^2}\,,
\end{align}
\begin{align}
    \frac{\textbf{q}_1\cdot\textbf{q}_2}{k^2} = & \frac{\cos\varphi_{12}}{4}\sqrt{(1-s_1^2)t_1(2+t_1)(1-s_2^2)t_2(2+t_2)}\nonumber\\
    &+
    \frac{1}{4}(1-s_1(1+t_1))(1-s_2(1+t_2))\,,
\end{align}
\begin{equation}
    \frac{\textbf{k}\cdot\textbf{q}_i}{k^2} = \frac{1}{2}(1-s_i(1+t_i))\,,
\end{equation}
and
\begin{equation}
    \begin{aligned}
    &\overline{I(|\textbf{k}-\textbf{q}_1|,q_1,\eta)I(|\textbf{k}-\textbf{q}_2|,q_2,\eta)} = \hspace{0.1cm}\frac{288}{x^2k^4}\frac{(-5+s_1^2+t_1(2+t_1))}{((1+t_1)^2-s_1^2)^3}\frac{(-5+s_2^2+t_2(2+t_2))}{((1+t_2)^2-s_2^2)^3}\\
    &\hspace{0.5cm}\times\Big[\Big(s_1^2-(1+t_1)^2+\frac{1}{2}(-5+s_1^2+t_1(2+t_1))\ln\Big|\frac{2-t_1(2+t_1)}{3-s_1^2}\Big|\Big)\\
    &\hspace{1.7cm}\times\Big(s_2^2-(1+t_2)^2+\frac{1}{2}(-5+s_2^2+t_2(2+t_2))\ln\Big|\frac{2-t_2(2+t_2)}{3-s_2^2}\Big|\Big)\\
    &\hspace{1.3cm}+\frac{\pi^2}{4}(-5+s_1^2+t_1(2+t_1))(-5+s_2^2+t_2(2+t_2))\Theta(1-\sqrt{3}+t_1)\Theta(1-\sqrt{3}+t_2)\Big]\,.
    \end{aligned}
\end{equation}
The expression for the polarization-summed, oscillation-averaged, ``t'' component of the power spectrum finally reads
\begin{equation}
    \begin{aligned}
    \sum_{\lambda=+,\times}\overline{P_{h,\lambda}(k,\eta)}\big|_{t} =& \hspace{0.1cm} \frac{2^3\pi}{k^3} f_{\rm NL}^2 \int_{0}^{\infty} d t_1 \int_{-1}^{1} ds_1 \int_{0}^{\infty} d t_2 \int_{-1}^{1} ds_2\int_{0}^{2\pi}d\varphi_{12} \cos(2\varphi_{12}) \hspace{0.1cm}u_1v_1u_2v_2\\
    &\hspace{1.5cm}\times\overline{\Tilde{J}(u_1,v_1,x)\Tilde{J}(u_2,v_2,x)} \frac{\Delta^2_{g}(v_2k)}{v_2^3}\frac{\Delta_g^2(u_2k)}{u_2^3}\frac{\Delta^2_g(w_{a,12} k)}{w_{a,12}^3}\,.
    \end{aligned}
\end{equation}

We additionally report as a further example the case of the ``u'' component of the power spectrum. Starting from equation \eqref{u} one obtains
\begin{equation}
    \begin{aligned}
    \sum_{\lambda=+,\times}\overline{P_{h,\lambda}(k,\eta)}\big|_{u} =& \hspace{0.1cm} 2^8 f_{\rm NL}^2 \int \frac{d^3 q_1}{(2\pi)^3} \int \frac{d^3 q_2}{(2\pi)^3}\sum_{\lambda=+,\times}\left[Q_{\lambda}(\textbf{k},\textbf{q}_1)Q_{\lambda}(\textbf{k},\textbf{q}_2)\right]\\
    & \times\overline{I(|\textbf{k}-\textbf{q}_1|,q_1,\eta)I(|\textbf{k}-\textbf{q}_2|,q_2,\eta)} P_{\mathcal{R}_g}(q_1)P_{\mathcal{R}_g}(q_2)P_{\mathcal{R}_g}(|\textbf{k}-(\textbf{q}_1+\textbf{q}_2)|)\,,
    \end{aligned}
\end{equation}
resulting, after the two change of variables, in
\begin{equation}
    \begin{aligned}
    \sum_{\lambda=+,\times}\overline{P_{h,\lambda}(k,\eta)}\big|_{u} =& \hspace{0.1cm} \frac{2^3\pi}{k^3} f_{\rm NL}^2 \int_{0}^{\infty} d t_1 \int_{-1}^{1} ds_1 \int_{0}^{\infty} d t_2 \int_{-1}^{1} ds_2\int_{0}^{2\pi}d\varphi_{12} \cos(2\varphi_{12}) \hspace{0.1cm}u_1v_1u_2v_2\\
    &\hspace{0.5cm}\times\overline{\Tilde{J}(u_1,v_1,x)\Tilde{J}(u_2,v_2,x)} \frac{\Delta^2_{g}(v_1k)}{v_1^3}\frac{\Delta_g^2(v_2k)}{v_2^3}\frac{\Delta^2_g(w_{b,12} k)}{w_{b,12}^3}\,.
    \end{aligned}
\end{equation}
Here we defined 
\begin{equation}
    \begin{aligned}
    w_{b,12} =& \hspace{0.1cm}\frac{|\textbf{k}-(\textbf{q}_1+\textbf{q}_2)|}{k} \\
    =&  \hspace{0.1cm}(1+v_1^2+v_2^2+2v_1v_2(\cos\theta_1\cos\theta_2+\sin\theta_1\sin\theta_2\cos\varphi_{12})-2v_1\cos\theta_1-2v_2\cos\theta_2)^{1/2}\\
    =& \left(1+v_1^2+v_2^2+2\frac{\textbf{q}_1\cdot\textbf{q}_2}{k^2}-2\frac{\textbf{k}\cdot\textbf{q}_1}{k^2}-2\frac{\textbf{k}\cdot\textbf{q}_2}{k^2}\right)^{1/2}\,.\\
\end{aligned}   
\end{equation}

\subsection*{Higher order nG contributions}\label{8}

The procedure for rewriting the integrals corresponding to higher order contributions coming from the 8-point correlation function, equations \eqref{fnl2-gnl}-\eqref{inl-disc} is similar to the one reported above, but with the additional vector $\textbf{q}_3$, defined as
\begin{equation}
\textbf{q}_3 = q_3(\cos\phi_3 \sin\theta_3,  \sin\phi_3 \sin\theta_3, \cos\theta_3)\,.
\end{equation}
Again, some convenient changes of variables are
\begin{equation}
    \begin{aligned}
        q_i =&\hspace{0.1cm} v_ik \\
        \cos \theta_i =&\hspace{0.1cm} \frac{1}{2v_i}(1+v_i^2-u_i^2)\,,
    \end{aligned}
\end{equation}
whose determinant of the Jacobian is $k^3u_1u_2u_3/(v_1v_2v_3)$, and
\begin{equation}
    \begin{aligned}
        u_i =& \hspace{0.1cm}\frac{1}{2}(t_i+s_i+1) \\
        v_i =&\hspace{0.1cm} \frac{1}{2}(t_i-s_i+1)\,,
    \end{aligned}
\end{equation}
with determinant of the Jacobian $1/8$. The projection factors $Q_{\lambda}(\textbf{k},\textbf{q})$ and the kernels $I(|\textbf{k}-\textbf{q}|,q,\eta)$ are unaffected by the presence of the additional integration. For easiness of notation, we introduce the following variables
\begin{equation}
    \begin{aligned}
    w_{a,ij} =&\hspace{0.1cm} \frac{|\textbf{q}_i-\textbf{q}_j|}{k} = \big(v_i^2+v_j^2-2v_iv_j(\cos\theta_i\cos\theta_j+\sin\theta_i\sin\theta_j\cos(\phi_i-\phi_j))\big)^{1/2}\\
    =& \left(v_i^2+v_j^2-2\frac{\textbf{q}_i\cdot\textbf{q}_j}{k^2}\right)^{1/2}
    \end{aligned}
\end{equation}
\begin{equation}
    \begin{aligned}
    w_{b,ij} =& \hspace{0.1cm}\frac{|\textbf{k}-\textbf{q}_i-\textbf{q}_j|}{k} \\
    =&  \hspace{0.1cm}\big(1+v_i^2+v_j^2+2v_iv_j(\cos\theta_i\cos\theta_j+\sin\theta_i\sin\theta_j\cos(\phi_i-\phi_j))-2v_i\cos\theta_i-2v_j\cos\theta_j\big)^{1/2}\\
    =& \left(1+v_i^2+v_j^2+2\frac{\textbf{q}_i\cdot\textbf{q}_j}{k^2}-2\frac{\textbf{k}\cdot\textbf{q}_i}{k^2}-2\frac{\textbf{k}\cdot\textbf{q}_j}{k^2}\right)^{1/2}\\
\end{aligned}   
\end{equation}
\begin{equation}
    \begin{aligned}
    w_{q123} =& \hspace{0.1cm}\frac{|\textbf{q}_1-\textbf{q}_2-\textbf{q}_3|}{k} = \hspace{0.1cm}\big(v_1^2+v_2^2+v_3^2-2v_1v_2(\cos\theta_1\cos\theta_2+\sin\theta_1\sin\theta_2\cos(\phi_1-\phi_2))\\ &\hspace{3.3cm}- 2v_1v_3(\cos\theta_1\cos\theta_3+\sin\theta_1\sin\theta_3\cos(\phi_1-\phi_3))\\
    &\hspace{3.3cm}+ 2v_2v_3(\cos\theta_2\cos\theta_3+\sin\theta_2\sin\theta_3\cos(\phi_2-\phi_3))\big)^{1/2}\\
    =& \left(v_1^2+v_2^2+v_3^2-2\frac{\textbf{q}_1\cdot\textbf{q}_2}{k^2}-2\frac{\textbf{q}_1\cdot\textbf{q}_3}{k^2}+2\frac{\textbf{q}_2\cdot\textbf{q}_3}{k^2}\right)^{1/2}\\
\end{aligned}   
\end{equation}
\begin{equation}
    \begin{aligned}
    w_{123} =& \hspace{0.1cm}\frac{|\textbf{k}-\textbf{q}_1-\textbf{q}_2+\textbf{q}_3|}{k} \\
    =&  \hspace{0.1cm}\big(1+v_1^2+v_2^2+v_3^2+2v_1v_2(\cos\theta_1\cos\theta_2+\sin\theta_1\sin\theta_2\cos(\phi_1-\phi_2))\\ &-2v_1v_3(\cos\theta_1\cos\theta_3+\sin\theta_1\sin\theta_3\cos(\phi_1-\phi_3))\\
    &-2v_2v_3(\cos\theta_2\cos\theta_3+\sin\theta_2\sin\theta_3\cos(\phi_2-\phi_3))\\
    &-2v_1\cos\theta_1-2v_2\cos\theta_2+2v_3\cos\theta_3\big)^{1/2}\\
    =& \left(1+v_1^2+v_2^2+v_3^2+2\frac{\textbf{q}_1\cdot\textbf{q}_2}{k^2}-2\frac{\textbf{q}_1\cdot\textbf{q}_3}{k^2}-2\frac{\textbf{q}_2\cdot\textbf{q}_3}{k^2}-2\frac{\textbf{k}\cdot\textbf{q}_1}{k^2}-2\frac{\textbf{k}\cdot\textbf{q}_2}{k^2}+2\frac{\textbf{k}\cdot\textbf{q}_3}{k^2}\right)^{1/2}\\
\end{aligned}   
\end{equation}

We conclude this appendix with some considerations on the coefficients associated to each integral. In order to obtain an expression for $\Omega_{\rm GW}(k,\eta)$ starting from the corresponding power spectra \eqref{fnl2-gnl}
- \eqref{inl-disc} there is a simple way to find the correct coefficient in front of every components in \eqref{Eq::fNL2gNL_Connected} - \eqref{Eq::iNL_Disconnected}, being the change of variables performed always the same. We consider as an example the third line of \eqref{fnl2-gnl}
\begin{equation} 
    \begin{aligned}
    \sum_{\lambda=+,\times}\overline{P_{h,\lambda}(k,\eta)}\big|_{\rm f^2_{\rm NL}g_{\rm NL}} =& \hspace{0.1cm} 3 \cdot 2^{9} f_{\rm NL}^2g_{\rm NL} \int \frac{d^3 q_1}{(2\pi)^3}\int \frac{d^3 q_2}{(2\pi)^3}\int\frac{d^3 q_3}{(2\pi)^3} \sum_{\lambda=+,\times}\left[Q_{\lambda}(\textbf{k},\textbf{q}_1)Q_{\lambda}(\textbf{k},\textbf{q}_2)\right] \\
    &\hspace{0.5cm}\times \overline{I(|\textbf{k}-\textbf{q}_1|,q_1,\eta)I(|\textbf{k}-\textbf{q}_2|,q_2,\eta)}\\
    &\hspace{0.5cm}\times P_{\mathcal{R}_g}(q_1)P_{\mathcal{R}_g}(|\textbf{q}_1-\textbf{q}_2|)P_{\mathcal{R}_g}(q_3)P_{\mathcal{R}_g}(|\textbf{k}-\textbf{q}_1-\textbf{q}_3|)\hspace{0.1cm}\,.
    \end{aligned}
\end{equation}
After the first change of variable it results
\begin{equation}
    \begin{aligned}
    \sum_{\lambda=+,\times}\overline{P_{h,\lambda}(k,\eta)}\big|_{\rm f^2_{\rm NL}g_{\rm NL}} =& \hspace{0.1cm} 3 \cdot 2^{9} f_{\rm NL}^2g_{\rm NL} \left(\frac{1}{2\pi}\right)^9\int d v_1 \int d v_2 \int d v_3 \int d u_1 \int d u_2 \int d u_3\\
    &\hspace{0.5cm}\times \int d\phi_1 \int d\phi_2 \int d\phi_3 (\hspace{1mm}v_1 k)^2 (v_2 k)^2 (v_3 k)^2 \hspace{0.1cm} k^3 \frac{u_1}{v_1} \frac{u_2}{v_2} \frac{u_3}{v_3}\\
    &\hspace{0.5cm}\times\left[\frac{(v_1 k)^2(v_2 k)^2}{2}\sin^2\theta_1\sin^2\theta_2\cos2(\phi_1-\phi_2)\right]\\
    &\hspace{0.5cm}\times \frac{1}{k^4}\overline{\Tilde{I}(|\textbf{k}-\textbf{q}_1|,q_1,\eta)\Tilde{I}(|\textbf{k}-\textbf{q}_2|,q_2,\eta)}\\
    &\hspace{0.5cm}\times\left(\frac{2\pi^2}{k^3}\right)^4 \frac{\Delta_g^2(v_1 k)}{v_1^2}\frac{\Delta_g^2(w_{a,12} k)}{w_{a,12}^2}\frac{\Delta_g^2(v_3 k)}{v_3^2}\frac{\Delta_g^2(w_{b,13} k)}{w_{b,13}^2}\,.
    \end{aligned}
\end{equation}

while the second change of variables leads to an additional numerical factor $1/2^3$ coming from the Jacobian, hence
\begin{equation}
    \begin{aligned}
    \sum_{\lambda=+,\times}\overline{P_{h,\lambda}(k,\eta)}\big|_{\rm f^2_{\rm NL}g_{\rm NL}} =& \hspace{0.1cm} 3 \cdot 2^{9} f_{\rm NL}^2g_{\rm NL} \left[\left(\frac{1}{2\pi}\right)^9\left(\frac{2\pi^2}{k^3}\right)^4k^6 k^3 \frac{k^4}{2}\frac{1}{k^4}\frac{1}{2^3}\right] \int d t_1 \int d t_2 \int d t_3 \\
    &\hspace{0.3cm}\times\int d s_1 \int d s_2 \int d s_3\int d\phi_1 \int d\phi_2 \int d\phi_3\hspace{0.1cm} v_1 v_2 v_3 u_1 u_2 u_3\\
    &\hspace{0.3cm}\cos2(\phi_1-\phi_2)\times\overline{\Tilde{J}(|\textbf{k}-\textbf{q}_1|,q_1,\eta)\Tilde{J}(|\textbf{k}-\textbf{q}_2|,q_2,\eta)}\\
    &\hspace{0.3cm}\times \frac{\Delta_g^2(v_1 k)}{v_1^2}\frac{\Delta_g^2(w_{a,12} k)}{w_{a,12}^2}\frac{\Delta_g^2(v_3 k)}{v_3^2}\frac{\Delta_g^2(w_{b,13} k)}{w_{b,13}^2}\,.
    \end{aligned}
    \label{es}
\end{equation}
The final numerical factor results $1/(2^9k^3\pi)$ and is \textit{independent} on the momenta enclosed in the primordial seeds. Recalling now equation \eqref{omega}
\begin{equation}
    \Omega_{\rm GW}(k,\eta)  = \frac{1}{48} \left(\frac{k}{a(\eta)H(\eta)}\right)^2 \frac{k^3}{2\pi^2}\sum_{\lambda=+,\times}\overline{P_{h,\lambda}(k,\eta)}\,,
\end{equation}
the exact coefficients for the various expressions of the $\Omega_{\rm GW}$ can be obtained from
\begin{equation}
     \Omega_{\rm GW}(k,\eta)  = \frac{1}{3\cdot2^{14}\pi^3} \left(\frac{k}{a(\eta)H(\eta)}\right)^2 \left(\sum_{\lambda=+,\times}\overline{P_{h,\lambda}(k,\eta)}\right)|_{t_i,s_i}\,,
\end{equation}
where the subscript $t_i$ and $s_i$, means that we are considering just the integrals written in terms of these new variables. In the example above, equation \eqref{es},
\begin{equation}
    \begin{aligned}
    \left(\sum_{\lambda=+,\times}\overline{P_{h,\lambda}(k,\eta)}\big|_{\rm f^2_{\rm NL}g_{\rm NL}} \right)|_{t_i,s_i} =& \hspace{0.1cm} 3 \cdot 2^{9} f_{\rm NL}^2g_{\rm NL} \int d t_1 \int d t_2 \int d t_3 \int d s_1 \int d s_2 \int d s_3\\
    &\hspace{0.3cm}\times\int d\phi_1 \int d\phi_2 \int d\phi_3\hspace{0.1cm} v_1 v_2 v_3 u_1 u_2 u_3\cos2(\phi_1-\phi_2)\\
    &\hspace{0.3cm}\times\overline{\Tilde{J}(|\textbf{k}-\textbf{q}_1|,q_1,\eta)\Tilde{J}(|\textbf{k}-\textbf{q}_2|,q_2,\eta)}\\
    &\hspace{0.3cm}\times \frac{\Delta_g^2(v_1 k)}{v_1^2}\frac{\Delta_g^2(w_{a,12} k)}{w_{a,12}^2}\frac{\Delta_g^2(v_3 k)}{v_3^2}\frac{\Delta_g^2(w_{b,13} k)}{w_{b,13}^2}\,,
    \end{aligned}
\end{equation}
The term $(k/aH)^2$ cancels out with the $1/x^2$ that is present in the expression of the oscillation average of the product of the kernels.
Starting from this simple relation it is possible to retain the coefficient of the various terms of the scalar-induced spectrum. In some cases an additional step to further simplify the expressions involves an extra integration over the internal angles $\phi_i$, when the integrands do not depend on them, giving rise to factors $2\pi$ (as seen for the ``t'' term above). 

\section{Feynman Diagrams}
\label{Sec::Feynman_Diagrams}
We report in this Appendix further details on the Feynman diagrams corresponding to the different contributions to the tensor GW spectrum we found at each order in $\mathcal{A}_{\mathcal{R}}$. We first briefly summarize the Feynman rules necessary to construct such diagrams and then we show how they can be used to predict the behaviour of the corresponding integrals.

Following \cite{2021JCAP...10..080A}, we draw transfer functions with dashed lines, while GW with wiggling lines. To ease the notation we do not indicate the polarization state $\lambda$ and the momentum flow relative to these latter, but in general to each GW a polarization state $\lambda$ or $\lambda'$ and a momentum directed outward (as expected from momentum conservation) are associated.
The general idea is that since by definition the tensor power spectrum is originated by the integration of a trispectrum of scalar perturbations, equation \eqref{source}, that in turn can be linked to products of power spectra, it has to be at leading order at 1-loop. We associate to each power spectrum a straight line, while the momentum flow is described by arrows. Furthermore, each diagram must contain 4 internal vertices, corresponding to each of the 4 nG scalar perturbations $\mathcal{R}^{\rm NG}$ generating the trispectrum. Each vertex relates a certain number of power spectra to a dashed line and the number of power spectra depends on the power of $\mathcal{R}_g$ in the local expansion that originates the contribution. For example, a term originating from a $\propto h_{\rm NL}\mathcal{R}_g^4$ contribution, a $\propto f_{\rm NL}\mathcal{R}_g^2$ one and two linear term, must contain a vertex linked to 4 lines, another one linked to only two lines and the remaining two linked to just one line each. Any integration over a fixed internal momentum leads to a loop and all the momenta are conserved at each vertex. The multiplicity of each diagram can be obtained considering the number of possible ways each vertex can be linked to the others, excluding contributions involving a propagator with vanishing momentum.

\subsection{Gaussian diagram}
\label{SubSec::Feynman_Gauss}
The Gaussian contribution, reported in Figure \ref{Fig::Feynman_Gaussian}, constitutes the leading order in the tensor power spectrum. Since it originates from the interaction of scalar perturbations (and consequently power spectra) it requires at least one integration over the internal momentum linking the different power spectra. This justifies why, as specified in \cite{PhysRevD.99.041301}, this contribution leads to a 1-loop diagram. The diagram reported corresponds to equation \eqref{g}. Since this contribution arises from the a trispectrum component originated by 4 linear contributions in each $\langle\mathcal{R}_g\mathcal{R}_g\mathcal{R}_g\mathcal{R}_g\rangle$, each vertex must be linked to only 1 solid line. Momentum conservation then leads to the diagram built in that way. 

\subsection{\texorpdfstring{$f_{\rm NL}^2$}{f} diagram}
\label{SubSec::Feynman_fNL2}
The diagrams corresponding to the next-to leading order corrections and proportional to $f_{\rm NL}^2$ are reported in Figure \ref{Fig::Feynman_fNL2}. The ones in the first row and the second in the second row are originated by connected contributions, while the first one in the second row is the only disconnected one. On the first line, we have respectively the C and Z contributions in \cite{2021JCAP...10..080A} or the ``t'' and ``u'' in \cite{2023JCAP...03..057G}, reported in equations \eqref{t} and \eqref{u}, the first contribution on the second line the hybrid one, corresponding to equation \eqref{ibrid}, while the last one is the ``s'', equation \eqref{s}. In order to build a trispectrum in this case, as also explained in Section \ref{SubSec::Local_nG}, we need two terms proportional to $f_{\rm NL}^2$ and two simple linear terms. Since each $f_{\rm NL}$ carries a factor $\mathcal{R}_g^2$, the corresponding vertices must be connected to two different lines. At the end, the diagrams need to have two vertices connected with two solid lines and two vertices connected to just a single one. 

It is interesting now to distinguish two different contributions, the ``connected'' and the ``disconnected'' ones, just by looking at the diagrams themselves. The former directly lead to a Dirac delta in the real momenta since the beginning, the latter need an intermediate integration that leads the $Q$ and $I$ functions to depend on the same internal momentum. Thus the connected or disconnected origin of these diagrams can be understood simply by the dependence on $q_2$ in the dashed lines on the right. 

Finally, the fourth diagram vanishes due to rotational invariance. This can be understood by the absence of any internal solid line depending on some combination of the momenta $\vec{q}_1$, $\vec{q}_2$. Mathematically this leads to the absence of any dependence on the the azimuthal angles in the power spectra and so the integration over them just leads to a vanishing contribution.

\subsection{\texorpdfstring{$g_{\rm NL}$}{g} diagrams}
\label{SubSec::Feynman_gNL}
Only two diagrams are present at order $g_{\rm NL}$. The first one in Figure \ref{Fig::Feynman_gNL} is a connected but vanishing contribution, for symmetry reasons. The second diagram, on the other hand, originates from a disconnected contribution and is not vanishing. The ``bubble'' originates from two of the three factors $\mathcal{R}_{g}$ in the $g_{\rm NL}$ term contracting with themselves. Such a contribution can be factorized, since there's no other spectra depending on the corresponding momentum. The remaining diagram has the exact same topology of the Gaussian one, further justifying that the term we call ``new'' can be written as proportional to the Gaussian one. 

\subsection{Higher order contributions}
\label{SubSec::Feynman_Higher}
We then report higher order diagrams corresponding to contributions at order $\mathcal{A}_{\mathcal{R}}^4$, in Figures \ref{Fig::Feynman_fNL2gNL}\ref{Fig::Feynman_gNL2}, \ref{Fig::Feynman_fNL4}, \ref{Fig::Feynman_fNLhNL} and \ref{Fig::Feynman_iNL}. Following the reasoning of the previous subsections, it is clear that the second and the fourth $g_{\rm NL}^2$ diagrams just vanish or that the fifth leads to a contribution proportional to the Gaussian. The remaining two correspond to the first and second line in square brackets of \eqref{gnl2} and the sixth to the last term in \eqref{gnl2-disc}. Again for symmetry reasons,  the first, the second and the ninth diagrams in Figure \ref{Fig::Feynman_fNL2gNL}, the first, third and fifth in Figure \ref{Fig::Feynman_fNLhNL} and the first in Figure \ref{Fig::Feynman_iNL} vanish. Furthermore the last $f_{\rm NL}^2g_{\rm NL}$ diagram and the last $f_{\rm NL}h_{\rm NL}$, corresponding to the disconnected contributions, share the same topology of the hybrid one, after the bubble is factorized. The only non vanishing $i_{\rm NL}$ contribution, again corresponding to the disconnected term, after the bubbles are factored out, shows the same topology of the Gaussian contribution and thus it is proportional to it. The three $f_{\rm NL}^4$ contributions, instead, correspond respectively to the ``planar'', ``non-planar'' and ``reducible'' in \cite{PhysRevD.99.041301,2021JCAP...10..080A}.

\bibliographystyle{JHEP}
\bibliography{bib}

\end{document}